\newif\iffigs\figstrue
\documentclass[12pt]{article}
\usepackage{latexsym,amssymb}
\usepackage[vsentermath]{youngtab}
\iffigs
             \input{epsf}
\else
\fi
\newcommand{\ft}[2]{{\textstyle\frac{#1}{#2}}}

\newsavebox{\uuunit}
\sbox{\uuunit}
                 {\setlength{\unitlength}{0.825em}
                      \begin{picture}(0.6,0.7)
                                      \thinlines
                                      \put(0,0){\line(1,0){0.5}}
                                      \put(0.15,0){\line(0,1){0.7}}
                                      \put(0.35,0){\line(0,1){0.8}}
                                     \multiput(0.3,0.8)(-0.04,-0.02){10}{\rule{0.5pt}{0.5pt}}
                      \end {picture}}

\makeatletter \@addtoreset{equation}{section} \makeatother

\newcommand{\be}{\begin{equation}}
\newcommand{\ee}{\end{equation}}
\newcommand{\ba}{\begin{eqnarray}}
\newcommand{\ea}{\end{eqnarray}}
\def\bfone{\relax{\rm 1\kern-.35em 1}}

\def\bfone{\relax{\rm 1\kern-.35em 1}}
 
\begin{document}
\begin{titlepage}
\begin{flushright}
DFTT 23/05\\
\end{flushright}
\vskip 0.5cm
\begin{center}
{\LARGE \bf  Cosmic Billiards with Painted Walls}\\
\vskip 0.5cm
{\LARGE \bf in Non Maximal Supergravities. } \\
\vskip 0.5cm
{\Large \bf A worked out example$^\dagger$ }\\ \vfill
{\large
Pietro Fr\'e$^1$,  \\
\vskip 0.2cm
Floriana Gargiulo$^2$ and Ksenya Rulik$^1$} \\
\vfill {\small
$^1$ Dipartimento di Fisica Teorica, Universit\'a di Torino, \\
$\&$ INFN -
Sezione di Torino\\
via P. Giuria 1, I-10125 Torino, Italy\\
$^2$ Dipartimento di Fisica Politecnico di Torino,\\
 C.so Duca degli Abruzzi,
24,
I-10129 Torino, Italy
 }
\end{center}
\begin{abstract}
{The derivation of smooth cosmic billiard solutions by means of the compensator
method, introduced by us sometimes ago, is extended to the case of supergravity with non maximal supersymmetry.
Here a new key feature is provided by the non-maximal split nature of the scalar coset manifold. To deal
with this, one has to consider the theory of Tits Satake projections leading to maximal split projected
algebras, where the compensator method can be successfully applied and interesting solutions that
display several smooth bounces can be derived. The generic bouncing feature of all exact solutions can thus be checked.
From the analysis of the Tits Satake projection emerges a regular scheme applicable to all non maximal
supergravity models and in particular a challenging so far unobserved structure, that of the paint group
$\mathbb{G}_{\mathrm{paint}}$. This latter, which is preserved through dimensional reduction, provides a powerful tool
to codify solutions of the exact supergravity theories in terms of solutions of their Tits Satake projected
partners,
which are much simpler and manageable. It appears that the dynamical walls on which
the cosmic ball bounces come actually in
\textit{painted copies} rotated into each other by the paint group. So the effective cosmic dynamics is that dictated
by the maximal split Tits Satake manifold \textit{plus paint}. In the present paper we work out in all minor details
the example provided by $N=6,D=4$ supergravity, whose scalar manifold is the special K\"ahlerian
$\mathrm{SO^\star(12)}/\mathrm{SU(6)\times U(1)}$ c-mapping in $D=3$ to the quaternionic
$\mathrm{E_{7(-5)}}/\mathrm{SO(12)\times SO(3)}$. This choice was not random. It is the next one after maximal
supergravity and at the same time can be reinterpreted in the context of $N=2$ supergravity. We plan
indeed, in a future publication, to apply the results we obtained here, to the discussion of the Tits Satake
projection within the  context of
generic special K\"ahler manifolds. We also comment on the merging of the Tits-Satake projection with the
affine Ka\v c--Moody extension originating in dimensional reduction to $D=2$ and relying on a general field--theoretical
mechanism illustrated by us in a separate paper.}
\end{abstract}
\vspace{2mm} \vfill \hrule width 3.cm {\footnotesize $^ \dagger $
This work is supported in part by the European Union network contract
MRTN-CT-2004-005104}
\end{titlepage}
\section{Introduction}
The cosmological implications of superstring theory have been under
attentive consideration in the last few years from various
viewpoints \cite{cosmicstringliteraturegeneral}. This involves the
classification and the study of possible time-evolving string
backgrounds which amounts to the construction, classification and analysis of supergravity solutions
depending only on time or, more generally, on a low number of coordinates
including time. In this context a quite challenging and potentially
highly relevant phenomenon for the overall interpretation of
extra--dimensions and string dynamics is provided by the so named
\textit{cosmic billiard}
phenomenon
\cite{cosmicprince},\cite{cosmicbilliardliterature1},\cite{cosmicbilliardliterature2},\cite{cosmicbilliardliterature3},
\cite{cosmicbilliardliterature4},\cite{Kassner}.
This is based on a profound link
between the features of time evolution of the cosmological scale factors
and the algebraic structure of  string
theory duality groups. As it is well known, the dualities that unify
the various perturbative quantum string models into a unique M-theory are
elements of a unified group $\mathrm{U}(\mathbb{Z})$ which is the suitable restriction
to integers of a corresponding Lie group $\mathrm{U}(\mathbb{R})$ encoded in
the low energy limit of superstrings, namely supergravity. The group
$\mathrm{U} \equiv \mathrm{U}(\mathbb{R})$ appears as isometry group of the scalar
manifold $\mathcal{M}_{scalar}$ emerging in compactifications of
$10$--dimensional supergravity to lower dimensions and
crucially depends on the geometry of the compact dimensions and on
the number of preserved supersymmetries $N_Q \le 32$.  For $N_Q > 8$ the scalar manifold is always a homogeneous
space $\mathrm{U/H}$ and what actually
happens is that the cosmological scale factors $a_i(t)$ associated with the
various  dimensions of space--time  can be interpreted as exponentials of those
scalar fields $h_i(t)$ which lie in the Cartan subalgebra of
$\mathbb{U}$, while the other scalar fields in $\mathrm{U/H}$ correspond to
positive roots $\alpha >0$ of the Lie algebra $\mathbb{U}$. In this
way the cosmological evolution is described by the motion of a
\textit{fictitious ball} in the CSA of $\mathbb{U}$. This space is
actually a billiard table whose walls are the hyperplanes orthogonal
to the various roots. The fictitious ball bounces on the billiard
walls and this means that there are inversions in the time evolution
of scale factors. Certain dimensions that were expanding almost
suddenly begin to contract and others do the reverse.
Such a scenario was introduced by Damour,
Henneaux, Julia and Nicolai in \cite{cosmicprince},
and in a series of papers with collaborators
\cite{cosmicbilliardliterature2},
\cite{cosmicbilliardliterature3},\cite{cosmicbilliardliterature4},
which generalize classical results obtained in the context of pure
General Relativity
\cite{Kassner},\cite{cosmicbilliardliterature1}.
In this approach the cosmic billiard
phenomenon is analyzed as an asymptotic regime in the neighborhood of
space-like singularities and the billiard walls are seen as delta
function potentials provided by the various $p$--forms of
supergravity localized at sharp instants of time.
\par
It was observed in \cite{noiconsasha} that the fundamental
mathematical setup underlying the appearance of the billiard
phenomenon is the so named \textit{Solvable Lie algebra
parametrization} of supergravity scalar manifolds, pioneered in \cite{primisolvi}
and later applied to the solution of a large variety of superstring/supergravity
problems, including the structure of supersymmetric black-hole solutions
\cite{noie7blackholes},\cite{otherBHpape}, the construction of gauged
SUGRA potentials\cite{gaugesugrafirst}, \cite{gaugedsugrapot} and several other issues (for
a comprehensive review see \cite{myparis}). Indeed we pointed out in
those  papers that, thanks to the solvable parametrization, one can
establish a precise algorithm to implement the following programme:
\begin{enumerate}
  \item Reduce the original supergravity in higher  dimensions $D \ge 4$ (for instance $D=10,11$) to
  a gravity-coupled $\sigma$--model in $D \le 3$ where gravity is
  non--dynamical and where the original higher dimensional bosonic field
  equations reduce to geodesic equations for a solvable
  group-manifold metrically equivalent to a non compact
  coset manifold $\exp \left[ Solv\left(\mathrm{U/H} \right)  \right]
  \cong \mathrm{U/H}$.
  \item Utilize the algebraic structure of the solvable Lie algebra
  $Solv\left(\mathrm{U/H} \right) $ in order to integrate analytically the geodesic
  equations. In particular we introduced in \cite{noiconsasha} a
  general method of integration, named the \textit{$H$-compensator
  method} which reduces the geodesic differential equations to a
  triangular form and hence to quadratures when $\mathrm{U/H}$ is
  maximally split, which is always the case when supersymmetry is
  maximal ($N_Q =32$)
  \item Dimensionally oxide the solutions obtained in this way to
  exact  time dependent solutions of  $D \ge 4$ supergravity. In
  particular we showed in \cite{noiconsasha} that the oxidation process is
  not unique but is algebraically classified by the embedding of \textit{Weyl
  orbits} of subalgebras $\mathbb{G} \subset \mathbb{U}$. Indeed the
  analytic structure of the solution is fully determined only by the algebraic
  structure of $\mathbb{G}$. Its physical interpretation varies and
  depends on the explicit embedding $ \mathbb{G} \mapsto \mathbb{U}$.
  In this way each solution in $D \le 3$ corresponds to an entire
  orbit of higher dimensional backgrounds, very different from one
  another, but dual to each other under transformations of the Weyl
  group $\mathcal{W}\equiv Weyl(\mathrm{U})$.
\end{enumerate}
Within this approach it was proved in \cite{noiconsasha} that the
\textit{cosmic billiard phenomenon} is indeed a general feature of exact time
dependent solutions of supergravity and has \textit{smooth
realizations}. Calling $\mathbf{h}(t)$ the $r$--component
vector of Cartan fields (where $r$ is the rank of $\mathbb{U}$) and
$\mathbf{h}_\alpha(t) \equiv \mathbf{\alpha} \, \cdot \,
 {\mathbf{h}}(t)$ its projection along any positive root $\alpha$,
a \textit{bounce} occurs at those instant of times $t_i$ such that:
\begin{equation}
\exists \,  \mathbf{\alpha} \, \in \, \Delta_+ \, \quad  \backslash
\quad \quad
 \dot{\mathbf{h}}_\alpha (t)\mid_{t=t_i} \, = \, 0
\label{bouncino}
\end{equation}
namely when the Cartan field in the direction of some root $\alpha$ inverts
its behaviour and begins to shrink if it was growing or viceversa
begins to grow if it was shrinking. Since all higher dimensional
bosonic fields (off-diagonal components of the metric $g_{\mu\nu}$ or
$p$--forms $A^{[p]}$) are, via the solvable parametrization of $\mathrm{U/H}$, in one-to-one correspondence with roots
$ \phi_\alpha \, \Leftrightarrow \, \alpha$, it follows that
the bounce on a \textit{wall} (hyperplane orthogonal to the root $\alpha$) is
caused by the sudden growing of that particular field $\phi_\alpha$.
Indeed we showed in \cite{noiconsasha} that in exact smooth solutions
which we were able to obtain by means of the compensator method,  each bounce is associated with
a typical bell-shaped behaviour of the root field $\phi_\alpha$ and
that the whole process can be interpreted as a temporary localization
of the Universe energy density in a \textit{lump} on a spatial brane associated with the
field $\phi_\alpha$.
\par
Although very much encouraging the analysis of \cite{noiconsasha} was still
limited in three respects:
\begin{description}
  \item[a] The dimensional reduction process which is responsible for
making manifest the duality algebra $ \mathbb{U}$ and hence for
creating the whole algebraic machinery utilized in deriving the
\textit{smooth cosmic billiard solutions} was stopped at $D=3$, namely
at the first point where all the bosonic degrees of freedom can be
represented by scalars. In $D=3$, $\mathbb{U}$ is still a finite
dimensional Lie algebra and the whole richness of the underlying
algebraic structure is not yet displayed. As it is well known \cite{Juliaetal_algebras}, in
$D=2$ and $D=1$, the algebra $\mathbb{U}$ becomes a Ka\v c--Moody
algebra, affine or hyperbolic, respectively. The smooth billiard
dynamics has to be reconsidered and extended in view of this.
  \item[b] The constructions of \cite{noiconsasha} depend, in some
  crucial points, on the assumption that the coset $\mathrm{U/H}$ corresponds
  to a pair $\left\{ \mathbb{U}, \mathbb{H} \subset \mathbb{U}\right
  \}$ of Lie algebra and Lie subalgebra which is \textit{maximally split}.
  This is always the case for maximal supersymmetry $N_Q = 32$ but it
  is not true for $N_Q \le 32$. Extending the $H$-compensator method
  to \textit{non maximally split pairs}  $\left\{ \mathbb{U}, \mathbb{H} \subset \mathbb{U}\right
  \}$ is necessary in order to discuss billiard dynamics in lower
  SUSY theories and hence in compactifications of string theory on
  internal manifolds $\mathcal{M}_{internal}$ with restricted holonomies and $G$-structures, with or without
  fluxes.
  \item[c] The solutions considered in \cite{noiconsasha} were
  solutions of the pure $\sigma$--model, namely of pure, ungauged
  supergravity. The extension also to \textit{gauged supergravities}
  is mandatory in order to make contact with potentially realistic
  models, in particular with currently considered flux compactifications.
  \cite{fluxcompacta}
\end{description}
In a recent paper \cite{noiKMpape} we have begun to address point
$\mathbf{a})$ of the above list. There we have shown that the
mechanism outlined several years ago by Nicolai \cite{NicolaiEHMM} as the origin of the
Ka\v c Moody extension of the duality algebra which appears in $D=3$
when you step down to $D=2$, namely the existence of two non-locally related
dimensional reduction schemes $D=4 \mapsto D=2$, the Ehlers reduction
and the Maztner Missner reduction, can be formulated in a general set
up which provides a regular scheme of analysis both at the algebraic and
at the field theoretical level and which applies to all supergravity
theories. In particular we have shown that the $\mathbb{U}_{D=3}$
algebra emerges from the Ehlers reduction and has the following
general decomposition with respect to the $\mathbb{U}_{D=4}$ algebra:
\begin{equation}
\mbox{adj}(\mathbb{U}_{D=3}) =
\mbox{adj}(\mathbb{U}_{D=4})\oplus\mbox{adj}(\mathrm{SL(2,\mathbb{R})_E})\oplus
W_{(2,\mathbf{W})}
\label{gendecompo}
\end{equation}
where $\mathbf{W}$ is a \textit{symplectic} representation of $\mathbb{U}_{D=4}$
determined by the vector fields in the parent $D=4$ supergravity and
$\mathrm{SL(2,\mathbb{R})_E}/\mathrm{O(2)}$ is the target space for a $\sigma$-model which encodes
the degrees of freedom of pure Einstein gravity. Continuing the
Ehlers reduction from $D=3$ to $D=2$ we obtain a Lagrangian with the
same symmetry
\begin{equation}
  \mathbb{U}^{[E]}_{D=2}=\mathbb{U}_{D=3}
\label{UED2}
\end{equation}
 Alternatively, following the Matzner Missner reduction scheme we obtain a \textit{twisted
$\sigma$-model} with symmetry
\begin{equation}
\mathbb{U}^{[MM]}_{D=2}=\mathbb{U}_{D=4} \, \otimes \,
\mathrm{SL(2,\mathbb{R})_{MM}}
\label{UMMD2}
\end{equation}
where
$\mathrm{SL(2,\mathbb{R})_{MM}}/\mathrm{O(2)}$ is the target space for a $\sigma$-model
also encoding the degrees of freedom of pure Einstein gravity. The Matzner
Missner $\mathrm{SL(2,\mathbb{R})_{[MM]}}$ group, however, is not the same as
the Ehlers one $\mathrm{SL(2,\mathbb{R})_{[E]}}$ and $\mathbb{U}^{[MM]}_{D=2}$,  $\mathbb{U}^{[E]}_{D=2}$
are just two different finite dimensional subalgebras
of the same infinite dimensional one $\mathbb{U}_{D=2}$, which is nothing else
but the affine Ka\v c--Moody extension of $\mathbb{U}_{D=3}$:
\begin{equation}
\left.  \begin{array}{cc}
   \mathbb{U}^{[E]}_{D=2}   & \subset \\
   \null& \null \\
   \mathbb{U}^{[MM]}_{D=2}  & \subset \
 \end{array} \right \}\mathbb{U}_{D=2} \, \equiv \, \mathbb{U}_{D=3}^\wedge
\label{wedgeU3}
\end{equation}
Understanding the general pattern for the Ka\v c--Moody extension and mastering its field
theoretical realization provides the necessary basis for the
construction of \textit{smooth billiard solutions} which rely on the
full fledged Lorentzian signature CSA, lying behind supergravity.
This we emphasized and begun to exploit in \cite{noiKMpape}.
\par
In the present paper we address point $\mathbf{b})$ of the list
mentioned above.
\par
Our starting point is \textit{ungauged supergravity} in $D=4$, whose bosonic
lagrangian takes the following general form:
\begin{eqnarray}
\mathcal{L}^{(4)} &=& \sqrt{\mbox{det}\, g}\left[-2R[g] - \frac{1}{6}
\partial_{\hat{\mu}}\phi^a\partial^{\hat{\mu}}\phi^b h_{ab}(\phi) \,
+ \,
\mbox{Im}\mathcal{N}_{\Lambda\Sigma}\, F_{\hat{\mu}\hat{\nu}}^\Lambda
F^{\Sigma|\hat{\mu}\hat{\nu}}\right] \nonumber\\
&&+
\frac{1}{2}\mbox{Re}\mathcal{N}_{\Lambda\Sigma}\, F_{\hat{\mu}\hat{\nu}}^\Lambda
F^{\Sigma}_{\hat{\rho}\hat{\sigma}}\epsilon^{\hat{\mu}\hat{\nu}\hat{\rho}\hat{\sigma}}
\label{d4generlag}
\end{eqnarray}
In eq.(\ref{d4generlag}) $\phi^a$ denotes the whole set of $n_S$ scalar fields
parametrizing the scalar manifold $ \mathcal{M}_{scalar}^{D=4}$
which, for $N_Q \ge 8$, is necessarily a coset manifold:
\begin{equation}
  \mathcal{M}_{scalar}^{D=4} \, =
  \,\frac{\mathrm{U_{D=4}}}{\mathrm{H}}
\label{cosettoquando}
\end{equation}
For $N_Q \le 8$, eq.(\ref{cosettoquando}) is not obligatory but it is
possible. Particularly in the $\mathcal{N}=2$ case, i.e. for $N_Q =8$, a large
variety of homogeneous special K\"ahler or quaternionic manifolds
\cite{specHomgeo} fall into the set up of the present general discussion.
The fields $\phi^a$ have $\sigma$--model interactions
dictated by the metric $h_{ab}(\phi)$ of $\mathcal{M}_{scalar}^{D=4}$. The theory includes also $n$ vector
fields $A_{\hat{\mu}}^\Lambda$ for which
\begin{equation}
  \mathcal{F}^{\pm| \Lambda}_{\hat{\mu}\hat{\nu}} \equiv \ft 12
  \left[{F}^{\Lambda}_{\hat{\mu}\hat{\nu}} \mp \, {\rm i} \, \frac{\sqrt{\mbox{det}\, g}}{2}
  \epsilon_{\hat{\mu}\hat{\nu}\hat{\rho}\hat{\sigma}} \, F^{\hat{\rho}\hat{\sigma}} \right]
\label{Fpiumeno}
\end{equation}
denote the self-dual (respectively antiself-dual) parts of the field-strengths. As displayed in
eq.(\ref{d4generlag}) they are non minimally coupled to the scalars via the symmetric complex matrix
\begin{equation}
  \mathcal{N}_{\Lambda\Sigma}(\phi)
  ={\rm i}\, \mbox{Im}\mathcal{N}_{\Lambda\Sigma}+ \mbox{Re}\mathcal{N}_{\Lambda\Sigma}
\label{scriptaenna}
\end{equation}
which transforms projectively under $\mathrm{U_{D=4}}$. Indeed the field strengths ${F}^{
\Lambda}_{\mu\nu}$ plus their magnetic duals fill up a $2\,
n$--dimensional symplectic representation of $\mathrm{\mathbb{U}_{D=4}}$
which we call by the name of $\mathbf{W}$.
\par
The main point in the analysis of billiard dynamics  for the lower
SUSY cases is that the pair $\left \{ \mathbb{U}_{D=4}, \mathbb{H}
\subset \mathbb{U}_{D=4} \right\} $ is generically \textit{not maximally
split}. This implies that  $\mathbb{U}_{D=3}$, whose decomposition with respect to $U_{D=4}$
is always given by eq.(\ref{gendecompo}) is also not maximally split.
This happens since, in these cases, $\mathbb{U}_{D=4,3}$ is a \textit{real section} of the corresponding
complex Lie algebra $\mathbb{U}(\mathbb{C})$ \textbf{different} from
\textit{the maximally non compact one}. Indeed it is only  for the
maximally non-compact real section that:
\begin{enumerate}
  \item All Cartan generators $\mathcal{H}_i$ are non compact and belong to the
  Solvable Lie algebra: $\forall \,i \quad $ $\mathcal{H}_i  \, \in \, Solv(\mathrm{U/H})$.
  \item All step operators $E^\alpha$ associated with positive roots
   belong to the solvable algebra: $\forall \alpha >0 \,\,$, $E^\alpha  \,
  \in \, Solv(\mathrm{U/H})$.
  \item The maximal compact subalgebra $\mathbb{H} \subset
  \mathbb{U}$ is the span of all generators $E^{\alpha} - E^{-\alpha}$, for all positive roots $\forall
  \alpha >0$.
\end{enumerate}
Since items 1-3 in the above list are essential ingredients in the
algorithm to derive exact solutions developed by us in
\cite{noiconsasha}, it is evident that our set-up has to be
reconsidered carefully in the more general case.
\par
In this paper we make an in depth analysis of a specific example of a non maximally split manifold
${\mathrm{U_{D=4}}}/{\mathrm{H}}$, that of $\mathcal{N}=6$ supergravity, from which we
extrapolate a general  elegant result which reduces the non-maximally
split cases to associated maximally split ones allowing, in this way, the extension
of the compensator method to all values of $N_Q$ and hence the derivation
of exact solutions in all instances.
\par
As we are going to see our present results concerning point $\mathbf{b}$
are quite relevant also for the appropriate discussion of point
$\mathbf{a}$ as well. Indeed the concept of \textit{painted walls} that
will emerge and that of \textit{paint  group} $G_{paint}$  are invariant by
dimensional reduction and apply also to the Ka\v c--Moody extensions.
\par
In the next subsection we summarize the main result of our paper.
\subsection{Tits Satake subalgebras and painted walls}
In the case of non maximally non-compact manifolds $\mathrm{U/H}$ the Lie
algebra $\mathbb{U}$ of the numerator group is some appropriate real form
\begin{equation}
  \mathbb{U} \, = \, \mathbb{G}_R
\label{realformaGR}
\end{equation}
of a complex Lie algebra $\mathbb{G}(\mathbb{C})$ of rank $r
=\mbox{rank}(\mathbb{G})$. The Lie algebra $\mathbb{H}$ of the denominator $\mathrm{H}$ is the
maximal compact subalgebra $\mathbb{H} \subset \mathbb{U}$ which has
typically rank $r_{compact} > r$. Denoting, as usual, by $\mathbb{K}$
the orthogonal complement of $\mathbb{H}$ in $\mathbb{G}_R$:
\begin{equation}
  \mathbb{G}_R = \mathbb{H} \, \oplus \,\mathbb{K}
\label{Grdecompo}
\end{equation}
and defining as non compact rank or rank of the coset $\mathrm{U/H}$ the dimension of the non
compact Cartan subalgebra:
\begin{equation}
  r_{nc}\, = \, \mbox{rank} \left( \mathrm{U/H}\right)  \, \equiv \, \mbox{dim} \,
  \mathcal{H}^{n.c.} \quad ; \quad \mathcal{H}^{n.c.} \, \equiv \,
  \mbox{CSA}_{\mathbb{G}(\mathbb{C})} \, \bigcap \, \mathbb{K}
\label{rncdefi}
\end{equation}
we obtain that $r_{nc} < r$. The manifold $\mathrm{U/H}$ is still
metrically equivalent to a solvable group manifold $M_{Solv} \equiv \exp [Solv(\mathrm{U/H})]$
and the field equations of supergravity still reduce to geodesic
equations in $M_{Solv}$, which can be reformulated as first order
equations by using the constant Nomizu connection (see \cite{noiconsasha}):
\begin{equation}
  \dot{Y}^A + \underbrace{\Gamma^A_{\phantom{A}BC}}_{Nomizu} \, {Y}^B \, {Y}^C \, =
  \, 0
\label{1storder}
\end{equation}
but it is the form of the Solvable Lie algebra
$Solv(\mathrm{U/H})$, whose structure constants define the Nomizu connection,  which is now more complicated and
apparently does not allow the immediate use of the compensator
method for the solution of equations (\ref{1storder}). Yet the system
(\ref{1storder}) can be reduced to an equivalent one which is
maximally split and can be solved with the methods of
\cite{noiconsasha}. This is a consequence of Tits-Satake theory of
non compact cosets and split subalgebras and, within such a
mathematical framework of a peculiar universal structure of the
solvable algebra $Solv(\mathrm{U/H})$ that, up to our knowledge, had not been observed before.
Explicitly we have the following scheme. Splitting the Cartan
subalgebra into its compact and non compact subalgebras:
\begin{equation}
\begin{array}{rcccc}
  \mathrm{CSA}_{\mathbb{G}_{R}} &=& {\rm i} \mathcal{H}^{comp} & \oplus
  &
  \mathcal{H}^{n.c.} \\
  \null & \null &\Updownarrow & \null & \Updownarrow  \\
  \mathrm{CSA}_{\mathbb{G}(\mathbb{C})} &=& \mathcal{H}^{comp} & \oplus
  & \mathcal{H}^{n.c.} \
  \end{array}
\label{hcompnoncom}
\end{equation}
every vector in the dual of the full Cartan subalgebra, in particular
every   root $\alpha$ can be decomposed into its parallel and
transverse part to $\mathcal{H}^{n.c.}$:
\begin{equation}
  \alpha = \alpha_{||} \, \oplus \, \alpha_{\bot}
\label{splittus}
\end{equation}
Setting all $\alpha_{\bot} = 0 $ corresponds to a projection:
\begin{equation}
  \Pi \quad : \quad \Delta_\mathbb{G} \, \mapsto  \, \overline{\Delta}
\label{projecto}
\end{equation}
of the original root system $\Delta_\mathbb{G}$ onto a new system of vectors
living in an euclidean space of dimension equal to the non compact
rank $r_{nc}$. A priori this is not obvious, but it is nonetheless
true that $\overline{\Delta}$ is by itself the root system of a simple Lie
algebra $\mathbb{G}_{TS}$, the Tits-Satake subalgebra of
$\mathbb{G}_R$:
\begin{equation}
  \overline{\Delta} = \mbox{root system of } \mathbb{G}_{TS} \,
  \subset \, \mathbb{G}_R
\label{TitsSatake}
\end{equation}
The Tits-Satake subalgebra $\mathbb{G}_{TS} \, \subset
\,\mathbb{G}_R$ is always the maximally non compact real section of
its own complexification. For this reason, considering its maximal
compact subalgebra $\mathbb{H}_{TS} \, \subset \, \mathbb{G}_{TS}$ we
have a new smaller coset $\mathrm{G_{TS} / H_{TS}}$ which is maximally split
and whose associated solvable algebra $Solv(\mathrm{G_{TS} / H_{TS}})$ has the
standard structure utilized in \cite{noiconsasha} to solve the
differential equations (\ref{1storder}). What is the relation between
the two solvable Lie algebras $Solv(\mathrm{G_{R} / H})$ and $Solv(\mathrm{G_{TS} /
H_{TS}})$? The explicit answer to this question and the illustration
of its relevance for the solution of the geodesic equations
(\ref{1storder}) is the key result of the present paper. It leads to
the concept of \textit{billiards with painted walls} and can be formulated
through the following statements.
\begin{itemize}
  \item {A] In a projection it can
occur that more than one higher dimensional vector maps to the same
lower dimensional one. This means that in general there will be several roots of
$\Delta_\mathbb{G}$ which have the same image in $\overline{\Delta}$.
Calling $\Delta^+_\mathbb{G}$ and $\overline{\Delta}^+$ the sets of
positive roots of the two root systems, it happens that both of them split
in two subsets with the following properties.
\begin{eqnarray}
 && \begin{array}{|rrcl|rrcl|}
  \hline
 \mathbb{G}_R &\null & \null & \null &  \mathbb{G}_{TS} &\null & \null &
 \null\\
 \hline
  \null& \null & \null& \null &\null& \null &\null& \null \\
   \Delta^+_\mathbb{G}  =  \Delta^\eta\, \bigcup \,\Delta^\delta
    & \null &\null & \null
    &\overline{\Delta}^+ = \overline{\Delta}^\ell \, \bigcup \, \overline{\Delta}^s &\null & \null&\null \\
    \forall \eta_1 ,\eta_2 \in \Delta^\eta \, ; \,& \eta_1 + \eta_2 & \in & \Delta^\eta &
    \forall \alpha^\ell_1 ,\alpha^\ell_2 \in \overline{\Delta}^\ell\, ; \, &   \alpha^\ell_1 + \alpha^\ell_2  &
    \in &  \overline{\Delta}^\ell \\
    \forall \eta \in \Delta^\eta\, , \,
    \forall \delta \in \Delta^\delta\, ; \, & \eta + \delta & \in & {\Delta}^\delta &
    \forall \alpha^\ell \in \overline{\Delta}^\ell\, , \, \forall \alpha^s \in \overline{\Delta}^s\, ;
    \, &   \alpha^\ell + \alpha^s &
    \in &  \overline{\Delta}^s \\
  \forall \delta_1 , \delta_2 \in \Delta^\delta \, ; \, & \delta_1 + \delta_2 & \in &
  \cases{ \Delta^\delta \cr
  \Delta^\eta \cr}  &
    \forall \alpha^s_1 ,\alpha^s_2 \in \overline{\Delta}^s
     \, ; \, &   \alpha^s_1 + \alpha^s_2 &
    \in &   \cases{ \overline{\Delta}^\ell \cr
  \overline{\Delta}^s \cr} \\
  \null& \null & \null& \null &\null& \null &\null& \null \\
  \hline
  \end{array}\nonumber\\
  \end{eqnarray}
  The projection acts on the two different sets in the following way:
  \begin{eqnarray}
  && \Pi \left [\Delta^\eta\,\right] = \overline{\Delta}^\ell
  \nonumber\\
&& \Pi \left [\Delta^\delta\,\right] = \overline{\Delta}^s
\nonumber\\
&& \forall \alpha^\ell \, \in \, \overline{\Delta}^\ell \quad ; \quad
\mbox{card} \, \Pi^{-1} \left[ \alpha^\ell \right] \, = \,  1 \nonumber\\
&& \forall \alpha^s \, \in \, \overline{\Delta}^s \quad ; \quad
\mbox{card} \, \Pi^{-1} \left[ \alpha^s \right] \, = \,  m \nonumber\\
&& \mbox{card} \, \Delta^+_{\mathbb{G}_R} \, = \, \mbox{card} \,
\overline{\Delta}^\ell \, + \, m \, \times \, \mbox{card} \,
\overline{\Delta}^s
\label{orpone}
\end{eqnarray}
It means that there are two type of roots those which have a distinct
image in the projected root system and those which arrange into
multiplets with the same projection. The possible multiplicities,
however, are only two, either $1$ or $m$. Because of that we can
enumerate the generators of the solvable algebra $Solv(\mathrm{G_{R} / H})$ in the following
way:
\begin{eqnarray}
H_i & \Rightarrow & \mbox{Cartan generators} \nonumber\\
\Phi_{\alpha^\ell} & \Rightarrow & \eta -\mbox{roots} \nonumber\\
\Omega_{\alpha^s | I} &\Rightarrow & \delta -\mbox{roots} \quad ;
\quad (I=1,\dots , m )
\label{enumerato}
\end{eqnarray}
The index $I$ enumerating the $m$--roots of $\Delta_{\mathbb{G}_R}$ that have  the
same projection in $\overline{\Delta}$ is named the \textit{paint
index}
}
  \item {B] There exists a \textit{compact subalgebra} $\mathbb{G}_{paint} \, \subset
  \, \mathbb{G}_R$ which acts as  an algebra of outer automorphisms ({\it i.e.} outer derivatives) on the  solvable
  algebra $Solv_{\mathbb{G}_R} \equiv Solv(\mathrm{G_{R} /
  H}) \subset \mathbb{G}_R$, namely:
\begin{eqnarray}
 && \left[ \mathbb{G}_{paint} \, , \, Solv_{\mathbb{G}_R} \right]  =
  Solv_{\mathbb{G}_R} \nonumber\\
\label{linearrepre}
\end{eqnarray}}
\item {C] The Cartan generators $H_i$ and the generators $\Phi_{\alpha^\ell} $ are singlets under the action
of  $\mathbb{G}_{paint}$, {\it i.e.} each of them  commutes with the whole of $\mathbb{G}_{paint}$:
\begin{equation}
  \left[ H_i \, , \,\mathbb{G}_{paint}\right] \, = \, \left[ \Phi_{\alpha^\ell} \, ,
  \,\mathbb{G}_{paint}\right]\, = \, 0
\label{hiphicommuti}
\end{equation}
On the other hand, each of the $m$-multiplets of generators $\Omega_{\alpha^s |
I}$ constitutes an orbit under the
action of the paint group ${G}_{paint}$, {\it i.e.} a linear
representation $\mathbf{D}{[\alpha^s]}$ which, for different roots $\alpha^s$ can be different,
but has always the same dimension $m$ :
\begin{equation}
  \forall \, X \, \in \, \mathbb{G}_{paint} \quad : \quad  \left[ X \, , \,
  \Omega_{\alpha^s |
I}\right] \, = \, \left( D^{[\alpha^s]}[X]\right) _I^{\phantom{I}J} \, \Omega_{\alpha^s |
J}
\label{replicas}
\end{equation}
}
\item{D] The \textit{paint algebra} $\mathbb{G}_{paint}$ contains a
subalgebra
\begin{equation}
  \mathbb{G}^0_{paint} \, \subset \, \mathbb{G}_{paint}
\label{paint0}
\end{equation}
such that with respect to $\mathbb{G}^0_{paint}$,  each $m$--dimensional representation
$\mathbf{D}{[\alpha^s]}$ branches in the same way as follows:
\begin{equation}
  \mathbf{D}{[\alpha^s]} \,
  \stackrel{\mathbb{G}^0_{paint}}{\Longrightarrow} \,
  \underbrace{{\mathbf{1}}}_{\mbox{singlet}}
  \, \oplus \, \underbrace{{\mathbf{J}}}_{(m-1)-\mbox{dimensional}
  }
\label{splittatonibus}
\end{equation}
Accordingly we can split the range of the paint index $I$ as follows:
\begin{equation}
  I = \left\{ 0, \underbrace{x}_{1,\dots,m-1}\right\}
\label{Irango}
\end{equation}
the index $0$ corresponding to the singlet, while $x$ ranges over the
representation $\mathbf{J}$
}
\item {E] The tensor product $\mathbf{J} \otimes \mathbf{J} $ contains both the identity
representation $\mathbf{1}$ and the representation $\mathbf{J}$ itself.
Furthermore, there exists, in the representation $\bigwedge ^3 \mathbf{J}$ a
$\mathbb{G}^0_{paint}$-invariant tensor $a^{xyz}$ such that the two
solvable Lie algebras $Solv_{\mathbb{G}_R}$ and
$Solv_{\mathbb{G}_{TS}}$ can be written as follows
\begin{eqnarray}
&&
  \begin{array}{|l|l|l|}
  \hline
\null &Solv_{\mathbb{G}_{TS}} & Solv_{\mathbb{G}_R} \\
\hline
 \null &   \left[ H_i \, , \, H_j \right] = 0 & \left[ H_i \, , \, H_j \right] =0 \\
\null &  \left [ H_i \, , \, E^{\alpha^\ell} \right ] = \alpha^\ell_i \, E^{\alpha^\ell} & \left[ H_i \, , \, \Phi_{\alpha^\ell}
  \right] = \alpha^\ell_i \, \Phi_{\alpha^\ell} \\
\null & \left [ H_i \, , \, E^{\alpha^s} \right ] = \alpha^s_i \, E^{\alpha^s} & \left[ H_i \, , \, \Omega_{\alpha^s|I}
  \right] = \alpha^s_i \, \Omega_{\alpha^s|I} \\
  \null & \null & \null \\
 \alpha^\ell + \beta^\ell \notin \overline{\Delta} &
 \left [ E^{\alpha^\ell} \, , \, E^{\beta^\ell} \right ] = 0 &
 \left [ \Phi_{\alpha^\ell} \, , \, \Phi_{\beta^\ell} \right ] =0 \\
 \null & \null & \null \\
 \alpha^\ell + \beta^\ell \in \overline{\Delta} &
 \left [ E^{\alpha^\ell} \, , \, E^{\beta^\ell} \right ] = N_{\alpha^\ell\beta^\ell} \,
E^{\alpha^\ell + \beta^\ell}  & \left [ \Phi_{\alpha^\ell} \, , \, \Phi_{\beta^\ell} \right ] =N_{\alpha^\ell\beta^\ell} \,
\Phi_{\alpha^\ell + \beta^\ell} \\
\null & \null & \null \\
 \alpha^\ell + \beta^s \notin \overline{\Delta} &
 \left [ E^{\alpha^\ell} \, , \, E^{\beta^s} \right ] = 0 &
 \left [ \Phi_{\alpha^\ell} \, , \, \Omega_{\beta^s|I} \right ] =0 \\
 \null & \null & \null \\
 \alpha^\ell + \beta^s \in \overline{\Delta}^s &
 \left [ E^{\alpha^\ell} \, , \, E^{\beta^s} \right ] = N_{\alpha^\ell\beta^s} \,
E^{\alpha^\ell + \beta^s}  & \left [ \Phi_{\alpha^\ell} \, , \, \Omega_{\beta^s|I} \right ] =N_{\alpha^\ell\beta^s} \,
\Omega_{\alpha^\ell + \beta^s|I} \\
\null & \null & \null \\
 \alpha^s + \beta^s \notin \overline{\Delta} &
 \left [ E^{\alpha^s} \, , \, E^{\beta^s} \right ] = 0 &
 \left [ \Omega_{\alpha^s|I} \, , \, \Omega_{\beta^s|J} \right ] =0 \\
  \null & \null & \null \\
 \alpha^s + \beta^s \in \overline{\Delta}^\ell &
 \left [ E^{\alpha^s} \, , \, E^{\beta^s} \right ] = N_{\alpha^s\beta^s} E^{\alpha^s +\beta^s} &
 \left [ \Omega_{\alpha^s|I} \, , \, \Omega_{\beta^s|J} \right ] =\delta^{IJ} \,
 N_{\alpha^s\beta^s} \Phi_{\alpha^s +\beta^s} \\
  \null & \null & \null \\
 \alpha^s + \beta^s \in \overline{\Delta}^s &
 \left [ E^{\alpha^s} \, , \, E^{\beta^s} \right ] = N_{\alpha^s\beta^s} E^{\alpha^s +\beta^s} &
\cases{ \left [ \Omega_{\alpha^s|0} \, , \, \Omega_{\beta^s|0} \right ]
=
 N_{\alpha^s\beta^s} \Omega_{\alpha^s +\beta^s|0} \cr
 \null \cr
 \left [ \Omega_{\alpha^s|0} \, , \, \Omega_{\beta^s|x} \right ]
=
 N_{\alpha^s\beta^s} \Omega_{\alpha^s +\beta^s|x} \cr
 \null \cr
 \left [ \Omega_{\alpha^s|x} \, , \, \Omega_{\beta^s|y} \right ]
=
 N_{\alpha^s\beta^s} \left( \delta^{xy} \Omega_{\alpha^s +\beta^s|0} \right. \cr
 \left. \, + \,  a^{xyz} \, \Omega_{\alpha^s +\beta^s|z} \right ) \cr}
  \\
   \hline
  \end{array}\nonumber\\
  && \null
\label{paragonando}
\end{eqnarray}
}
\end{itemize}
The existence of the paint group $G_{paint}$ and the structure of the
solvable Lie algebra displayed in eq. (\ref{paragonando}) imply that
we can reduce the geodesic problem on $G_{R}/H$ and hence the supergravity field
equations to the geodesic problem on $G_{TS}/H_{TS}$ which is
maximally split and can be solved with the compensator method
introduced in \cite{noiconsasha}. It suffices to observe that by
setting all the components of the tangent vectors in the directions
of the generators $\Omega_{\alpha^s |x}$ to zero we simply reproduce
a copy of the solvable Lie algebra of the Tits Satake manifold. Once
we have found a solution for this latter, we can extend it to
a full fledged solution of the original system by applying rotations
of the paint group $G_{paint}$ with constant parameters. Physically
this means that indeed the billiard table is just the Weyl chamber of the Tits
Satake algebra as observed by Damour et al
\cite{cosmicbilliardliterature4}, yet, in the smooth billiard
realization the raising and lowering of the walls occurs in
\textit{paints} which specify the precise correspondence with the
supergravity fields and hence with the oxidation to higher
dimensions.
\subsection{Content of the paper}
In the sequel of this paper we illustrate these general structures by
working out in all details a specific example, that arising from
$D=4,\mathcal{N}=6$ supergravity. Our choice is motivated as follows. On one
hand, the case $N_Q=24$ is the next simplest apart from that of
maximal supersymmetry $N_Q=32$. Indeed there is just the graviton
multiplet, the number of fields is completely fixed and so is the
geometric structure of the lagrangian. On the other hand the scalar manifold
of $\mathcal{N}=6$ supergravity is an instance of a special K\"ahler manifold and
the bosonic lagrangian can be reinterpreted as the lagrangian of a
particular $\mathcal{N}=2$ model. In other words we could also reconsider our
constructions from an $\mathcal{N}=2$ viewpoint and interpret the
scalar fields we deal with as moduli of an abstract Calabi-Yau compactification.
Indeed in a subsequent paper we shall extrapolate from the
present example general considerations on billiard dynamics and
painted walls in the context of special geometries.
\par
Our paper is organized as follows.
\par
In section \ref{casestudy} we present the in depth analysis of the
$\mathrm{E_{7(-5)}}$ real section: how generators are constructed, how they
are subdivided into compact and non compact ones, how the Tits Satake
projection works in this case, what is the structure of the solvable
Lie algebra generating the coset manifold $\mathrm{E_{7(-5)}}/\mathrm{SO(12)
\times SO(3)}$ and what is the structure of the paint group. Then in section
\ref{firstorder} we derive the Nomizu connection for both the
original manifold and its Tits Satake projection and we compare the
structure of the two systems of first order equations for the tangent
vectors. In section \ref{compensator} we derive explicit smooth solutions
for the maximally split $\mathrm{F}_4$ system and we show that they display
several bounces: smooth cosmic billiards.
In section \ref{uplift} we uplift the previously found solutions to
the original $\mathrm{E}_{7(-5)}$ system by means of the paint group. Then we
discuss the general features of the Tits Satake projection, how it
commutes with dimensional reduction and how the paint group is
preserved in the reduction. We illustrate these concepts on the
specific example. Section \ref{concludo} contains our conclusions.
Then we have two appendices. The first, appendix \ref{benissimo1},
contains the listing and ordering of $\mathrm{E}_7$ roots utilized throughout
the paper. The second, appendix \ref{appendoF4}, is devoted to the
explicit construction of the fundamental $26$--dimensional
representation of $\mathrm{F_{4(4)}}$ which we used in the paper to
calculate the needed $N_{\alpha\beta}$ matrix.
\section{The example of $\mathcal{N}$=6 supergravity}
\label{casestudy}
In  $\mathcal{N}=6,D=4$ supergravity there are $30$ scalars which span the special K\"ahler manifold:
\begin{equation}
  \mathcal{M}_{scalar}^{D=4,\mathcal{N}=6} = \frac{\mathrm{SO^\star(12)}}{\mathrm{SU(6)} \times
  \mathrm{U(1)}}
\label{so12starmanif}
\end{equation}
and the relevant duality algebra is therefore:
\begin{equation}
  \mathbb{G}_{D=4} = \mathrm{SO^\star(12)}
\label{uppa1}
\end{equation}
The $16$ graviphotons give rise to $16$ electric plus $16$ magnetic
field strengths that  organize into the $32$ spinor representation of
$\mathrm{SO^\star(12)}$ which is symplectic as it should be.
\par
After reduction to $D=3$ dimensions and dualization of all the vector
fields to scalars we obtain a 3D-gravity coupled $\sigma$--model
based on the quaternionic symmetric space:
\begin{equation}
  \mathcal{M}_{scalar}^{D=3,N=12} = \frac{\mathrm{E_{7(-5)}}}{\mathrm{SO(3)_R \times SO(12)} }
\label{e7-5starmanif}
\end{equation}
which is the $c$-map of the special K\"ahler manifold
\ref{so12starmanif}.
In this section we study the structure of the solvable Lie algebra
describing the non maximally split non-compact manifold
(\ref{e7-5starmanif}) and how it is related to its Tits Satake
submanifold:
\begin{equation}
  \mathcal{M}_{scalar}^{Tits\,\,\,Satake} = \frac{\mathrm{F_{4(4)}}}{\mathrm{SU(2)_R \times Usp(6)} }
\label{f4starmanif}
\end{equation}
which is instead maximally split and it is the relevant submanifold
defining the cosmic billiard. Our main goal is to show how the
solution of the first order equations for the system
(\ref{f4starmanif}) can be used to obtain solutions for the system
(\ref{e7-5starmanif}). In particular we shall appropriately study how
the dynamic walls of the billiard (\ref{e7-5starmanif}) are painted
copies of the walls associated with the billiard
(\ref{f4starmanif}).
\par
To this effect we have to develop all the algebraic machinery
associated with the real form $\mathrm{E}_{7(-5)}$ of the $\mathrm{E}_7$ complex Lie
algebra. We begin by spelling out the particular form of the
decomposition (\ref{gendecompo})
\begin{equation}
\mbox{adj}(\mathrm{E_{7(-5)}}) =
\mbox{adj}(\mathrm{SO^\star(12)})\oplus\mbox{adj}(\mathrm{SL(2,\mathbb{R})})\oplus
{(\mathbf{2,32_s})}
\label{e7compo}
\end{equation}
where $\mathbf{32_s}$ denotes the spinor representation of
$\mathbf{\mathrm{SO^\star(12)}}$, while $\mathbf{2}$ denotes the
fundamental representation of $\mathrm{SL(2,R)}$. The subgroup
$\mathrm{SO^\star(12)} \times \mathrm{SL(2,\mathbb{R})}$ is regularly
embedded and non compact. There is another similar decomposition of
the adjoint of $\mathrm{E}_{7(-5)}$ with respect to its maximal compact
subgroup:
\begin{equation}
\mbox{adj}(\mathrm{E_{7(-5)}}) =
\mbox{adj}(\mathrm{SO(12)})\oplus\mbox{adj}(\mathrm{SO(3)_R})\oplus
{(\mathbf{2,32_s})}
\label{e7compo2}
\end{equation}
where, once again $\mathbf{32_s}$ denotes the spinor representation
of the compact $\mathrm{SO(12)}$, this time.
\par
The non compact symmetric space (\ref{e7-5starmanif}) has $rank =4$.
This means that of the seven Cartan generators of $\mathrm{E}_{7(-5)}$, four
are non compact and belong to the coset, while three are compact and
belong to the compact subalgebra. We proceed to the explicit
construction of the involutive automorphism of the complex $\mathrm{E^\mathbb{C}_7}$ algebra
\begin{equation}
  \sigma \, : \, \mathrm{E^\mathbb{C}_7} \, \rightarrow \, \mathrm{E^\mathbb{C}_7}
\label{sigmaauto}
\end{equation}
which defines the real form $\mathrm{E_{7(-5)}}$. This given we
obtain also the compact subalgebra $\mathbb{H}$, the complementary
non compact subspace $\mathbb{K}$ and the solvable Lie algebra
$Solv_{\mathrm{E}7(-5)}$ whose corresponding solvable group manifold is
isometrical to the coset manifold (\ref{e7-5starmanif}).
\subsection{The $\mathrm{E}_7$ root system, and its projection onto the
$\mathrm{F}_{4}$ root system}
In order to realize the programme we have just outlined, we begin by choosing an explicit basis of simple roots
for $\mathrm{E_7}$. In an Euclidean orthonormal basis they are the
following ones:
\begin{eqnarray}
\alpha _1 & = & \{1, -1, 0, 0, 0, 0, 0 \} \nonumber\\
\alpha _2 & = & \{0, 1, -1, 0, 0, 0, 0 \} \nonumber\\
\alpha _3 & = & \{0, 0, 1, -1, 0, 0, 0 \} \nonumber\\
\alpha _4 & = & \{0, 0, 0, 1, -1, 0, 0 \}, \nonumber\\
\alpha _5 & = & \{0, 0, 0, 0, 1, -1, 0\}, \nonumber\\
\alpha _6 & = & \{0, 0, 0, 0, 1, 1, 0\}, \nonumber\\
\alpha _7 & = & \{ - \frac{1}{2}  ,
  - \frac{1}{2}  ,
  - \frac{1}{2}  ,
  - \frac{1}{2}  ,
  -\frac{1}{2}  ,
  -\frac{1}{2}  ,
  \frac{1}{{\sqrt{2}}}\}
\label{e7simple}
\end{eqnarray}
and they are associated with the $\mathrm{E_{7}}$ Dynkin diagram labeled as
it is displayed in fig.\ref{except7}:
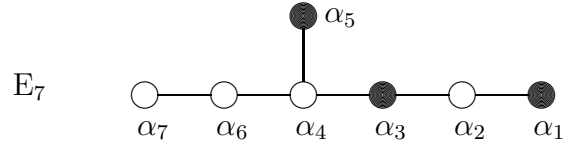
\begin{figure}
\centering
\begin{picture}(100,100)
     \put (-70,35){$\mathrm{E}_7$} \put (-20,35){\circle {10}} \put
(-23,20){$\alpha_7$} \put (-15,35){\line (1,0){20}} \put
(10,35){\circle {10}} \put (7,20){$\alpha_6$} \put (15,35){\line
(1,0){20}} \put (40,35){\circle {10}} \put (37,20){$\alpha_4$}
\put (40,65){\circle {10}}
\put (40,65){\circle {9}}
\put (40,65){\circle {8}}
\put (40,65){\circle {7}}
\put (40,65){\circle {6}}
\put (40,65){\circle {5}}
\put (40,65){\circle {4}}
\put (40,65){\circle {3}}
\put (40,65){\circle {2}}
\put (40,65){\circle {1}}
\put (48,62.8){$\alpha_5$} \put
(40,40){\line (0,1){20}} \put (45,35){\line (1,0){20}} \put
(70,35){\circle {10}}
\put(70,35){\circle {9}}
\put(70,35){\circle {8}}
\put(70,35){\circle {7}}
\put(70,35){\circle {6}}
\put(70,35){\circle {5}}
\put(70,35){\circle {4}}
\put(70,35){\circle {3}}
\put(70,35){\circle {2}}
\put(70,35){\circle {1}}
\put (67,20){$\alpha_{3}$} \put
(75,35){\line (1,0){20}} \put (100,35){\circle {10}} \put
(97,20){$\alpha_{2}$} \put (105,35){\line (1,0){20}} \put
(130,35){\circle {10}}
\put(130,35){\circle {9}}
\put(130,35){\circle {8}}
\put(130,35){\circle {7}}
\put(130,35){\circle {6}}
\put(130,35){\circle {5}}
\put(130,35){\circle {4}}
\put(130,35){\circle {3}}
\put(130,35){\circle {2}}
\put(130,35){\circle {1}}
\put (127,20){$\alpha_1$}
\end{picture}
\vskip 1cm \caption{The Dynkin diagram of $\mathrm{E}_{7}$ and the labeling of simple roots. The three orthogonal roots
$\alpha_1$, $\alpha_3$ and $\alpha_5$ are marked
black, since they are used to define the part of the Cartan subalgebra which is compact in the $\mathrm{E}_{7(-5)}$ real form.}
\label{except7}
\end{figure}
Next we list all the positive roots of $\mathrm{E}_{7}$ arranged according to
their height. They are $63$ and they are listed in Appendix \ref{benissimo1} .

The real section $\mathrm{E}_{7(-5)}$ of the complex Lie algebra $\mathrm{E}_{7}^\mathbb{C}$ is
identified by the Tits Satake diagram depicted in fig. \ref{except7}
where the simple roots $\alpha_1$, $\alpha_3$, $\alpha_5$ are
black. This means that in the chosen real form the Cartan generators
dual to these three roots $\mathcal{H}_{\alpha_{1,3,5}}$ are compact,
while non compact are the Cartan generators in the complementary
$4$--dimensional subspace.  It is fairly easy to describe the space of
non-compact Cartan generators $\mathcal{H}^{n.c.}$. It is the span of
the four weight vectors $\lambda_{2,4,5,7}$ which, by construction, are
orthogonal to the roots $\alpha_{1,3,5}$. Thus in the chosen euclidean
basis we obtain:
\begin{eqnarray}
\mathcal{H}^{n.c.} & = & \mbox{span} \,\left \{\lambda_{2},\lambda_4,\lambda_6 , \lambda_7 \right \} \nonumber\\
\lambda_2 & = &\left \{ 1,1,0,0,0,0,\sqrt{2} \right \} \nonumber\\
\lambda_4 & = &\left \{ 1,1,1,1,0,0,2\sqrt{2} \right \} \nonumber\\
\lambda_6 & = & \left \{ \ft 12,\ft 12,\ft 12,\ft 12,\ft 12,\ft 12,\frac 3{\sqrt{2}} \right \} \nonumber\\
\lambda_7 & = & \left \{ 0,0,0,0,0,0, \sqrt{2}\right \} \nonumber\\
\label{Hspan2467}
\end{eqnarray}
It is now easy to construct an orthogonal basis of four length-two $7$-vectors
for the space $\mathcal{H}^{n.c.}$ defined by eq.(\ref{Hspan2467}).
It is given by:
\begin{eqnarray}
e_1 & = & \{1,1,0,0,0,0,0 \} \nonumber\\
e_2 & = & \{0,0,1,1,0,0,0 \} \nonumber\\
e_3  & = & \{0,0,0,0,1,1,0 \} \nonumber\\
e_4 & = & \{0,0,0,0,0,0,\sqrt{2}\} \nonumber\\
\label{ortonchbas}
\end{eqnarray}
Indeed, equivalently to eq.(\ref{Hspan2467}) we can also write:
\begin{equation}
  \mathcal{H}^{n.c.} = \mbox{span} \,\left \{e_1,e_2,e_3 , e_4 \right \}
\label{ncspane1234}
\end{equation}
We can complete the basis (\ref{ortonchbas}) with other three vectors
also of length 2, which are orthogonal to $e_{1,2,3,4}$ and also orthogonal among
themselves:
\begin{eqnarray}
e_5 & = & \{1,-1,0,0,0,0,0\} \nonumber\\
e_6 & = & \{0,0,1,-1,0,0,0\}  \nonumber\\
e_7 & = & \{0,0,0,0,1,-1,0\}  \nonumber\\
\label{compbas}
\end{eqnarray}
The compact Cartan subalgebra is provided by the span of these three
vectors:
\begin{equation}
  \mathcal{H}^{ c}=\mbox{span} \left\{e_5,e_6,e_7 \right \}
\label{compaspan}
\end{equation}
The $\mathrm{E}_7$ roots are  vectors in the dual of the $7$-dimensional space which is
the direct sum of the four dimensional space $\mathcal{H}^{n.c.}$ plus the three dimensional space
$\mathcal{H}^c$:
\begin{equation}
  \mathcal{H} = \mathcal{H}^{n.c.} \oplus \mathcal{H}^c
\label{Hsumma}
\end{equation}
Hence every root $\alpha \in \Delta_{E_7}$ can be decomposed as follows:
\begin{equation}
  \alpha = \alpha_{||} \oplus \alpha_{\bot}
\label{decompa}
\end{equation}
where $\alpha_{||}$ lies in $\mathcal{H}^{n.c.}$ and $\alpha_{\bot}$
is orthogonal to it. The essential point in Tits Satake theory of
real forms is that the parallel projections of the roots, namely
$\alpha_{||}$, are not just arbitrary vectors, rather they are roots
of a Lie algebra of rank equal to the dimension of the non compact
Cartan Lie algebra which is actually a subalgebra of the original algebra.
In our case we have $rank \, =\,  4$ and the
relevant subalgebra (Tits Satake) is $\mathrm{F}_{4(4)} \subset E_{7(-5)}$.
Indeed the parallel projections $\alpha_{||}$ fill the cardinality
$24$ root-system $\Delta_{F_4}$.
\par
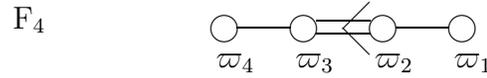
\begin{figure}
\centering
\begin{picture}(100,100)
     \put (-70,35){$\mathrm{F}_4$}  \put
(10,35){\circle {10}} \put (7,20){$\varpi_4$} \put (15,35){\line
(1,0){20}} \put (40,35){\circle {10}} \put (37,20){$\varpi_3$}\put (45,38){\line (1,0){20}}
\put (55,35){\line (1,1){10}} \put (55,35){\line (1,-1){10}}\put (45,33){\line (1,0){20}} \put
(70,35){\circle {10}} \put (67,20){$\varpi_{2}$} \put
(75,35){\line (1,0){20}} \put (100,35){\circle {10}} \put
(97,20){$\varpi_{1}$}
\end{picture}
\vskip 1cm \caption{The Dynkin diagram of $\mathrm{F}_4$ and the labeling of simple roots.}
\label{F4dynk}
\end{figure}

The actual construction of the real form $\mathrm{E}_{7(-5)}$ involves the
careful analysis of the \textbf{onto} projection:
\begin{equation}
  \Delta_{E_7} \, \stackrel{\pi}{\Longrightarrow} \, \Delta_{F_4}
\label{proiezioneDE7toF4}
\end{equation}
Explicitly, if we decompose the $63$ positive roots of $\mathrm{E}_7$ along the new orthogonal basis
$e_{1,2,3,4,5,7}$  we discover the following:
\begin{description}
\item[1] There are just three roots that are orthogonal to the
  subspace spanned by $e_{1,2,3,4}$, namely such that $\alpha_{||} = 0$. They are precisely the simple
  roots $\alpha_1$, $\alpha_3$ and $\alpha_5$.
  \item[2] The remaining $60$ roots have a projection onto the space
  spanned by $e_{1,2,3,4}$ which takes the form of one of the $24$ roots of
  $\mathrm{F}_4$ and all such 24 roots are reproduced in the projection.
  Namely $\alpha_{||} \in \Delta_{F_4}$.
  \item[3] The set of $24$ roots of $\mathrm{F}_4$ is subdivided in two
  subsets of 12 roots each. The \textit{long} and the \textit{short} roots. Each long root  appears only
  once in the projection of $\mathrm{E}_7$ roots. Each of the 12 short roots, instead,
  appears  exactly  four times as image of four distinct $\mathrm{E}_{7}$ roots. So that we count $4 \times 12 + 12 = 60$
\end{description}
To understand this pattern we have to introduce the $\mathrm{F}_4$ root
system. The Dynkin diagram of $\mathrm{F}_4$ is given in fig. \ref{F4dynk}
and calling $y_{1,2,3,4}$ a basis of orthonormal vectors:
\begin{equation}
  y_i \, \cdot \, y_j = \delta_{ij}
\label{yiyj}
\end{equation}
a possible choice of simple roots $\varpi_i$ which reproduces the
Cartan matrix encoded in the Dynkin diagram (\ref{F4dynk}) is the
following:
\begin{eqnarray}
\varpi_1 & = & -y_1 -y_2 - y_3 + y_4\nonumber\\
\varpi_2 & = & 2\, y_3 \nonumber\\
\varpi_3 & = & y_2 - y_3 \nonumber\\
\varpi_4 & = & y_1 -y_2
\label{simplef4}
\end{eqnarray}
With this basis of simple roots the full root system  composed
of $48$ vectors is given by:
\begin{equation}
\Delta_{F_4} \quad  \equiv \quad \underbrace{\pm y_i \pm y_j}_{\mbox{24 roots}} \quad ; \quad
  \underbrace{\pm y_i}_{\mbox{8 roots}}\quad ; \quad \underbrace{\pm y_1 \pm y_2 \pm y_3 \pm
  y_4}_{\mbox{16 roots}}
\label{tutterutte}
\end{equation}
and one can list the positive roots by height as displayed in table
\ref{tablettas}.
\begin{table}
  \centering
  $$
\begin{array}{|lcl|l|}
\hline
\varpi[ 1 ] & = &
 \{ 1, 0, 0,
      0\} & -{y_1} - {y_2} - {y_3}  +
{y_4}\nonumber\\
\varpi[ 2 ] & = &
 \{ 0, 1, 0,
      0\} & 2\, {y_3}
\nonumber\\
\varpi[ 3 ] & = &
 \{ 0, 0, 1,
      0\} & {y_2} - {y_3}\
\nonumber\\
\varpi[ 4 ] & = &
 \{ 0, 0, 0,
      1\} & {y_1} - {y_2}\
\nonumber\\
\varpi[ 5 ] & = &
 \{ 1, 1, 0,
      0\} & -{y_1} - {y_2} + {y_3}  +
{y_4}\nonumber\\
\varpi[ 6 ] & = &
 \{ 0, 1, 1,
      0\} & {y_2} + {y_3}\
\nonumber\\
\varpi[ 7 ] & = &
 \{ 0, 0, 1,
      1\} & {y_1} - {y_3}\
\nonumber\\
\varpi[ 8 ] & = &
 \{ 1, 1, 1,
      0\} & -{y_1} + {y_4}\
\nonumber\\
\varpi[ 9 ] & = &
 \{ 0, 1, 2,
      0\} & 2\, {y_2}
\nonumber\\
\varpi[ 10 ] & = &
 \{ 0, 1, 1,
      1\} & {y_1} + {y_3}\
\nonumber\\
\varpi[ 11 ] & = &
 \{ 1, 1, 2,
      0\} & -{y_1} + {y_2} - {y_3}  +
{y_4}\nonumber\\
\varpi[ 12 ] & = &
 \{ 1, 1, 1,
      1\} & -{y_2} + {y_4}\
\nonumber\\
\varpi[ 13 ] & = &
 \{ 0, 1, 2,
      1\} & {y_1} + {y_2}\
\nonumber\\
\varpi[ 14 ] & = &
 \{ 1, 2, 2,
      0\} & -{y_1} + {y_2} + {y_3}  +
{y_4}\nonumber\\
\varpi[ 15 ] & = &
 \{ 1, 1, 2,
      1\} & -{y_3} + {y_4}\
\nonumber\\
\varpi[ 16 ] & = &
 \{ 0, 1, 2,
      2\} & 2\, {y_1}
\nonumber\\
\varpi[ 17 ] & = &
 \{ 1, 2, 2,
      1\} & {y_3} + {y_4}\
\nonumber\\
\varpi[ 18 ] & = &
 \{ 1, 1, 2,
      2\} & {y_1} - {y_2} - {y_3}  +
{y_4}\nonumber\\
\varpi[ 19 ] & = &
 \{ 1, 2, 3,
      1\} & {y_2} + {y_4}\
\nonumber\\
\varpi[ 20 ] & = &
 \{ 1, 2, 2,
      2\} & {y_1} - {y_2} + {y_3}  +
{y_4}\nonumber\\
\varpi[ 21 ] & = &
 \{ 1, 2, 3,
      2\} & {y_1} + {y_4}\
\nonumber\\
\varpi[ 22 ] & = &
 \{ 1, 2, 4,
      2\} & {y_1} + {y_2} - {y_3}  +
{y_4}\nonumber\\
\varpi[ 23 ] & = &
 \{ 1, 3, 4,
      2\} & {y_1} + {y_2} + {y_3}  +
{y_4}\nonumber\\
\varpi[ 24 ] & = &
 \{ 2, 3, 4,
      2\} & 2\, {y_4}
\nonumber\\
\hline
\end{array}
$$
\caption{Listing of all positive roots of $\mathrm{F}_4$. The second column gives the Dynkin labels, while the second column gives
the form of the root in an euclidean basis}
  \label{tablettas}
\end{table}
If we identify the $\mathrm{E}_{7}$ roots with their progressive number as it
is defined by their listing in Appendix \ref{benissimo1}, we can
reorganize them into the following three subsets according to their projection onto the $\mathrm{F}_4$ root space.
\begin{description}
  \item[1] First we have the  $ \Delta_{\beta}$ set:
  \begin{equation}
 \beta_1 = \alpha_1 \quad ; \quad \beta_2 =
  \alpha_3 \quad ; \quad \beta_3 =
  \alpha_5
\label{betarut}
\end{equation}
which contains the three roots with vanishing projection onto the $\mathrm{F}_4$
root space. As we are going to see, together with their negative and
with the compact Cartan generators, these roots define a compact
subalgebra $\mathrm{SO(3)}_1 \times \mathrm{SO(3)}_2 \times \mathrm{SO(3)}_3$
with respect to which the generators of the solvable Lie algebra of
$\mathrm{E_{7(-5)}}/\mathrm{SO(12)} \times \mathrm{SU(2)}$ transform covariantly
and arrange into representations. Indeed this $\mathrm{SO(3)^3}$ is,
for the present case, the \textit{paint group}
$\mathbb{G}_\mathrm{{paint}}$  mentioned in eq. (\ref{linearrepre}).
The subgroup $\mathbb{G}_{paint}^0 \subset
\mathbb{G}_\mathrm{{paint}}$ mentioned in eq.s (\ref{paint0}) and
(\ref{splittatonibus}) is actually the diagonal subgroup
$\mathrm{SO(3)}_{diag} \subset \mathrm{SO(3)}^3$.
\item[2] {Secondly we have the $\Delta_\eta$ set containing those twelve roots
  whose projection onto the $\mathrm{F}_4$ root space is unique. We organize
  them according to the height of the $\mathrm{F}_4$ root on which they project.
  The result is displayed in  table \ref{etaset}}
\item[3] {Thirdly we have the $\Delta_\delta$ set of those $48 $ $\mathrm{E}_7$ roots which
arrange into quadruplets having  the same projection
onto the $\mathrm{F}_4$ system. We denote these roots by $\delta_I^i$ where $I=1,\dots ,12$ and $i=1,2,3,4$.
They are displayed in table \ref{deltaset}}
\end{description}
\begin{table}
\centering
 $$
  \begin{array}{|c|ll|c|}
  \hline
 \null & \mbox{$\mathrm{F}_4$ root Dynkin labels} & \mbox{$\mathrm{F}_4$ root in eucl. basis}
 & \mbox{corresp. root of $\mathrm{E}_{7}$} \nonumber\\
  \hline
\alpha^\ell_1 &  \{ 0, 1, 0,
      0\}    &   2\, {y_3}   \
&   {{\eta }_1}   =   {\alpha_6}\
 \nonumber\\
\alpha^\ell_2 & \{ 1, 0, 0,
      0\}    &   -{y_1} - {y_2} - {y_3} + \
{y_4}   &   {{\eta }_2}   =  \
 {\alpha_7} \nonumber\\
\alpha^\ell_3 & \{ 1, 1, 0,
      0\}    &   -{y_1} - {y_2} + {y_3} + \
{y_4}   &   {{\eta }_3}   =  \
 {\alpha_{13}} \nonumber\\
\alpha^\ell_4 & \{ 0, 1, 2,
      0\}    &   2\, {y_2}   \
&   {{\eta }_4}   = \
  {\alpha_{31}} \nonumber\\
\alpha^\ell_5 & \{ 1, 1, 2,
      0\}    &   -{y_1} + {y_2} - {y_3} + \
{y_4}   &   {{\eta }_5}   =  \
 {\alpha_{36}} \nonumber\\
\alpha^\ell_6 & \{ 1, 2, 2,
      0\}    &   -{y_1} + {y_2} + {y_3} + \
{y_4}   &   {{\eta }_6}   =  \
 {\alpha_{41}} \nonumber\\
\alpha^\ell_7 & \{ 0, 1, 2,
      2\}    &   2\, {y_1}   \
&   {{\eta }_7}   = \
  {\alpha_{48}} \nonumber\\
\alpha^\ell_8 &\{ 1, 1, 2,
      2\}    &   {y_1} - {y_2} - {y_3} + \
{y_4}   &   {{\eta }_8}   =  \
 {\alpha_{51}} \nonumber\\
\alpha^\ell_9 &\{ 1, 2, 2,
      2\}    &   {y_1} - {y_2} + {y_3} + \
{y_4}   &   {{\eta }_9}   =  \
 {\alpha_{54}} \nonumber\\
\alpha^\ell_{10} &\{ 1, 2, 4,
      2\}    &   {y_1} + {y_2} - {y_3} + \
{y_4}   &   {{\eta }_{10}}   \
=   {\alpha_{61}} \nonumber\\
\alpha^\ell_{11} &\{ 1, 3, 4,
      2\}    &   {y_1} + {y_2} + {y_3} + \
{y_4}   &   {{\eta }_{11}}   \
=   {\alpha_{62}} \nonumber\\
\alpha^\ell_{12} &\{ 2, 3, 4,
      2\}    &   2\, {y_4}   \
&   {{\eta }_{12}}   = \
  {\alpha_{63}} \nonumber\\
  \hline
\end{array}
$$
\caption{The $\Delta_\eta$ set of those twelve $\mathrm{E}_7$ roots whose projection on the $\mathrm{F}_4$ root system is unique.
As it is evident from the table, from the point of view of $\mathrm{F}_4$ the $\Delta_\eta$
set is composed by the long roots. The first column gives the name by means of which these roots will be referred to
within the $\mathrm{F}_{4(4)}$ algebra. The last column gives the name of the corresponding root in $\mathrm{E}_7$ calculations}
\label{etaset}
\end{table}
\begin{table}
\centering
 $$
  \begin{array}{|c|c|ll|cccc|}
  \hline
  \null & \null & \mbox{$\mathrm{F}_4$ root Dynkin labels} & \mbox{$\mathrm{F}_4$ root in eucl. basis}
  & \delta_I^1 & \delta_I^2 & \delta_I^3 & \delta_I^4 \nonumber\\
  \hline
I=1 & \alpha^s_1 & \{ 0, 0, 0,
      1\}    &   {y_1} - {y_2}\
   &   {\alpha_2}   &  \
 {\alpha_8}   &   {\alpha_9}\
   &   {\alpha_{14}}   \nonumber\\
I=2 &\alpha^s_2 & \{ 0, 0, 1,
      0\}    &   {y_2} - {y_3}\
   &   {\alpha_4}   &  \
 {\alpha_{10}}   &   {\alpha_{11}}\
   &   {\alpha_{16}}   \nonumber\\
I=3 &\alpha^s_3 &\{ 0, 1, 1,
      0\}    &   {y_2} + {y_3}\
   &   {\alpha_{12}}   &  \
 {\alpha_{17}}   &   {\alpha_{19}}\
   &   {\alpha_{24}}   \nonumber\\
I=4 & \alpha^s_4 &\{ 0, 0, 1,
      1\}    &   {y_1} - {y_3}\
   &   {\alpha_{20}}   &  \
 {\alpha_{15}}   &   {\alpha_{26}}\
   &   {\alpha_{21}}   \nonumber\\
I=5 & \alpha^s_5 &\{ 1, 1, 1,
      0\}    &   -{y_1} + {y_4}\
   &   {\alpha_{18}}   &  \
 {\alpha_{23}}   &   {\alpha_{25}}\
   &   {\alpha_{30}}   \nonumber\\
I=6 &\alpha^s_6 & \{ 0, 1, 1,
      1\}    &   {y_1} + {y_3}\
   &   {\alpha_{27}}   &  \
 {\alpha_{22}}   &   {\alpha_{33}}\
   &   {\alpha_{29}}   \nonumber\\
I=7 & \alpha^s_7 &\{ 1, 1, 1,
      1\}    &   -{y_2} + {y_4}\
   &   {\alpha_{32}}   &  \
 {\alpha_{28}}   &   {\alpha_{37}}\
   &   {\alpha_{34}}   \nonumber\\
I=8 & \alpha^s_8 &\{ 0, 1, 2,
      1\}    &   {y_1} + {y_2}\
   &   {\alpha_{35}}   &  \
 {\alpha_{38}}   &   {\alpha_{40}}\
   &   {\alpha_{43}}   \nonumber\\
I=9 &\alpha^s_9 & \{ 1, 1, 2,
      1\}    &   -{y_3} + {y_4}\
   &   {\alpha_{39}}   &  \
 {\alpha_{42}}   &   {\alpha_{45}}\
   &   {\alpha_{47}}   \nonumber\\
I=10 &\alpha^s_{10} & \{ 1, 2, 2,
      1\}    &   {y_3} + {y_4}\
   &   {\alpha_{44}}   &  \
 {\alpha_{46}}   &   {\alpha_{49}}\
   &   {\alpha_{50}}   \nonumber\\
I=11 & \alpha^s_{11} &\{ 1, 2, 3,
      1\}    &   {y_2} + {y_4}\
   &   {\alpha_{53}}   &  \
 {\alpha_{52}}   &   {\alpha_{56}}\
   &   {\alpha_{55}}   \nonumber\\
I=12 & \alpha^s_{12} &\{ 1, 2, 3,
      2\}    &   {y_1} + {y_4}\
   &   {\alpha_{58}}   &  \
 {\alpha_{57}}   &   {\alpha_{60}}\
   &   {\alpha_{59}}   \nonumber\\
  \hline
\end{array}
$$
\caption{The $\Delta_\delta$ set of those $48$ $\mathrm{E}_7$ roots which arrange into 12 quadruplets having
the same $\mathrm{F}_4$ projection. As it is evident from the table, from the point of view of $\mathrm{F}_4$ the $\Delta_\delta$
set is composed by the short roots. The second column gives the name of the projected root within the $\mathrm{F}_{4(4)}$
algebra context. In the last four columns, $\alpha_i$ denotes an $\mathrm{E}_7$-root numbered according to the order listed in Appendix A}
\label{deltaset}
\end{table}
\subsection{The real form $\mathrm{E}_{7(-5)}$ and its associated solvable Lie
algebra}
Given these preliminaries we can now introduce the real form
$\mathrm{E}_{7(-5)}$   which follows from the action of a suitable
involutive automorphism (\ref{sigmaauto}) of the complex Lie algebra $\mathrm{E}_{7}^\mathbb{C}$.
\par
Following the general definitions presented in most textbooks on Lie
algebra theory (see for instance \cite{Helgason}), the real form
$\mathbb{G}_\mathbb{R}$ is defined as the subspace of eigenvalue $1$
of the relevant automorphism $\sigma$, namely we have:
\begin{equation}
  \sigma \left ( \mathbb{G}_\mathbb{R} \right
)=\mathbb{G}_\mathbb{R}
\label{autosigone}
\end{equation}
On the other hand, $\sigma$ is  completely identified by the Tits
Satake diagram depicted in  fig.\ref{except7}. Indeed the action of
$\sigma$ is originally defined on the Cartan subalgebra and
corresponds to changing the signs of all vectors lying in the
compact part while keeping unchanged those lying in the non
compact part:
\begin{equation}
  \sigma \, : \, \mathcal{H}^{n.c.} \, \rightarrow \, \mathcal{H}^{n.c.}
  \quad ; \quad  \sigma \, : \, \mathcal{H}^{c} \, \rightarrow \, - \mathcal{H}^{c}
\label{perotto1}
\end{equation}
From the Cartan algebra, the action of $\sigma$ is canonically
extended to the root space. Decomposing each root in its parallel and
transverse parts we have:
\begin{equation}
  \sigma (\alpha) = \sigma \left( \alpha_{||} + \alpha_{\bot}
  \right)=\alpha_{||} - \alpha_{\bot}
\label{sigmaonalpa}
\end{equation}
If we rewrite all the sixty three $\mathrm{E}_7$ roots in the  $e_i$ basis defined by
eq.s (\ref{ortonchbas}) and (\ref{compbas}) we unveil the meaning of
our regrouping from the point of view of the automorphism $\sigma$.
The set $\Delta_\eta$ is composed by all those roots whose transverse
part vanishes, namely the components of each $\eta$-root along
$e_{5,6,7}$ are zero:
\begin{equation}
  \forall \,\eta \, \in \, \Delta_\eta \, : \, \eta_{\bot} = 0
\label{etaortonul}
\end{equation}
which implies
\begin{equation}
 \sigma \left (\eta_I \right )\, =  \, \eta_I
\label{etaortonullius}
\end{equation}
On the other hand the roots in the set $\Delta_\delta$ arrange into
pairs such that the transverse part of $\delta_I^1$ is the opposite
of that of $\delta_I^4$ and similarly that of $\delta_I^2$ is the
opposite of that of $\delta_I^3$. Hence we have:
\begin{equation}
  \sigma \left( \delta_I^1\right) = \delta_I^4 \quad ; \quad \sigma \left( \delta_I^4\right) =
  \delta_I^1 \quad ; \quad  \sigma \left( \delta_I^2\right) =
  \delta_I^3 \quad ; \quad  \sigma \left( \delta_I^3\right) =
  \delta_I^2
\label{sigondel}
\end{equation}
Finally from the root space the automorphism $\sigma$ can be lifted to
the step operators and hence to the whole algebra. This last step involves
the introduction of a set of sign factors. To see this,
let $\mathcal{H}_i $  and $ E^\alpha$ be the Cartan and the step
operators,
respectively, realized in the maximally non compact, split, real
section $\mathbb{G}_{split}$ of the complex Lie algebra $\mathbb{G}^\mathbb{C}$ . If regarded as matrices in any of its
irreducible representations both  $\mathcal{H}_i $ and $ E^\alpha$
are \textbf{real matrices}. In our case the complex Lie algebra is
$\mathrm{E_7}^\mathbb{C}$ and the maximally non compact split real section is
$\mathrm{E_{7(7)}}$. The representation we can focus on is the
fundamental $56$-dimensional representation and the explicit form of
the matrices $\mathcal{H}_i$ and $E^\alpha$ we shall utilize was constructed by us in
1997 and it is described in \cite{noie7blackholes}. This fixes the
conventions, which is a necessary step, since the definition of the step operators
is up to choices of some arbitrary signs. This being set, the lifting
of the automorphism $\sigma$ from the root space to the complex Lie algebra is
defined in the following way. Firstly the complex Lie algebra is
defined as the complex span (linear combinations with complex coefficients) of
the generators $\mathcal{H}_i $ and $ E^\alpha$:
\begin{equation}
 \mathbb{G}^\mathbb{C} = \mbox{complex span}\left\{\mathcal{H}_i, E^\alpha\right\}
\label{complexspan}
\end{equation}
Secondly, for each element $g \in \mathbb{G}^\mathbb{C}$,  we require:
\begin{equation}
  \sigma \left(\mathrm{i} \, g\right) = -{\mathrm{i}} \, \sigma\left( g\right)
\label{ibehave}
\end{equation}
where $ \mathrm{i}= \sqrt{-1}$ denotes the imaginary unit.
Thirdly the automorphism is fixed by writing its action on the generators:
\begin{equation}
  \sigma \left( \mathcal{H}_i^{||} \right) = \mathcal{H}_i^{||} \quad ; \quad \sigma \left( \mathcal{H}_i^{\bot}
  \right)= -\mathcal{H}_i^{\bot}\quad ; \quad \sigma \left( E^\alpha \right) = a_\alpha \,
  E^{\sigma(\alpha)}
\label{sigmongene}
\end{equation}
In the above equation $a_\alpha$ is a real number whose absolute is immediately fixed to one by
consistency with Jacobi identities.
Hence $a_\alpha = \pm 1$. Yet the choice of these signs is not immediately obvious.
Indeed it follows from the original choice of normalizations of the step operators for the split algebra
$\mathbb{G}_{split}$ and therefore it is convention dependent.
In a moment we shall resolve this ambiguity relative to the already
mentioned choice of conventions, namely those of \cite{noie7blackholes}.
First let us observe that once the $a_\alpha$ are fixed, the complex
linear combinations of \textbf{split generators} forming a complete
basis for the real Lie algebra $\mathbb{G}_R$ are also fixed. As an
example let us consider the maximally compact real section
$\mathbb{G}_{compact}$ for which, as it is well known, we always have:
\begin{equation}
  \mathbb{G}_{compact} =\mbox{real span}\left\{{\rm i}H_i \, , \, {\rm i}\ft 12  \left (E^\alpha +
  E^{-\alpha}
   \right) \, , \, \ft 12  \left(E^\alpha - E^{-\alpha}
   \right) \right\}
\label{compatex}
\end{equation}
In view of the previous theory this is easily explained as follows.
In this case the whole Cartan subalgebra is compact and hence
$\sigma(\mathcal{H}_i) = -\mathcal{H}_i$ for all Cartan generators.
From this it follows that $\alpha_{||}=0$ for all roots and
therefore $\sigma(\alpha) = -\alpha$. The actual linear combinations
displayed in eq.(\ref{compatex}) follow from the choice $a_\alpha =
-1,\,  \forall \alpha$ which implies:
\begin{equation}
 \sigma \left(  {\rm i}\ft 12  \left( E^\alpha +
 E^{-\alpha} \right)  \right) = {\rm i}\ft 12  \left( E^\alpha +
 E^{-\alpha} \right)  \quad; \quad  \sigma \left(  \ft 12  \left(E^\alpha - E^{-\alpha}
 \right)\right) = \ft 12  \left(E^\alpha - E^{-\alpha}
 \right)
\label{exemplum}
\end{equation}
Had we chosen $a_\alpha = 1$ we would have obtained the same linear
combinations but with the $i$-factors interchanged: $\ft 12  \left (E^\alpha +
E^{-\alpha} \right)$ , $ {\rm i}\ft 12  \left (E^\alpha -
  E^{-\alpha}\right)$. Such a choice, however, would be wrong since it
  does not define an algebra. Indeed the commutator of two generators
  of type ${\rm i}\ft 12  \left (E^\alpha -
  E^{-\alpha}\right)$ produces a generator of the same type, but
  without the $i$-factor in front. On the contrary the opposite
  choice of $a_\alpha$, which amounts to the well known choice
  (\ref{compatex}) of $i$-prefactors consistently defines a
  subalgebra.
This discussion shows that:
\begin{enumerate}
  \item {The choice of the $a_\alpha$ factors which completely
  determines the action the automorphism $\sigma$ is fully equivalent
  to deciding the position of the $i$-factors, namely to deciding
  whether, for each pair $\alpha$ and $\sigma(\alpha)$ of roots
  mapped into each other by the automorphism it is
\begin{equation}
  \ft 12  \left(E^\alpha - E^{\sigma(\alpha)}\right)\quad  ;
  \quad {\rm i} \, \ft 12  \left(E^\alpha + E^{\sigma(\alpha)}\right)
\label{primascelta}
\end{equation}
or
\begin{equation}
  \ft 12  \left(E^\alpha + E^{\sigma(\alpha)}\right)\quad  ;
  \quad {\rm i} \, \ft 12  \left(E^\alpha - E^{\sigma(\alpha)}\right)
\label{secondascelta}
\end{equation}
which appear as generators of the algebra $\mathbb{G}_\mathbb{R}$.}
  \item { The decision whether  (\ref{primascelta})   or
  (\ref{secondascelta}) is the right choice is determined by the
  commutation relations and the closure of the algebra $\mathbb{G}_\mathbb{R}$ and can
  be different for different pairs of related roots.}
\end{enumerate}
In the case of the $\mathrm{E_{7(-5)}}$ real section of the $\mathrm{E_7^\mathbb{C}}$
complex Lie algebra, using the normalization of step operators
derived in \cite{noie7blackholes} we have carefully inspected by
computer calculations all the commutation relations and we have
derived the  assignment of $i$-factors displayed in the  explicit
enumeration of generators of $\mathrm{E_{7(-5)}}$ displayed in table \ref{pirillone}.
\begin{table}
  \centering
  $$
  \begin{array}{|ll|}
  \hline
    \mbox{Cartan generators} & \left \{\begin{array}{ll}
      \mbox{compact} & \left \{ \mathcal{H}^c_i = {\rm i}\frac{1}{\sqrt{2}}\,
       (H_{2i-1} -H_{2i}) =  {\rm i}\frac{1}{\sqrt{2}}\,
      H_{\beta_i} \quad (i=1,2,3)  \right. \\
      \mbox{non compact} & \left \{\begin{array}{l}
        \mathcal{H}^{n.c.}_i=\frac{1}{\sqrt{2}}\, (H_{2i-1} + H_{2i}) \quad (i=1,2,3) \\
        \mathcal{H}^{n.c.}_4=H_7 \\
\end{array} \right. \\
    \end{array} \right.\\
    \hline
    \mbox{Step operators} & \left \{ \begin{array}{ll}
      \mbox{$\beta$-set} & \left \{ \begin{array}{lcl}
        E^+_{\beta_i} & = & {\rm i} \frac{1}{\sqrt{2}} \, \left( E^{\beta_i} +E^{-\beta_i}\right) \\
         E^-_{\beta_i} & = & \frac{1}{\sqrt{2}} \, \left( E^{\beta_i} -E^{-\beta_i}\right) \\
      \end{array} \right \} (i=1,2,3)  \\
     \mbox{$\eta$-set} & \left \{\begin{array}{lcl}
       \mathbf{E}^+_I & = & \frac{1}{\sqrt{2}} \left(E^{\eta^I} + E^{-\eta^I} \right)  \\
       \mathbf{E}^-_I & = & \frac{1}{\sqrt{2}}\left(E^{\eta^I} - E^{-\eta^I} \right) \\
     \end{array} \right \}  (I=1,\dots,12)\\
 \mbox{$\delta$-set} & \left \{ \begin{array}{lcl}
   \mathcal{E}^+_{2I-1} & = & \frac{1}{\sqrt{2}}\left( E^{\delta^1_I} + E^{\delta^4_I}\right)  \\
   \left(  \mathcal{E}^+_{2I-1}\right) ^\dagger & = & \frac{1}{\sqrt{2}} \left(E^{-\delta^1_I} + E^{-\delta^4_I} \right)\\
  \mathcal{E}^+_{2I} & = & {\rm i} \frac{1}{\sqrt{2}}\left(E^{\delta^2_I} + E^{\delta^3_I} \right) \\
   \left(  \mathcal{E}^+_{2I}\right) ^\dagger & = & - {\rm i} \frac{1}{\sqrt{2}}\left(E^{-\delta^2_I}
   + E^{-\delta^3_I} \right)\\
   \mathcal{E}^-_{2I-1} & = & {\rm i} \frac{1}{\sqrt{2}}\left( E^{\delta^1_I} - E^{\delta^4_I} \right) \\
   \left(  \mathcal{E}^-_{2I-1}\right) ^\dagger & = & - {\rm i} \frac{1}{\sqrt{2}}\left(E^{-\delta^1_I}
    - E^{-\delta^4_I} \right) \\
  \mathcal{E}^-_{2I} & = & \frac{1}{\sqrt{2}}\left( E^{\delta^2_I} + E^{\delta^3_I} \right) \\
   \left(  \mathcal{E}^-_{2I}\right) ^\dagger & = & \frac{1}{\sqrt{2}}\left(E^{-\delta^2_I} + E^{-\delta^3_I} \right) \\
 \end{array}\right \} (I=1,\dots,12)  \\
 \end{array} \right.\\
 \hline
  \end{array} $$
  \caption{Explicit enumeration of the generators of the real Lie algebra $\mathrm{E}_{7(-5)}$\label{pirillone}}
\end{table}
\subsection{The maximal compact subalgebra $\mathrm{SO(3)_R \times SO(12)}$ and the basis of coset generators}
Having explicitly constructed the  real  Lie algebra $\mathbb{G}_\mathbb{R} =
\mathrm{E_{7(-5)}}$ we can now consider its decomposition with
respect to its maximal compact subalgebra $ \mathbb{H} \equiv \mathrm{SO(3)_R \times SO(12)}$
 which is to us the most relevant issue, since the final goal of our study is the construction of
geodesic motions in the manifold (\ref{e7-5starmanif}). Being
interested in the splitting:
\begin{equation}
  \mathbb{G}_\mathbb{R} = \mathbb{H} \oplus \mathbb{K}
\label{GHK}
\end{equation}
 we proceed to establishing a canonical basis of generators for
 $\mathbb{G}_\mathbb{R}$ organized as it follows:
\begin{equation}
(A=1,\dots,133) \quad ; \quad  T_A = \left \{ \begin{array}{lcll}
T_i & =& \mathbb{H}_i & (i=1, \dots, 69) \\
T_{i+69} & = & \mathbb{K}_i & (i=1,\dots, 64) \
\end{array} \right.
\label{Tagener}
\end{equation}
where $\mathbb{H}_i$ is a basis of generators for the maximal compact
subalgebra $\mathrm{SO(3)_R \times SO(12)} $ and $\mathbb{K}_i$ is a
basis of generators for its orthogonal complement, namely for the
tangent space to the manifold (\ref{e7-5starmanif}).
With reference to table \ref{pirillone} our choice and ordering of
the basis $\mathbb{H}_i$ is the following:
\begin{eqnarray}
\mathbb{H}_{3i-2} & = & {\rm i} \, \frac{1}{\sqrt{2}} \, H_{\beta_i} \quad (i=1,2,3)\nonumber\\
\mathbb{H}_{3i-1} & = & E^+_{\beta_i} \quad (i=1,2,3)\nonumber\\
\mathbb{H}_{3i} & = & E^-_{\beta_i} \quad (i=1,2,3)\nonumber\\
\mathbb{H}_{9+I} & = & \mathbf{E}^-_I \quad ( I =1,\dots,12)\nonumber\\
\mathbb{H}_{21+A} & = & \frac{1}{\sqrt{2}}\left( \mathcal{E}^+_A -\left(\mathcal{E}^+_A \right)^\dagger \right)
 \quad ( A =1,\dots,24)\nonumber\\
\mathbb{H}_{45+A} & = & \frac{1}{\sqrt{2}}\left( \mathcal{E}^-_A -\left(\mathcal{E}^-_A \right)^\dagger \right)
 \quad ( A =1,\dots,24) \quad ( A =1,\dots,24)\nonumber\\
\label{Hbasis}
\end{eqnarray}
Correspondingly, our choice and ordering for the coset generators
$\mathbb{K}_i$ is displayed below:
\begin{eqnarray}
\mathbb{K}_i & = &  \mathcal{H}^{n.c.}_i \quad (1,2,3)\nonumber\\
\mathbb{K}_4 & = & \mathcal{H}_4 \nonumber\\
\mathbb{K}_{4+I} & = & \mathbf{E}^+_I \quad ( I =1,\dots,12) \nonumber\\
\mathbb{K}_{16+A} & = & \frac{1}{\sqrt{2}}\left( \mathcal{E}^+_A +\left(\mathcal{E}^+_A \right)^\dagger \right)
 \quad ( A =1,\dots,24) \nonumber\\
\mathbb{K}_{40+A} & = & \frac{1}{\sqrt{2}}\left( \mathcal{E}^-_A +\left(\mathcal{E}^-_A \right)^\dagger \right)
 \quad ( A =1,\dots,24)  \nonumber\\
\label{Kbasis}
\end{eqnarray}
Let us make a few comments. In our ordering of the compact  generators
$\mathbb{H}_i$, the first nine generate a special subgroup, which we
have already identified as the \textit{paint group}:
\begin{equation}
 \mathbb{G}_{\mathrm{paint}} = \mathrm{SO(3)}^3_{\beta}\equiv \mathrm{SO(3)}_{\beta_1} \times
  \mathrm{SO(3)}_{\beta_2}\times \mathrm{SO(3)}_{\beta_3}
  \, \subset  \,\mathrm{ \mathrm{SO(3)_R \times SO(12)}}
\label{so3cub}
\end{equation}
This latter is associated with the three "\textbf{compact roots}" defining
the real section and  will play an important role as
automorphism algebra of the Solvable Lie Algebra $Solv_{7(-5)}$
associated with the coset (\ref{e7-5starmanif}). It is appropriate to
stress that the subgroup $\mathrm{SO(3)}_R$ is none of these three
$\mathrm{SO(3)}_{\beta_i}$. On the contrary the subgroup
$\mathrm{SO(3)}^3_{\beta}$ sits inside $\mathrm{SO(12)}$.
The subgroup $\mathrm{SO(3)}_R$, which commutes with the whole $\mathrm{SO(12)}$, is
instead generated by the following uniquely determined linear combinations of the
generators
$\mathbb{H}_i$:
\begin{eqnarray}
  J_1^R &=& \frac{1}{2\sqrt{2}} \, \left(\mathbb{H}_{10} -\mathbb{H}_{13}+\mathbb{H}_{16}-\mathbb{H}_{21}
  \right)\nonumber\\
J_2^R &=& -\frac{1}{2\sqrt{2}} \, \left(\mathbb{H}_{12} -\mathbb{H}_{14}+\mathbb{H}_{17}+\mathbb{H}_{20}
  \right)\nonumber\\
  J_3^R &=& \frac{1}{2\sqrt{2}} \, \left(\mathbb{H}_{11} -\mathbb{H}_{15}+\mathbb{H}_{18}-\mathbb{H}_{19}
  \right)\nonumber\\
\label{Jcbasa}
\end{eqnarray}
which close the standard commutation relations:
\begin{equation}
  \left [ J_i^R \, , \, J_j^R \right ] \, = \, \epsilon_{ijk} \,
  J^R_k
\label{standaso3c}
\end{equation}
The ordering of the coset generators is obvious from equation
(\ref{Kbasis}). First we have have listed the four non-compact
Cartans, then non-compact combinations associated with the 12 roots
that project onto the long roots of $\mathrm{F}_4$ with multiplicity one.
Finally  we have listed the non-compact combinations associated
with the roots that project onto the short roots of $\mathrm{F}_4$ in exactly
the same order as their compact analogues appear in the listing of
$\mathbb{H}$-generators. From the point of view of representation
theory we know that the $\mathbb{K}$-space   transforms as follows under
$\mathrm{SO(3)}_R \times \mathrm{SO(12)}$:
\begin{equation}
 \mathbb{ K} = (\mathbf{2},\mathbf{32}_s)
\label{2times32}
\end{equation}
and we could arrange the generators into linear combinations
corresponding to the weights of the representation (\ref{2times32}),
yet this is not essential for our present purposes.
\subsection{Structure of the Solvable Lie algebra}
We can now come to the main point of our construction which relates
to the solvable Lie algebra $Solv_{\mathrm{E7}(-5)}$ whose corresponding group
manifold is metrically equivalent to the coset manifold
(\ref{e7-5starmanif}) and to its relation with the solvable Lie
algebra $Solv_{\mathrm{F}4(4)}$ whose corresponding group manifold is
metrically equivalent to the coset manifold (\ref{f4starmanif}).
\par
First we define the solvable Lie algebra $Solv_{\mathrm{E}7(-5)}$. This is
easily done. Following the general theory recalled, for instance in
\cite{primisolvi},\cite{noie7blackholes}, we know that $Solv_{\mathrm{E}7(-5)}$
is the linear span of the non-compact Cartan generators plus those
linear combinations of the \textit{positive root} step operators
which pertain to the considered real section. In practice this means:
\begin{eqnarray}
  Solv_{E7(-5)} & =& \mbox{real span}\left\{\mathcal{H}^{n.c.}_i \, , \, E^{\eta_I} \, , \, \mathcal{E}^+_A \, , \,
  \mathcal{E}^-_A \right\} \nonumber\\
   && (i=1,\dots,4 \, ; \,
  I=1,\dots , 12 \, ; \, A=1, \dots , 24)
\label{soldefi}
\end{eqnarray}
As we know the solvable algebra $Solv_{G/H}$ associated with any
non-compact coset $G/H$ has the great advantage that by exponentiation it provides a
polynomial parametrization of the coset representative and hence of
the scalar fields of supergravity. With respect to the traditional
parametrization of cosets in terms of $\exp (\mathbb{K})$ the
advantages of the solvable parametrization are obtained at one price:
while $\mathbb{K}$ is a representation of $\mathbb{H}$, the solvable
algebra $Solv_{G/H}$ is not. In the non maximally split case
something very useful, however, occurs. Although $Solv_{E7(-5)}$ is
not a representation of the full compact group $\mathrm{SO(3)}_R \times
\mathrm{SO(12)}$ yet it transforms covariantly under the action of a
proper compact subgroup, the \textit{paint group}, specifically $\mathbb{G}_{paint}=\mathrm{SO(3)}^3_\beta$, defined in
eq.(\ref{so3cub}).
Indeed the decomposition of $Solv_{E7(-5)}$ with respect to
$\mathrm{SO(3)}^3_\beta$ is:
\begin{equation}
  Solv_{E7(-5)} = 16 \, \times \, (\mathbf{1,1,1})\,  \oplus  \,  4\, \times
  \, (\mathbf{2,2,0}) \,  \oplus  \, 4\, \times
  \, (\mathbf{2,0,2}) \,\oplus \, 4\, \times
  \, (\mathbf{0,2,2}) \,
\label{decomposo3}
\end{equation}
and $\mathrm{SO(3)}^3_\beta$ works as an automorphism group of the
solvable Lie algebra. The sixteen singlets are the four  Cartan generators
plus the twelve $E^{\eta_I}$ step operators associated with long roots of
$\mathrm{F}_4$. The forty-eight non-singlets, distributed in irrepses as
described in eq.(\ref{decomposo3}) are instead the generators
associated with the $\delta$--roots that project onto the short ones
of $\mathrm{F}_4$.
\par
This covariant structure of the solvable Lie algebra $Solv_{E7(-5)}$
is responsible for its relation with $Solv_{F4(4)}$ and for the
\textit{painted billiard} phenomenon.
Let us see how.
\par
We are interested in the structure constants of the Solvable Lie
algebra which in turn determine the Nomizu connection and hence the
$1st$ order equations for the tangent vector to the geodesic
\cite{noiconsasha}.  Calling $T_\Lambda$ a
set of generators in the adjoint representation of the algebra we
read off the structure constants from the equation:
\begin{equation}
\left[ T_\Sigma \, , \, T_\Pi\right ] =  C^{\Lambda}_{\phantom{\Lambda}\Sigma\Pi} \, T_\Lambda \,
\label{Mcequegen}
\end{equation}
Let us first consider the solvable Lie algebra $Solv_{F4(4)}$
associated with the coset (\ref{f4starmanif}). This is a maximally
split case and the structure of $Solv_{F4(4)}$ is the canonical one
discussed in \cite{noiconsasha}. We have 4 Cartan generators ${H}_a$ and
$24$ positive roots that split into two subsets of $12$ long roots
$\alpha^\ell$ and $12$ short roots $\alpha^{s}$. Calling
$\Delta_\ell$ and $\Delta_s$ the two subsets we have the following
structure:
\begin{eqnarray}
  \forall \alpha^\ell \, , \, \beta^\ell \, \in \, \Delta_\ell & : & \alpha^\ell + \beta^\ell
  \,= \, \cases{\mbox{not a root or} \cr
  \gamma^\ell \, \in \, \Delta_\ell \cr}\nonumber\\
\forall \alpha^\ell \, \in \, \Delta_\ell \,\,\, \mbox{and}\,\,\, \forall \beta^s \, \in \, \Delta_s  & : &
\alpha^\ell + \beta^s
  \,= \, \cases{\mbox{not a root or} \cr
  \gamma^s \, \in \, \Delta_s \cr}\nonumber\\
  \forall \alpha^s \, , \, \beta^s \, \in \, \Delta_s & : & \alpha^s +
  \beta^s
  \,= \, \cases{\mbox{not a root or} \cr
  \gamma^s \, \in \, \Delta_s \, \, \, \mbox{or}\cr
  \gamma^\ell \, \in \, \Delta_\ell \cr}
\label{organo}
\end{eqnarray}
Consequently, let $H_i$ be the Cartan
generators and $E^{\alpha^\ell}$, $E^{\alpha^s}$ be the  step operators
respectively associated with positive long and short roots.
This set of operators completes a basis of 28 generators for the
solvable Lie algebra $Solv_{F_{4(4)}}$. In view of eq.s (\ref{organo})
the possible structure constants are:
\begin{equation}
  C^\Lambda_{\phantom{\Lambda}\Sigma\Pi}  \,\equiv \, \left \{
  C^{\alpha^\ell}_{\phantom{\alpha_\ell}j \beta^\ell}\, , \, C^{\alpha^\ell}_{\phantom{\alpha^\ell}\beta^\ell \,
\gamma^\ell} \, , \, C^{\alpha^\ell}_{\phantom{\alpha^\ell}\beta^s \,
\gamma^s} \,  , \, C^{\alpha^s}_{\phantom{\alpha^s}j\beta^s} \, , \, C^{\alpha^s}_{\phantom{\alpha^s}\beta^s \,
\gamma^s} \,  , \, C^{\alpha^s}_{\phantom{\alpha^s}\beta^\ell \,\gamma^s} \, \right \}
\label{struttureconst}
\end{equation}
and we further have:
\begin{eqnarray}
 C^{\alpha^\ell}_{\phantom{\alpha_\ell}j \beta^\ell}\, & = & \delta^{\alpha^\ell}_{\beta^\ell} \, \alpha^\ell_j \nonumber\\
C^{\alpha^s}_{\phantom{\alpha_\ell}j \beta^s}\, & = & \delta^{\alpha^s}_{\beta^s} \, \alpha^s_j \nonumber\\
C^{\alpha^\ell}_{\phantom{\alpha^\ell}\beta^\ell \,
\gamma^\ell} \, & = & \delta^{\alpha^\ell}_{\beta^\ell + \gamma^\ell} \, N_{\beta^\ell\gamma^\ell}\nonumber\\
C^{\alpha^s}_{\phantom{\alpha^\ell}\beta^\ell \,
\gamma^s} \, & = & \delta^{\alpha^s}_{\beta^\ell + \gamma^s} \, N_{\beta^\ell\gamma^s}\nonumber\\
C^{\alpha^\ell}_{\phantom{\alpha^\ell}\beta^s \,
\gamma^s} \, & = & \delta^{\alpha^\ell}_{\beta^s + \gamma^s} \, N_{\beta^s\gamma^s}\nonumber\\
C^{\alpha^s}_{\phantom{\alpha^\ell}\beta^s \,
\gamma^s} \, & = & \delta^{\alpha^s}_{\beta^s + \gamma^s} \, N_{\beta^s\gamma^s}\nonumber\\
\label{cunstanti}
\end{eqnarray}
where the matrix $N_{\beta \gamma }$, defined by the standard Cartan-Weyl commutation relations
as given in eq.(\ref{cartaweila}) of the appendix, or in table \ref{paragonando}, differs from zero only when
the sum of the two roots $\beta$ and $\gamma$ is a root. Hence it
suffices to know $N_{\beta \gamma }$ and the solvable Lie algebra
structure constants are completely determined. In the following three
tables (\ref{nalbeLL}), (\ref{nalbeLS}), (\ref{nalbeSS}) we
exhibit the values of $N_{\beta \gamma }$ for the $\mathrm{F}_{4(4)}$ Lie
algebra.
\begin{equation}
 \underbrace{\begin{array}{|c|c|c|c|c|c|c|c|c|c|c|c||l|}
 \hline
     \alpha^\ell_{1} & \alpha^\ell_{2} & \alpha^\ell_{3} &
     \alpha^\ell_{4} & \alpha^\ell_{5} & \alpha^\ell_{6} & \alpha^\ell_{7} &
     \alpha^\ell_{8} & \alpha^\ell_{9} & \alpha^\ell_{10} &\alpha^\ell_{11} & \alpha^\ell_{12} & \null \\
 \hline
 \hline
 0 & -{\sqrt{2}} & 0 & 0 & -{\sqrt{2}} & 0 & 0 & -{\sqrt{2}} & 0 & {
     \sqrt{2}} & 0 & 0 & \alpha^\ell_1\cr
     \hline
     {\sqrt{2}} & 0 & 0 & -{\sqrt{2}} & 0 & 0 & -{
      \sqrt{2}} & 0 & 0 & 0 & -{\sqrt{2}} & 0 & \alpha^\ell_2\cr
      \hline
      0 & 0 & 0 & -{\sqrt{2}} & 0 & 0 & -{
      \sqrt{2}} & 0 & 0 & -{\sqrt{2}} & 0 & 0 & \alpha^\ell_3\cr
      \hline
      0 & {\sqrt{2}} & {
     \sqrt{2}} & 0 & 0 & 0 & 0 & -{\sqrt{2}} & {\sqrt{2}} & 0 & 0 & 0 & \alpha^\ell_4\cr
     \hline
     {\sqrt{2}} & 0 & 0 & 0 & 0 & 0 & {\sqrt{2}} & 0 & {
     \sqrt{2}} & 0 & 0 & 0 & \alpha^\ell_5\cr
     \hline
     0 & 0 & 0 & 0 & 0 & 0 & -{\sqrt{2}} & -{
      \sqrt{2}} & 0 & 0 & 0 & 0 & \alpha^\ell_6 \cr
      \hline
      0 & {\sqrt{2}} & {\sqrt{2}} & 0 & -{\sqrt{2}} & {
     \sqrt{2}} & 0 & 0 & 0 & 0 & 0 & 0 & \alpha^\ell_7 \cr
     \hline
     {\sqrt{2}} & 0 & 0 & {\sqrt{2}} & 0 & {
     \sqrt{2}} & 0 & 0 & 0 & 0 & 0 & 0 & \alpha^\ell_8\cr
     \hline
     0 & 0 & 0 & -{\sqrt{2}} & -{
      \sqrt{2}} & 0 & 0 & 0 & 0 & 0 & 0 & 0 & \alpha^\ell_9\cr
      \hline
      -{\sqrt{2}} & 0 & {
     \sqrt{2}} & 0 & 0 & 0 & 0 & 0 & 0 & 0 & 0 & 0 & \alpha^\ell_{10}\cr
     \hline
     0 & {
     \sqrt{2}} & 0 & 0 & 0 & 0 & 0 & 0 & 0 & 0 & 0 & 0 & \alpha^\ell_{11}\cr
     \hline
     0 & 0 & 0 & 0 & 0 & 0 & 0 & 0 &
   0 & 0 & 0 & 0 & \alpha^\ell_{12}\cr
   \hline
   \end{array}}_{N_{\alpha^\ell\beta^\ell}}
\label{nalbeLL}
\end{equation}
\begin{equation}
  \underbrace{\begin{array}{|c|c|c|c|c|c|c|c|c|c|c|c||l|}
    \hline
     \alpha^s_{1} & \alpha^s_{2} & \alpha^s_{3} &
     \alpha^s_{4} & \alpha^s_{5} & \alpha^s_{6} & \alpha^s_{7} &
     \alpha^s_{8} & \alpha^s_{9} & \alpha^s_{10} &\alpha^s_{11} & \alpha^s_{12} & \null \\
 \hline
 \hline
 0 & {\sqrt{2}} & 0 & -{\sqrt{2}} & 0 & 0 & 0 & 0 & {
     \sqrt{2}} & 0 & 0 & 0 & \alpha^\ell_1 \cr 0 & 0 & -{\sqrt{2}} & 0 & 0 & -{\sqrt{2}} & 0 & {
     \sqrt{2}} & 0 & 0 & 0 & 0 & \alpha^\ell_2 \cr 0 & -{\sqrt{2}} & 0 & {\sqrt{2}} & 0 & 0 & 0 & -{
      \sqrt{2}} & 0 & 0 & 0 & 0 & \alpha^\ell_3 \cr {\sqrt{2}} & 0 & 0 & 0 & 0 & 0 & {
     \sqrt{2}} & 0 & 0 & 0 & 0 & 0 & \alpha^\ell_4 \cr -{\sqrt{2}} & 0 & 0 & 0 & 0 & -{
      \sqrt{2}} & 0 & 0 & 0 & 0 & 0 & 0 & \alpha^\ell_5 \cr {\sqrt{2}} & 0 & 0 & {
     \sqrt{2}} & 0 & 0 & 0 & 0 & 0 & 0 & 0 & 0 & \alpha^\ell_6 \cr 0 & 0 & 0 & 0 & -{
      \sqrt{2}} & 0 & 0 & 0 & 0 & 0 & 0 & 0 & \alpha^\ell_7 \cr 0 & 0 & {
     \sqrt{2}} & 0 & 0 & 0 & 0 & 0 & 0 & 0 & 0 & 0 & \alpha^\ell_8 \cr 0 & {
     \sqrt{2}} & 0 & 0 & 0 & 0 & 0 & 0 & 0 & 0 & 0 & 0 & \alpha^\ell_9 \cr 0 & 0 & 0 & 0 & 0 & 0 & 0 & 0 &
   0 & 0 & 0 & 0 & \alpha^\ell_{10} \cr 0 & 0 & 0 & 0 & 0 & 0 & 0 & 0 & 0 & 0 & 0 & 0 & \alpha^\ell_{11} \cr 0 & 0 & 0 & 0 & 0 &
   0 & 0 & 0 & 0 & 0 & 0 & 0 & \alpha^\ell_{12} \cr
  \end{array}}_{N_{\alpha^\ell\beta^s}}
\label{nalbeLS}
\end{equation}
\begin{equation}
  \underbrace{\begin{array}{|c|c|c|c|c|c|c|c|c|c|c|c||l|}
    \hline
     \alpha^s_{1} & \alpha^s_{2} & \alpha^s_{3} &
     \alpha^s_{4} & \alpha^s_{5} & \alpha^s_{6} & \alpha^s_{7} &
     \alpha^s_{8} & \alpha^s_{9} & \alpha^s_{10} &\alpha^s_{11} & \alpha^s_{12} & \null \\
 \hline
  0 & 1 & -1 & 0 & -1 & 0 & 0 & {\sqrt{2}} & -{\sqrt{2}} & {\sqrt{2}} & 1 & 0 &\alpha^s_1 \cr
  \hline
    -1 & 0 & {\sqrt{2}} & 0 & {\sqrt{2}} & 1 & -1 & 0 & 0 & -1 & 0 & {
     \sqrt{2}} &\alpha^s_2 \cr
\hline
      1 & -{\sqrt{2}} & 0 & 1 & -{\sqrt{2}} & 0 & -1 & 0 & 1 & 0 & 0 & {
     \sqrt{2}} &\alpha^s_3 \cr
\hline
      0 & 0 & -1 & 0 & 1 & {\sqrt{2}} & {\sqrt{2}} & 0 & 0 & 1 & -{
      \sqrt{2}} & 0 &\alpha^s_4 \cr
\hline
       1 & -{\sqrt{2}} & {\sqrt{2}} &
    -1 & 0 & 1 & 0 & 1 & 0 & 0 & 0 & {\sqrt{2}} &\alpha^s_5 \cr
\hline
     0 & -1 & 0 & -{\sqrt{2}} &
    -1 & 0 & {\sqrt{2}} & 0 & 1 & 0 & {\sqrt{2}} & 0 &\alpha^s_6 \cr
\hline
     0 & 1 & 1 & -{
      \sqrt{2}} & 0 & -{\sqrt{2}} & 0 & 1 & 0 & 0 & {\sqrt{2}} & 0 &\alpha^s_7 \cr
\hline
       -{
      \sqrt{2}} & 0 & 0 & 0 & -1 & 0 & -1 & 0 & {\sqrt{2}} & {\sqrt{2}} & 0 & 0 &\alpha^s_8 \cr
\hline
       {
     \sqrt{2}} & 0 & -1 & 0 & 0 & -1 & 0 & -{\sqrt{2}} & 0 & -{\sqrt{2}} & 0 & 0 &\alpha^s_ 9\cr
\hline
      -
     {\sqrt{2}} & 1 & 0 & -1 & 0 & 0 & 0 & -{\sqrt{2}} & {\sqrt{2}} & 0 & 0 & 0 &\alpha^s_{10} \cr
\hline
    -1 & 0 & 0 & {\sqrt{2}} & 0 & -{\sqrt{2}} & -{
      \sqrt{2}} & 0 & 0 & 0 & 0 & 0 &\alpha^s_{11} \cr
\hline
       0 & -{\sqrt{2}} & -{\sqrt{2}} & 0 & -{
      \sqrt{2}} & 0 & 0 & 0 & 0 & 0 & 0 & 0 &\alpha^s_{12} \cr
  \end{array}}_{N_{\alpha^s\beta^s}}
\label{nalbeSS}
\end{equation}
The ordering of long and short roots of the $\mathrm{F}_4$ system is that used in
tables: \ref{etaset} and \ref{deltaset}. On the other hand the  explicit determination
of the tensor $N_{\alpha\beta}$ which appears in the standard Cartan-Weyl
commutation relations was performed via the explicit construction
of the fundamental $26$-dimensional representation of this Lie
algebra. This construction is described in appendix \ref{appendoF4}.
\par
Now the exciting point about the solvable Lie algebra of the full non
split coset (\ref{e7-5starmanif}) which contains all the degrees of freedom of
 supergravity is that it can be exhibited in terms of
the structure constants of its split Tits Satake submanifold
(\ref{f4starmanif}) by utilizing also the covariance with respect to the
compact \textbf{paint subgroup} $\mathbb{G}_{paint} = \mathrm{SO(3)}^3_\beta$.
This is the result of an essential interplay of impressive
elegance between the graded structure of the split Tits Satake
algebra, which is non simply laced and for that reason contains a
distinction between \textbf{short} and \textbf{long} roots, and the
rearrangement of those roots of $\mathrm{E}_7$ which project onto the short ones
of $\mathrm{F}_4$ into representations of the compact \textbf{paint group}
$\mathrm{SO(3)}^3_\beta$.
\par
\begin{table}
  \centering
  $$
\begin{array}{|lclclcl|}
\hline
  H[1] &=& \frac{1}{\sqrt{2}} \,\mathcal{H}^{n.c.}_1 & ; & H[2] &=&
 \frac{1}{\sqrt{2}} \,
  \mathcal{H}^{n.c.}_2 \\ H[3] &=& \frac{1}{\sqrt{2}}  \, \mathcal{H}^{n.c.}_3 &
  ; & H[4] &=&\frac{1}{\sqrt{2}}  \, \mathcal{H}^{n.c.}_4 \\
  \hline
\end{array}
  $$
  \caption{Listing of Cartan generators of $Solv_{E_{7(-5)}}$ which exactly correspond to the Cartan generators of
  $\mathrm{F}_{4(4)}$ \label{carticorri}}
\end{table}

\begin{table}
 \centering
$$
  \begin{array}{|lclclclclcl|}
  \hline
    \Phi[\alpha^\ell_1] & = &  \frac{1}{\sqrt{2}}  \, E^{\eta_1} & ; &  \Phi[\alpha^\ell_2] & = &
    \frac{1}{\sqrt{2}}  \, E^{\eta_2} & ; &  \Phi[\alpha^\ell_3] &
    = & -\frac{1}{\sqrt{2}}  \, E^{\eta_3} \\
     \Phi[\alpha^\ell_4] & = & \frac{1}{\sqrt{2}}  \, E^{\eta_4} & ; &  \Phi[\alpha^\ell_5] & = &\frac{1}{\sqrt{2}}  \, E^{\eta_5} & ; &
     \Phi[\alpha^\ell_6] &
     = & \frac{1}{\sqrt{2}} \, E^{\eta_6} \\
     \Phi[\alpha^\ell_7] & = &\frac{1}{\sqrt{2}}  \, E^{\eta_7} & ; &  \Phi[\alpha^\ell_8] & = & \frac{1}{\sqrt{2}}  \, E^{\eta_8} & ; &
     \Phi[\alpha^\ell_9] &
      = &\frac{1}{\sqrt{2}}  \, E^{\eta_9} \\
     \Phi[\alpha^\ell_{10}] & = & \frac{1}{\sqrt{2}}  \, E^{\eta_{10}} & ; &  \Phi[\alpha^\ell_{11}] & = &
     -\frac{1}{\sqrt{2}}  \, E^{\eta_{11}} &
     ; &  \Phi[\alpha^\ell_{12}] & = & - \frac{1}{\sqrt{2}}  \, E^{\eta_{12}} \\
     \hline
  \end{array}
$$
  \caption{\label{Philisting} {Listing of generators of $Solv_{E7(-5)}$ which correspond to long roots of $\mathrm{F}_4$
  and appear in one copy. The order of $\eta$-roots is that listed in table \ref{etaset}.}}
\end{table}

\begin{table}
 \centering
{\small $$
\begin{array}{|c|l|l|l|l|}
\hline
i & \Omega_0[\alpha^s_i]\, = \, &  \Omega_X[\alpha^s_i]\, = \, & \Omega_Y[\alpha^s_i]\, = \, & \Omega_Z[\alpha^s_i]\, = \,
\nonumber\\
\hline
 1 &  {\frac{\rm i}{2} \, \left( {E }^{{{\delta }_2}(1)} + {E }^{{{\delta
}_3}(1)} \right)
}  &   {\frac{\rm i}{2} \, \left(
{E }^{{{\delta }_1}(1)} - {E }^{{{\delta }_4}(1)} \right)
}  &  -\frac{1}{2}\left(  {{E }^{{{
\delta }_1}(1)} + {E }^{{{\delta }_4}(1)}} \right)
  &  \frac{1}{2}( {{E }^{{{\delta }_2}(1)} -
{E }^{{{\delta }_3}(1)}}) \nonumber\\
 2 &
 {\frac{\rm i}{2} \, \left( {E }^{{{\delta }_2}(2)} + {E }^{{{\delta
}_3}(2)} \right)
}  &   {-\frac{\rm i}{2} \, \left(
{E }^{{{\delta }_1}(2)} - {E }^{{{\delta }_4}(2)} \right)
}  &   {\frac{1}{2}({E }^{{{\delta
}_1}(2)} + {E }^{{{\delta }_4}(2)})}  &
  {\frac{1}{2}(-{E }^{{{\delta }_2}(2)} + {E }^{{{\delta
}_3}(2)})} \nonumber\\
  3 &
 {-\frac{\rm i}{2} \, \left( {E }^{{{\delta }_2}(3)} + {E }^{{{\delta
}_3}(3)} \right)
}  &   {\frac{\rm i}{2} \, \left(
{E }^{{{\delta }_1}(3)} - {E }^{{{\delta }_4}(3)} \right)
}  &  -\frac{1}{2} ({{E }^{{{\delta
}_1}(3)} + {E }^{{{\delta }_4}(3)}})  &
 \frac{1}{2} ({{E }^{{{\delta }_2}(3)} - {E }^{{{\delta
}_3}(3)}}) \nonumber\\
  4 &
-\frac{1}{2}\left(  {{E }^{{{\delta }_1}(4)} + {E }^{{{\delta
}_4}(4)}} \right)
  &  \frac{1}{2} {{E }^{{{\delta }_2}(4)} -
{E }^{{{\delta
}_3}(4)}}  &   {\frac{\rm i}{2} \,
\left( {E }^{{{\delta }_2}(4)} + {E }^{{{\delta }_3}(4)} \right)
}  &   {-\frac{\rm i}{2} \, \left(
{E }^{{{\delta }_1}(4)} - {E }^{{{\delta }_4}(4)} \right)
} \nonumber\\
  5 &
 {-\frac{\rm i}{2} \, \left( {E }^{{{\delta }_2}(5)} + {E }^{{{\delta
}_3}(5)} \right)
}  &   {\frac{\rm i}{2} \, \left(
{E }^{{{\delta }_1}(5)} - {E }^{{{\delta }_4}(5)} \right)
}  & - \frac{1}{2}( {{E }^{{{\delta
}_1}(5)} + {E }^{{{\delta }_4}(5)}} ) &
 -\frac{\rm i}{2}( {-{E }^{{{\delta }_2}(5)} + {E }^{{{\delta
}_3}(5)}}) \nonumber\\
  6 &
-\frac{1}{2}\left(  {{E }^{{{\delta }_1}(6)} + {E }^{{{\delta
}_4}(6)}} \right)
  &  \frac{1}{2}( {{E }^{{{\delta }_2}(6)} -
{E }^{{{\delta
}_3}(6)}} ) &   {\frac{\rm i}{2} \,
\left( {E }^{{{\delta }_2}(6)} + {E }^{{{\delta }_3}(6)} \right)
}  &   {-\frac{\rm i}{2} \, \left(
{E }^{{{\delta }_1}(6)} - {E }^{{{\delta }_4}(6)} \right)
} \nonumber\\
  7 &-\frac{1}{2}\left(  {{E }^{{{\delta }_1}(7)} + {E }^{{{\delta
}_4}(7)}} \right)
  &  \frac{1}{2}( {{E }^{{{\delta }_2}(7)} -
{E }^{{{\delta
}_3}(7)}})  &   {\frac{\rm i}{2} \,
\left( {E }^{{{\delta }_2}(7)} + {E }^{{{\delta }_3}(7)} \right)
}  &   {-\frac{\rm i}{2} \, \left(
{E }^{{{\delta }_1}(7)} - {E }^{{{\delta }_4}(7)} \right)
} \nonumber\\
  8 &
 {-\frac{\rm i}{2} \, \left( {E }^{{{\delta }_2}(8)} + {E }^{{{\delta
}_3}(8)} \right)
}  &   {-\frac{\rm i}{2} \, \left(
{E }^{{{\delta }_1}(8)} - {E }^{{{\delta }_4}(8)} \right)
}  &  \frac{1}{2}\left(  {{E }^{{{
\delta }_1}(8)} + {E }^{{{\delta }_4}(8)}} \right)
  & - \frac{1}{2}( {{E }^{{{\delta }_2}(8)} -
{E }^{{{\delta }_3}(8)}}) \nonumber\\
  9 &
 {\frac{\rm i}{2} \, \left( {E }^{{{\delta }_2}(9)} + {E }^{{{\delta
}_3}(9)} \right)
}  &   {\frac{\rm i}{2} \, \left(
{E }^{{{\delta }_1}(9)} - {E }^{{{\delta }_4}(9)} \right)
}  & -\frac{1}{2} \left(  {{E }^{{{
\delta }_1}(9)} + {E }^{{{\delta }_4}(9)}} \right)
  &  \frac{1}{2}\left( + {{E }^{{{\delta }_2}(9)} -
{E }^{{{\delta }_3}(9)}}\right)  \nonumber\\
  10 &
 {-\frac{\rm i}{2} \, \left( {E }^{{{\delta }_2}(10)} + {E }^{{{\delta
}_3}(10)} \right)
}  &   {-\frac{\rm i}{2} \, \left(
{E }^{{{\delta }_1}(10)} - {E }^{{{\delta }_4}(10)} \right)
}  &  \frac{1}{2}\left(  {{E }^{{{
\delta }_1}(10)} + {E }^{{{\delta }_4}(10)}} \right)
  &  -\frac{1}{2}( {{E }^{{{\delta }_2}(10)} -
{E }^{{{\delta }_3}(10)}}) \nonumber\\
  11 &
\frac{1}{2}\left(  {{E }^{{{\delta }_1}(11)} + {E }^{{{\delta
}_4}(11)}} \right)
  &  -\frac{1}{2}\left(  {{E }^{{{\delta }_2}(11)} -
{E }^{{{\delta }_3}(11)}} \right)  &
  -{\frac{\rm i}{2} \, \left( {E }^{{{\delta }_2}(11)} +
{E }^{{{\delta }_3}(11)} \right) }  &
  {\frac{\rm i}{2} \, \left( {E }^{{{\delta }_1}(11)} -
{E }^{{{\delta }_4}(11)} \right)
} \nonumber\\
  12 &
 {\frac{\rm i}{2} \, \left( {E }^{{{\delta }_1}(12)} - {E }^{{{\delta
}_4}(12)} \right)
}  &   {-\frac{\rm i}{2} \, \left(
{E }^{{{\delta }_2}(12)} + {E }^{{{\delta }_3}(12)} \right)
}  &   {\frac{1}{2}({E }^{{{\delta
}_2}(12)} - {E }^{{{\delta }_3}(12)})}  &
 \frac{1}{2}\left(  {{E }^{{{\delta }_1}(12)} + {E
}^{{{\delta }_4}(12)}} \right)  \nonumber\\
\hline
\end{array}
$$}
  \caption{Listing of generators of $Solv_{E7(-5)}$ which correspond to short roots of $\mathrm{F}_4$
  and appear in four copies. The order of $\delta roots$ is that displayed in table \ref{deltaset}}\label{Omegalisting}
\end{table}
\par
We already emphasized that, under the action of the paint group
(\ref{so3cub}), the generators of the Solvable Lie algebra
$Solv_{E_{7(-5)}}$ decompose into the irreducible representations
mentioned in eq. (\ref{decomposo3}). Let us define the diagonal
subgroup of the three $\mathrm{SO(3)}_\beta$:
\begin{equation}
 \mathbb{G}^0_{\mathrm{paint}} \equiv \mathrm{SO(3)}^{diag}_\beta = \mbox{diagonal} \, \left[\mathrm{SO(3)}_{\beta_1}
 \times
  \mathrm{SO(3)}_{\beta_2} \times \mathrm{SO(3)}_{\beta_3} \right ]
\label{so3diagodefi}
\end{equation}
since
\begin{equation}
   \mathbf{2 }\otimes \mathbf{2} = \mathbf{1} \oplus \mathbf{3}
\label{2x2=1+3}
\end{equation}
holds true for  $\mathrm{SO(3)}$ representations, it follows that
under $ \mathrm{SO(3)}^{diag}_\beta$ the Lie algebra $Solv_{E_{7(-5)}}$
decomposes as follows:
\begin{equation}
  Solv_{E_{7(-5)}} \stackrel{\mathrm{SO(3)}^{diag}_\beta}{\Longrightarrow}
  \left (\underbrace{4}_{\mbox{Cartan}} + \underbrace{12}_{ \mbox{long roots}}
  + \underbrace{12}_{\mbox{short roots}} \right ) \times \mathbf{1}
  \bigoplus \left( \underbrace{12}_{\mbox{short roots}}\right) \times \mathbf{3}
\label{decomposo3diago}
\end{equation}
Hence the representation $\mathbf{J}$ mentioned in
eq.s (\ref{splittatonibus}) and (\ref{Irango}) is $\mathbf{J=3}$, the
triplet of $\mathrm{SO(3)}_{diag}$.
The decomposition (\ref{decomposo3diago}) is explicitly exhibited in tables
\ref{carticorri}, \ref{Philisting} and \ref{Omegalisting}. In table
\ref{carticorri}, modulo some changes in normalization we list the non compact Cartan
generators of $\mathrm{E_{7(7)}}$ which correspond to the full set of Cartan generators
of $\mathrm{F}_{4(4)}$. In table \ref{Philisting} we define a set of generators
$\Phi[\alpha^s]$ associated with the long roots of $\mathrm{F}_{4(4)}$, which
are obviously given by the step operators $E^\eta$ of $\mathrm{E}_{7(-5)}$
since the $\eta$--roots project on such short roots of $\mathrm{F}_4$. There
are just some suitable $\pm \sqrt{2}$ factors in the normalization
which are purposely chosen in order to make the relation between the
two solvable Lie algebras clean. All these generators are singlets
under $\mathrm{SO(3)}^3_\beta$ and therefore also under
$\mathrm{SO(3)}^{diag}_\beta$. Finally in table \ref{Omegalisting} we
list  a set of four $\mathrm{E}_{7(-5)}$
generators $\Omega_{A}[\alpha^s]$, ($A=0,X,Y,Z$),  associated with each of the short roots $\alpha^s$ of
$\mathrm{F}_{4(4)}$. Indeed each such root is the image, in the projection, of four
different $\mathrm{E_{7(-5)}}$ roots, namely,  the $\delta_I^i$
roots, displayed in table \ref{deltaset}. Hence the four
$\Omega_{A}[\alpha^s]$ operators
are, with convenient normalization factors, step operators of $\mathrm{E_{7(-5)}}$
corresponding to the $\delta$-roots. The normalization factors and the
precise correspondence is chosen in such a way that the $\Omega_0$
are singlets under $\mathrm{SO(3)}^{diag}_\beta$, while
$\Omega_{x=X,Y,Z}$ form a triplet.
\par
If we use these generators and, in order to avoid proliferation of
symbols, we denote by the same letter the generator of the algebra
and its dual one-form appearing in the Maurer Cartan equations:
\begin{equation}
  dT^\Lambda = \ft 12 \, C^\Lambda_{\Sigma\Pi} T^\Sigma \, \wedge \,
  T^\Pi
\label{MCGE}
\end{equation}
the
structure constants of $Solv_{E_{7(-5)}}$ can be exhibited by writing the
following Maurer Cartan equations, which just contain the structure constants of
the $\mathrm{F}_{4(4)}$ solvable algebra, discussed before, plus the
quaternionic structure anticipated in table (\ref{paragonando}) of the
introduction.
\begin{eqnarray}
dH_i & = & 0 \nonumber\\
d\Phi[\alpha^\ell] & = & \, C^{\alpha^\ell}_{\phantom{\alpha^\ell}j \,
\beta^\ell} \, H_j \, \wedge \, \Phi[\beta^\ell] +\ft 12 \, C^{\alpha^\ell}_{\phantom{\alpha^\ell}\beta^\ell \,
\gamma^\ell} \, \Phi[\beta^\ell] \, \wedge \, \Phi[\gamma^\ell] \, +
\, \nonumber\\
&&
\ft 12 \, C^{\alpha^\ell}_{\phantom{\alpha^\ell}\beta^s \,
\gamma^s} \, \left ( \Omega_0[\beta^s] \, \wedge \, \Omega_0[\gamma^s] +
\Omega_x[\beta^s] \, \wedge \, \Omega_x[\gamma^s]\right )\nonumber\\
d\Omega_0[\alpha^s] & = & \, C^{\alpha^s}_{\phantom{\alpha^s}j \,
\beta^s} \, H_j \, \wedge \, \Omega_0[\beta^s] \nonumber\\
&& +\ft 12 \, C^{\alpha^s}_{\phantom{\alpha^s}\beta^s \,
\gamma^s} \, \left ( \Omega_0[\beta^s] \, \wedge \, \Omega_0[\gamma^s] \, +
\, \Omega_x[\beta^s] \, \wedge \, \Omega_x[\gamma^s] \,\right ) \nonumber\\
&& +
\,
 \, C^{\alpha^s}_{\phantom{\alpha^s}\beta^\ell \,
\gamma^s} \, \Phi[\beta^\ell] \, \wedge \, \Omega_0[\gamma^s]\nonumber\\
d\Omega_x[\alpha^s] & = & \, C^{\alpha^s}_{\phantom{\alpha^s}j \,
\beta^s} \, H_j \, \wedge \, \Omega_x[\beta^s] \nonumber\\
&& +\ft 12 \, C^{\alpha^s}_{\phantom{\alpha^s}\beta^s \,
\gamma^s} \,  \left ( \Omega_0[\beta^s] \, \wedge \, \Omega_x[\gamma^s] \,
- \Omega_x[\beta^s] \, \wedge \, \Omega_0[\gamma^s] -
\, \epsilon^{xyz} \, \Omega_y[\beta^s] \, \wedge \, \Omega_z[\gamma^s] \,\right ) \nonumber\\
&& +
\,
 \, C^{\alpha^s}_{\phantom{\alpha^s}\beta^\ell \,
\gamma^s} \, \Phi[\beta^\ell] \, \wedge \, \Omega_x[\gamma^s]\nonumber\\
\label{E7MC}
\end{eqnarray}
\par
Equations (\ref{E7MC}) are just a short way of writing all commutation relations and exhibit  the
interplay between the graded structure of $Solv_{F_4(4)}$ and the
structure of the \textbf{paint group} representation. Indeed we see
the announced
\textbf{quaternionic structure}! What actually happens is that the
\textbf{Cartan} and \textbf{long root generators} are isomorphic in the two algebras,
while the \textbf{short root generators} are promoted to quaternions
while going from $\mathrm{F}_{4(4)}$ to $\mathrm{E}_{7(-5)}$. We can write
\begin{equation}
F_{4(4)} \ni E^{\alpha^s} \, \Longrightarrow \, \Omega_0[\alpha^s] +
\Omega_X[\alpha^s]\, j^X + \Omega_Y[\alpha^s]\, j^Y +
\Omega_Z[\alpha^s]\, j^Z \, \in \, E_{7(-5)}
\label{promoquater}
\end{equation}
where $j^X,j^Y,j^Z$ are the three quaternionic imaginary units.
\par
This structure has very relevant consequences for the solution of the
differential equations and for the billiard phenomenon.  Since the
Nomizu connection determining the first order equations for tangent
vectors is completely determined by the structure constants of the
solvable Lie algebra we can just adopt the following strategy:
\begin{description}
  \item[1] Rather than considering the original problem associated
  with the non split manifold (\ref{e7-5starmanif}) we consider the
  problem associated with the split Tits Satake manifold
  (\ref{f4starmanif}), which can be solved along the lines of paper
  \cite{noiconsasha} using, in particular, the compensator method to
  integrate the 1st order differential equations.
  \item[2] Once we have obtained a solution for the system described
  by the structure constants (\ref{struttureconst}-\ref{cunstanti}) we also posses a
  particular solution for the system (\ref{E7MC}). It just correspond
  to setting the fields associated with $\Omega_{X,Y,Z}$ to zero.
  \item[3] A large class of solutions of the system (\ref{E7MC}) can be
  obtained from the general solution of the $\mathrm{F}_4$ system with structure constants
  (\ref{struttureconst}-\ref{cunstanti}) by
  means of \textbf{global rotations} of the \textbf{paint group}
  $\mathbb{G}_{paint} = \mathrm{SO(3)}^3_\beta$.
\end{description}
From the point of view of the \textit{billiard picture} we know that
switching on root fields correspond to the introduction of dynamical
walls on which the fictitious cosmic ball will bounce. The structure
of the solvable algebra implies that the billiard chamber is just the
Weyl chamber of $\mathrm{F}_{4(4)}$, yet certain dynamical walls are
\textit{painted}, namely occur in four copies constituting a
quaternion. In explicit solutions we see also the color of the actual
wall which is excited.
\section{The first order equations for the tangent vectors}
\label{firstorder}
As we showed in \cite{noiconsasha}, the field equations of the purely
time dependent $\sigma$-model, which is what we are supposed to solve
in our quest for  time dependent solutions of supergravity, can be
written as follows:
\begin{equation}
\label{D=3feqn} \dot{Y}^A \,+\, \Gamma^A_{BC} Y^B Y^C
\,=\,0
\end{equation}
where ${Y}^A$ denotes the purely time dependent tangent vectors to the geodesic  in an anholonomic
basis:
\begin{equation}
Y^A \,=\,  \cases{Y^i = {V}^i_I\left(\phi\right) \dot{\phi}^I  \quad \quad \quad i\in \mbox{CSA} \cr
Y^\alpha = \sqrt{2} \, {V}^\alpha_I\left(\phi\right) \dot{\phi}^I   \quad i\in \mbox{positive root system
$\Delta_+$}\cr}
\label{anholotang}
\end{equation}
${V}^A_I\left(\phi\right) d{\phi}^I$ being the vielbein  of
the target manifold we are considering.
In eq. (\ref{D=3feqn}) the symbol $\Gamma^A_{BC}$ denotes the components of the Levi-Civita
connection in the chosen anholonomic basis. Explicitly they are
related to the components of the Levi Civita connection in an
arbitrary holonomic basis by:
\begin{equation}
\Gamma^A_{BC}\,= \Gamma^I_{JK}V^A_I V^J_B V^K_C - \partial_K
(V^A_J)V^J_B V^K_C, \label{capindgamma}
\end{equation}
where the inverse vielbein is defined in the usual way:
\begin{equation}
V^A_I \, V^I_B \,=\, \delta^A_B \label{invervielb}
\end{equation}
The basic idea of \cite{noiconsasha}, which was exploited together with the \textit{compensator method}
in order to construct
explicit solutions,  is the following. As already recalled in eq.(\ref{1storder}), the connection
$\Gamma^A_{BC}$ can be identified with the \textit{Nomizu
connection} defined on a solvable Lie algebra, if the coset
representative $\mathbb{L}$ from which we construct the vielbein
is solvable, namely if it is represented as the exponential of the associated solvable Lie algebra $Solv(\mathrm{U/H})$.
 In fact, as we can read in \cite{alekseevskii} once we
have defined over $Solv$ a non degenerate, positive definite
symmetric form:
\begin{eqnarray}
\langle \,,\, \rangle & \; : \; Solv \otimes Solv \longrightarrow \mathbb{R} \nonumber \\
\langle X \,,\, Y \rangle & \; = \; \langle Y \,,\, X \rangle
\end{eqnarray}
whose lifting to the manifold produces the metric, the covariant
derivative is defined through the \textbf{Nomizu operator}:
\begin{equation}
\forall X \in Solv \,:\, \mathbb{L}_X: Solv \longrightarrow Solv
\end{equation}
so that
\begin{equation}
\forall X,Y,Z \in Solv \,:\, 2 \langle Z \,,\, \mathbb{L}_X Y
\rangle \,=\, \langle Z, \left[ X,Y \right] \rangle \,-\, \langle
X, \left[ Y,Z \right] \rangle \,-\, \langle Y, \left[ X,Z \right]
\rangle \label{Nomizuoper}
\end{equation}
while the Riemann curvature 2-form is given by the commutator of
two Nomizu operators:
\begin{equation}
R^W_{\phantom{W}Z} \left( X,Y \right) \,=\, \langle W \,,\,
\left\{ \left[ \mathbb{L}_X , \mathbb{L}_Y \right] \,-\,
\mathbb{L}_{\left[X,Y\right]} \right\} Z \rangle
\label{Nomizucurv}
\end{equation}
This implies that the covariant derivative explicitly reads:
\begin{equation}
\mathbb{L}_X \,Y \,=\, \Gamma_{XY}^Z \,Z \label{Gammonedefi}
\end{equation}
where
\begin{equation}
\Gamma_{XY}^Z \,=\,\frac{1}{2}\left( \langle Z, \left[ X,Y \right]
\rangle \,-\, \langle X, \left[ Y,Z \right] \rangle \,-\, \langle
Y, \left[ X,Z \right] \rangle\right) \, \frac{1}{<Z,Z>} \,\qquad\,
\forall X,Y,Z \in Solv \label{Nomizuconne}
\end{equation}
\par
Eq.(\ref{Nomizuconne}) is true for any solvable Lie algebra. In
the case of \textbf{maximally non-compact, split algebras} we can
write a general form for $\Gamma_{XY}^Z$, namely:
\begin{eqnarray}
\Gamma^i_{jk} & \,=\, & 0 \nonumber \\
\Gamma^i_{\alpha\beta} & \,= \, & \ft 12 \left(-\langle
E_\alpha,\,\left[E_\beta,\,H^i\right]\rangle - \langle
E_\beta,\,\left[E_\alpha,\,H^i\right]\rangle\right)\,=\,  \, \ft 12
\,
\alpha^i \delta_{\alpha\beta}
\nonumber \\
\Gamma^{\alpha}_{ij}  & \,=
\, & \Gamma^{\alpha}_{i\beta} \,=\, \Gamma^i_{j\alpha} \,=\,0 \nonumber \\
\Gamma^{\alpha}_{\beta i} & \,= \, & \frac{1}{2}\left(\langle
E^\alpha,\,\left[E_\beta,\,H_i\right]\rangle -
\langle E_\beta,\,\left[H_i,\,E^\alpha\right]\rangle\right)\,=\, -\alpha_i\,\delta^{\alpha}_{\beta}\nonumber\\
\Gamma^{\alpha+\beta}_{\alpha\beta}  & \,=
\, &-\Gamma^{\alpha+\beta}_{\beta\alpha}\,=\,\frac{1}{2} N_{\alpha\beta}\nonumber\\
 \Gamma^{\alpha}_{\alpha+\beta\,\beta}  & \,=
 \, & \Gamma^{\alpha}_{\beta\,\alpha+\beta} \,=\,\frac{1}{2} N_{\alpha\beta}
 \label{Nomizuconne2}
\end{eqnarray}
where $N^{\alpha\beta}$ is defined by the commutator
$\left[ E_\alpha \,,\, E_\beta \right] \,=\,
N_{\alpha\beta}\,E_{\alpha + \beta} \label{nalfabeta}$, as usual. In
the case of $\mathrm{F}_{4(4)}$, the coefficients $N^{\alpha\beta}$ are
read--off from the eq.s (\ref{nalbeLS},\ref{nalbeLL},\ref{nalbeSS}).
The explicit form (\ref{Nomizuconne2}) follows from the choice of the
non degenerate metric:
\begin{eqnarray}
\langle \mathcal{H}_i \,,\, \mathcal{H}_j \rangle & \,=\, & \, 2 \, \delta_{ij} \nonumber \\
\langle \mathcal{H}_i \,,\, E_\alpha \rangle & \,=\, & 0 \nonumber \\
\langle E_\alpha \,,\, E_\beta \rangle & \,=\, &
\delta_{\alpha,\beta}
\end{eqnarray}
$\forall \mathcal{H}_i ,\, \mathcal{H}_j \,\in\,
\mathrm{CSA} $ and $\forall E_\alpha$, step
operator associated with a positive root $\alpha \in \Delta_+$.
For any other  \textbf{non split case}, as that  of $Solv\left(\mathrm{E_{7(-5)}}\right)$, the
Nomizu connection exists nonetheless although it does not take the
form (\ref{Nomizuconne2}). It follows from eq.(\ref{Nomizuconne}) upon the choice of an invariant
positive metric on $Solv$ and the use of the structure constants of $Solv$ . Given the list of
generators in tables \ref{carticorri}, \ref{Philisting} and
\ref{Omegalisting}, the positive metric on $Solv$ is easily
defined in full analogy with the definition of \cite{noiconsasha}.  The metric is diagonal and normalized as
it follows:
\begin{equation}
  \begin{array}{rclrclrcl}
    \langle H_i \, , \, H_j \rangle &=& 2 \, \delta_{ij} & \langle H_i \, ,
    \, \Phi[\alpha^\ell] \rangle &=& 0  & \langle H_i \, , \, \Omega_I[\alpha^s] \rangle &=& 0 \\
    \langle   \Phi[\alpha^\ell] \, , \, H_i  \rangle &=& 0  & \langle   \Phi[\alpha^\ell] \, ,
    \, \Phi[\beta^\ell] \rangle &=& \delta_{\alpha^\ell\beta^\ell} &
    \langle \Phi[\alpha^\ell] \, , \, \Omega_I[\alpha^s] \rangle &=& 0 \\
    \langle \Omega_I[\alpha^s]  \, , \, H_i\rangle &=& 0  & \langle \Omega_I[\alpha^s]  \, ,
    \, \Phi[\beta^\ell]\rangle &=& 0  &
    \langle \Omega_I[\alpha^s]  \, , \, \Omega_J[\beta^s]\rangle &=& \delta_{IJ} \, \delta_{\alpha^s \beta^s} \
  \end{array}
\label{metraSolv7}
\end{equation}
The Nomizu connection can be explicitly calculated from
eq.(\ref{Nomizuconne}) reading the structure constants from the
Maurer Cartan equations (\ref{E7MC}). In the case of all \textbf{split
algebras}, the first order equations take the general form:
\begin{eqnarray}
\label{D=3feqn_2} \dot{Y}^i & \,+\, & \ft 12
\,\sum_{\alpha\in \Delta_+}  \alpha^i Y_\alpha^2 \,=\, 0
\nonumber \\
\dot{Y}^\alpha &\,+\, &  \,\sum_{\beta\in
\Delta_+}\,N_{\alpha\beta} Y^{\beta}  Y^{\alpha+\beta} -
\alpha_i\,Y^i Y^\alpha\,=\,0
\end{eqnarray}
which follows from eq.(\ref{Nomizuconne2}). For the solvable Lie
algebra of $\mathrm{F_{4(4)}}$ eq.(\ref{D=3feqn_2}) takes the form:
\begin{eqnarray}
 \dot{H}^i \, + \, \ft 12  \sum_{\alpha_\ell \in \Delta^\ell} \, \alpha^i_\ell \, \Phi[\alpha_\ell]^2
\, + \, \, \ft 12  \sum_{\alpha_s \in \Delta^s} \, \alpha^i_s \, \Omega[\alpha_s]^2 & = &  0 \nonumber\\
\dot{\Phi}[\alpha_\ell] \, - \, \alpha_\ell \cdot H \, \Phi[\alpha_\ell] \, + \,\sum_{\beta_\ell \in \Delta^\ell}
\, N_{\alpha_\ell \beta_\ell}\Phi[\beta_\ell] \, \Phi[\alpha_\ell + \beta_\ell] &&\nonumber\\
\, +\,\sum_{\beta_s \in \Delta^s}
\, N_{\alpha_\ell \beta_s}\,\Omega[\beta_s] \, \Omega[\alpha_\ell +
\beta_s] &=&0 \nonumber\\
\dot{\Omega}[\alpha_s] \, - \, \alpha_s \cdot H \, \Omega[\alpha_s] \,
+ \,\sum_{\beta_s / \alpha_s + \beta_s \in \Delta^\ell}
\, N_{\alpha_s \beta_s}\Omega[\beta_s] \, \Phi[\alpha_s + \beta_s] &&\nonumber\\
\, +\,\sum_{\beta_s /\alpha_s +\beta_s\in \Delta^s}
\, N_{\alpha_s \beta_s}\,\Omega[\beta_s] \, \Omega[\alpha_s +
\beta_s] &=&0
\label{F41st}
\end{eqnarray}
where for notation simplicity we have given to the component $Y^A$ of
the tangent vector $\overrightarrow{Y}$ along a generator $\mathrm{T_A}$ of the solvable Lie algebra the
same name as the generator itself.
\par
In the case of the $\mathrm{E_{7(-5)}}$ Lie algebra the first order
equations for the tangent vector take the following form:
\begin{eqnarray}
 \dot{H}^i \, + \, \ft 12  \sum_{\alpha_\ell \in \Delta^\ell} \, \alpha^i_\ell \, \Phi[\alpha_\ell]^2
\, + \, \, \ft 12  \sum_{\alpha_s \in \Delta^s} \, \alpha^i_s \, \left (
\Omega_0[\alpha_s]^2 \,
+  \, \sum_{i=1}^3\Omega_i[\alpha_s]^2 \right) & = &  0 \nonumber\\
\dot{\Phi}[\alpha_\ell] \, - \, \alpha_\ell \cdot H \, \Phi[\alpha_\ell] \, + \,\sum_{\beta_\ell \in \Delta^\ell}
\, N_{\alpha_\ell \beta_\ell}\Phi[\beta_\ell] \, \Phi[\alpha_\ell + \beta_\ell] &&\nonumber\\
\, +\,\sum_{\beta_s \in \Delta^s}
\, N_{\alpha_\ell \beta_s}\,\left( \Omega_0[\beta_s] \, \Omega_0[\alpha_\ell +
\beta_s] \, + \, \sum_{i=1}^3\Omega_i[\beta_s] \, \Omega_i[\alpha_\ell +
\beta_s] \right)&=&0 \nonumber\\
\dot{\Omega}_0[\alpha_s] \, - \, \alpha_s \cdot H \, \Omega_0[\alpha_s] \,
+ \,\sum_{\beta_s / \alpha_s + \beta_s \in \Delta^\ell}
\, N_{\alpha_s \beta_s}\Omega_0[\beta_s] \, \Phi[\alpha_s + \beta_s] &&\nonumber\\
\, +\,\sum_{\beta_s /\alpha_s +\beta_s\in \Delta^s}
\, N_{\alpha_s \beta_s}\,\left( \Omega_0[\beta_s] \, \Omega_0[\alpha_s +
\beta_s] \, + \, \sum_{i=1}^3 \, \Omega_i[\beta_s] \, \Omega_i[\alpha_s +
\beta_s]\right)&=&0 \nonumber\\
\dot{\Omega}_i[\alpha_s] \, - \, \alpha_s \cdot H \, \Omega_i[\alpha_s] \,
+ \,\sum_{\beta_s / \alpha_s + \beta_s \in \Delta^\ell}
\, N_{\alpha_s \beta_s}\Omega_i[\beta_s] \, \Phi[\alpha_s + \beta_s] &&\nonumber\\
\, +\,\sum_{\beta_s /\alpha_s +\beta_s\in \Delta^s}
\, N_{\alpha_s \beta_s}\,\left( \ft 12 \Omega_0[\beta_s] \,
\Omega_i[\alpha_s+
\beta_s] \, + \, \ft 12 \Omega_i[\beta_s] \, \Omega_0[\alpha_s+
\beta_s]\, \right. &&\nonumber\\
\left. + \, \epsilon_{ijk} \,  \Omega_j[\beta_s] \, \Omega_k[\alpha_s
 + \beta_s]\right)&=&0 \nonumber\\
\label{E7st}
\end{eqnarray}
As one sees from the above equations the $\mathrm{E_{7(-5)}}$ differential
system (\ref{E7st}) is consistently truncated to the $\mathrm{F}_{4(4)}$
system (\ref{F41st}) by setting $ \Omega_i[\alpha_s] =0$, ($i=1,2,3$)
and identifying $\Omega[\alpha_s] =\Omega_0[\alpha_s]$. Hence any
solution of the $\mathrm{F_{4(4)}}$ equations is also a particular solution of
the $\mathrm{E_{7(-5)}}$ ones. On the other hand eq.s (\ref{E7st})
are invariant under the action of the paint group $G_{paint} =
\mathrm{SO(3)}^3$.
\par
In the next section, we utilize the \textit{compensator method} to
solve the first order equations (\ref{D=3feqn_2}) in the case of
$Solv_{ F_{4(4)}}$ and then we use the \textit{paint
group} to rotate these solutions to  general solutions of the first order equations of
$Solv_{\mathrm{E_{7(-5)}}}$.
\section{Solutions of the $\mathrm{F}_{4(4)}$ system by means of the compensator
method}
\label{compensator}
As we showed in \cite{noiconsasha} in the split case the first order
equations for the tangent vectors can be solved in the following way.
First one considers the decomposition (\ref{GHK}) of the full algebra
and recalls that, in this case, the compact subalgebra is generated by
$E^\alpha - E^{-\alpha}$ for all $\alpha \, \in \, \Delta_+$.
Secondly, one writes the decomposition of the left-invariant one--form
on the coset manifold $\mathrm{U/H}$ along the compact and
non-compact generators:
\begin{equation}
\Omega = \mathbb{L}^{-1} d \mathbb{L} \,=\, V^A \mathbb{K}_A +
\omega^{\alpha}t_{\alpha} . \label{solvodecompo}
\end{equation}
where  $V = V^A\, \mathbb{K}_A $ corresponds
to the coset manifold vielbein, while $\omega=\omega^\alpha\, t_\alpha$ corresponds to the
coset manifold $\mathrm{H}$--connection. One notes that the condition for the
coset representative $ \mathbb{L}$ to be solvable (namely to be the exponential of
the solvable algebra) is expressed very simply by:
\begin{equation}
  V^\alpha = \sqrt{2} \, \omega^\alpha
\label{solvocondo}
\end{equation}
Thirdly one derives the condition to be fulfilled by an H-gauge transformation:
\begin{eqnarray}
  \mathbb{L} & \mapsto & \mathbb{L} \, h = \overline{\mathbb{L}}
  \nonumber \\
   h &=& \exp \, \left[  \theta^\alpha  \, t_\alpha  \right]
\label{hgauga}
\end{eqnarray}
in order for the solvable gauge (\ref{solvocondo}) to be preserved. This latter reads as follows:
\begin{equation}
 \frac{\sqrt{2}}{\mbox{tr}(t_\alpha ^2)} \, \,\mbox{tr} \left( h^{-1}(\theta)\, dh(\theta) \, t_\alpha
  \right)= V^\beta \, \left( - A(\theta)_\beta^{\phantom{\beta}\alpha}
  \, +\, D(\theta)_\beta^{\phantom{\beta}\alpha} \right) \, + \,  V^i \,
  D(\theta)_i^{\phantom{i}\alpha}
\label{daequa}
\end{equation}
In the above equation the matrix $A(\theta)$ is the adjoint representation of $h \in \mathrm{H}$
and $D(\theta)$ is the $D$--representation of the same group
element which acts on the complementary space $\mathbb{K}$ and which depends case to
case:
\begin{eqnarray}
h^{-1} \, t_\alpha  \, h & = & A(\theta)_\alpha^{\phantom{\alpha }\beta} \,t_\beta  \nonumber\\
h^{-1} \, \mathbb{K}_A  \, h &  = & D(\theta)_A^{\phantom{A }B} \,\mathbb{K}_B
\label{duematrici}
\end{eqnarray}
In our example of $\mathrm{F_{4(4)}}$ the compact group is $\mathrm{H=
SU(2)_R\times Usp(6) }$ and the representation $D$ is the $\left ( \mathbf{14} , \mathbf{2}\right )$
\par
A simple solution of the first order equations (\ref{D=3feqn_2}) is
easily obtained by setting $Y^\alpha=0$ and $Y^i=c^i=\mbox{const}$,
namely we can begin with a constant vector in the direction of the $\mathrm{CSA}$.
Such a solution is named the \textit{normal form} of the tangent
vector. In the language of billiard dynamics it corresponds to a
\textit{fictitious ball} that moves on a straight line with a
constant velocity. All other solutions of eq.s (\ref{D=3feqn_2}) can
be obtained from the normal form solution by means of successive
rotations of the compact group, with parameters $\theta[t]$
satisfying the differential equation (\ref{daequa}). The advantage of
this method, emphasized in \cite{noiconsasha} where we introduced it,
is that at each successive rotation we obtain an equation which is
fully integrable in terms of the integral of the previous ones.
\par
In this paper we just present one solution of the $\mathrm{F_{4(4)}}$ system
which is fully analytical and already sufficiently complicated to
display the billiard dynamics with several bounces. Our solution is
obtained by applying $5$ successive rotations to a normal form vector
that we parametrize in terms of $4$ constants. We use an
\textit{intelligent} parametrization which is the following one:
\begin{equation}
\mathbf{Y}_{\mbox{nf}} \, = \, \{ \frac{-{{\omega }_5}}{2} - {{\omega }_6},
  \frac{{{\omega }_5}}{2},
  \frac{-{{\omega }_5}}{2} - {{\omega }_6} + {{\omega }_7},
  \frac{-{{\omega }_{24}}}{4}\}
\label{Ynormalform}
\end{equation}
\par
The way $\mathbf{Y}_{\mbox{nf}}$ is parametrized and the name given
to the constants $\omega_{24,7,6,5}$ anticipate their physical
interpretation in the solution we are going to derive. Indeed we
obtain our solution by writing:
\begin{equation}
  \mathbf{Y}(t) =\mathbf{Y}_{\mbox{nf}} \, \exp \left[ \mathbb{H}^{24}
  \theta_{24}(t) \right] \cdot \exp \left[ \mathbb{H}^{7}
  \theta_{7}(t) \right]\cdot \exp \left[ \mathbb{H}^{6}
  \theta_{6}(t) \right]\cdot \exp \left[ \mathbb{H}^{5}
  \theta_{5}(t) \right]\cdot \exp \left[ \mathbb{H}^{4}
  \theta_{4}(t) \right]
\label{orpusso}
\end{equation}
where
\begin{equation}
  \mathbb{H}^{n} = \ft 12 \left (E^{\varpi[n]} -E^{-\varpi[n]}\right)
\label{urp}
\end{equation}
are the compact generators associated with the $\mathrm{{F_4}}$ roots
numbered as in table \ref{tablettas}. The explicit form of the
solution of the differential equations (\ref{daequa}) for the five
rotation angles is given below:
\begin{eqnarray}
&&{{{{\theta }_4}(t)} =
   {\arccos \left [\frac{e^
         {\left( t - {{\tau }_4} \right) \,
           \left( {{\omega }_5} + 2\,{{\omega }_6} -
             {{\omega }_7} \right) }\,
        {\sqrt{1 + e^
             {2\,\left( t - {{\tau }_7} \right) \,
               {{\omega }_7}}}}}{{\sqrt{1 +
           e^
            {2\,\left( t - {{\tau }_6} \right) \,
              {{\omega }_6}} +
           e^
            {2\,\left( t - {{\tau }_7} \right) \,
              {{\omega }_7}} +
           e^
             {2\,\left( t - {{\tau }_4} \right) \,
               \left( {{\omega }_5} + 2\,{{\omega }_6} -
                 {{\omega }_7} \right) }\,
            \left( 1 + e^
               {2\,\left( t - {{\tau }_7} \right) \,
                 {{\omega }_7}} \right) }}}\right ]}}\nonumber\\
&&{{{{\theta }_5}(t)}=
   {\frac{\mbox{arccsc} \left [{\sqrt{1 +
           e^
             {\left( t - {{\tau }_5} \right) \,
               {{\omega }_5}}\,
            \cosh (\frac{\left( t - {{\tau }_{24}} \right) \,
                {{\omega }_{24}}}{2})\,
            {\mbox{sech}(\left( t - {{\tau }_7} \right)
                  \,{{\omega }_7})}^2}}\right ]}{{\sqrt{2}}}}}\nonumber\\
&&{{{{\theta }_6}(t)}=
   {\mbox{arccot}\left [\frac{e^
        {\left( t - {{\tau }_6} \right) \,{{\omega }_6}}}
        {{\sqrt{1 + e^
            {2\,\left( t - {{\tau }_7} \right) \,
              {{\omega }_7}}}}}\right ]}}\nonumber\\
&&{{{{\theta }_7}(t)}=
   {\arccos \left [\frac{e^
        {\left( t - {{\tau }_7} \right) \,{{\omega }_7}}}
        {{\sqrt{1 + e^
            {2\,\left( t - {{\tau }_7} \right) \,
              {{\omega }_7}}}}}\right]}}\nonumber\\
&&{{{{\theta }_{24}}(t)}\rightarrow
   {\frac{\mbox{arccsc}\left [{\sqrt{1 +
           e^
            {\left( t - {{\tau }_{24}} \right) \,
              {{\omega }_{24}}}}}\right ]}{{\sqrt{2}}}}}\nonumber\\
\label{soluzie}
\end{eqnarray}
Let us briefly mention the explicit form of the eq.s (\ref{daequa})
from which we obtained the above result. As specified in
eq.(\ref{orpusso}) we perform the compact rotations in the order $24\rightarrow
\, 7 \rightarrow\, 6 \rightarrow \, 5\rightarrow \, 4$. This is not a
random choice but it is motivated by the fact that in this way the
differential equations (\ref{daequa}) come up triangular: in other
words at each step we just obtain a differential equation for the
angle $\theta_i(t)$ that depends only on the previously determined
angles $\theta_j(t)$. The systematic study of triangulization of the
differential system (\ref{daequa}) for general algebras is postponed
to a later publication. We  just note that one typically has to
perform rotations along roots arranged in reverse order with respect
to their height but this criterion, although necessary is not yet
sufficient in full generality. A complete solution requires a more
systematic analysis. It is however fairly easy, by computer
calculations, to obtain ordered lists of root-rotations that have the
triangular property and hence lead to exact analytic solutions by
quadratures. In the case of the present algebra we have already found
lists of up to eight successive such rotations and the solution we
present with five rotations has just been chosen as an illustrative
example of the physical and analytical mechanisms occurring in the
differential system (\ref{daequa}).
\par
This being clarified, we present  the differential equations  obtained for the
angles $\theta_i(t)$ in succession. Performing the first rotation
around the highest root $\varpi_{24}$ we obtain:
\begin{equation}
  \frac{\sin \left [2\,{\sqrt{2}}\,{{\theta }_{24}}(t)\right ]\,
     {{{\omega }}_{24}}}{4\,{\sqrt{2}}} +
  {{\dot{\theta} }_{24}}(t) = 0
\label{primeq24}
\end{equation}
Performing the second rotation around the root $\varpi_7$ we get:
\begin{equation}
  \frac{\sin \left [2\,{{\theta }_7}(t)\right ]\,
     {{{\omega }}_7}}{2}
   + {{\dot{\theta} }_7}(t)
\label{primeq7}
\end{equation}
which is still an independent equation. Performing  the third
rotation around the root $\varpi_6$ we get a differential equation that depends
on the solution of the previous two:
\begin{equation}
  \frac{\sin  [2\,{{\theta }_6}(t)]\,
     {{{\omega }}_6}}{2}
   - \frac{{\cos^2 \left [{{\theta }_7}(
         t)\right ]}\,
     \sin \left [2\,{{\theta }_6}(t)\right]\,
     {{{\omega }}_7}}{2}
   + {{\dot{\theta} }_6}(t)
\label{primeq6}
\end{equation}
The same happens when we introduce the fourth rotation around
$\varpi_5$. We obtain:
\begin{eqnarray}
&&2\,{\sqrt{2}}\,
   \sin \left[2\,{\sqrt{2}}\,
     {{\theta }_5}(t)\right]\,
   {{{\omega }}_5} -
  4\,{\sqrt{2}}\,
   \cos \left [2\,{{\theta }_7}(t)\right ]\,
   \sin \left [2\,{\sqrt{2}}\,
     {{\theta }_5}(t)\right ]\,
   {{{\omega }}_7}  \nonumber\\
&&
 + {\sqrt{2}}\,
   \cos \left[2\,{\sqrt{2}}\,
     {{\theta }_{24}}(t)\right ]\,
   \sin \left [2\,{\sqrt{2}}\,
     {{\theta }_5}(t) \right ]\,
   {{{\omega }}_{24}} +
  16\,{{\dot{\theta} }_5} (t) \, = \,0
\label{primeq5}
\end{eqnarray}
Finally, when we perform the $5$-th rotation, we get:
\begin{eqnarray}
&& 8\,\sin \left [2\,{{\theta }_4}(t)\right ]\,{{ {\omega }}_5} -
  4\,\left( -3 + \cos \left[2\,{{\theta }_6}(t)\right] \right ] \,\sin \left[2\,{{\theta }_4}(t)\right]\,
   {{ {\omega }}_6} - 6\,\sin \left[2\,{{\theta }_4}(t)\right]\,{{{\omega }}_7} \nonumber\\
   && +
  2\,\cos \left[2\,{{\theta }_6}(t)\right]\,\sin \left[2\,{{\theta }_4}(t)\right]\,{{ {\omega }}_7} +
  \cos \left[2\,\left( {{\theta }_6}(t) - {{\theta }_7}(t) \right) \right ]\,
   \sin \left[2\,{{\theta }_4}(t)\right]\,{{ {\omega }}_7}\nonumber\\
   && +
  2\,\cos \left[2\,{{\theta }_7}(t)\right ]\,\sin \left[ 2\,{{\theta }_4}(t)\right]\,{{ {\omega }}_7} +
  \cos \left[2\, {{\theta }_6}(t) + 2\,  {{\theta }_7}(t)  \right]\,
   \sin \left[2\,{{\theta }_4}(t)\right ]\,{{ {\omega }}_7} \nonumber\\
   && + 16\,{{\dot{\theta} }_4}(t)\, = \, 0
\label{primeq4}
\end{eqnarray}
The explicit functions $\theta_i(t)$ displayed in eq.(\ref{soluzie})
are the general integral of the system of eq.s
(\ref{primeq24}-\ref{primeq4}) where the integration constants are
represented by the fixed instants of time $\tau_i$.  The physical
interpretation of these constants becomes clear when we investigate
the properties of the scalar fields $h_i(t)$ lying in the Cartan
subalgebra of $\mathrm{F_{4(4)}}$ and eventually representing, after
dimensional oxidation the logarithms of the scale factors in the
various available dimensions. Following the discussion of
\cite{noiconsasha} we can write
\begin{eqnarray}
  h_i(t) &=& \int \, H_i(t^\prime) dt^\prime\nonumber\\
  H_i(t) & \equiv & Y^i(t) \quad \mbox{in CSA}
\label{piccolacca}
\end{eqnarray}
where $H_i(t)$ are obtained by inserting the explicit solutions (\ref{soluzie}) into eq.(\ref{orpusso})
and then extracting the first four components of such a vector.
We have an analytic although cumbersome expression for  the
$H_i(t)$ in the case of all the considered rotations, yet the
next integration to $h_i(t)$ can no longer be  analytically done if we include all
the thetas $\theta_{24}(t), \theta_{7}(t), \theta_{6}(t), \theta_{5}(t)$ and
$\theta_4(t)$. For this reason we prefer to discuss the features of billiard dynamics
by considering the simpler solution obtained by including only the first three
rotations $\theta_{24}(t), \theta_{7}(t)$ and $\theta_{6}(t)$. This
solution is already complicated enough to display the phenomena we
want to illustrate yet it still leads to manageable analytic
formulae. Explicitly we obtain:
\begin{eqnarray}
    H_1(t) & = &\frac{-{{\omega }_5}}{2} - {{\omega }_6} +
  {\sin^2 \left[{\theta _7}(t)\right]}\,{{\omega }_7} \nonumber\\
  \null & \null & \null \nonumber\\
  \null & = & -\frac{{{\omega }_5}}{2} - {{\omega }_6} +
  \frac{{{\omega }_7}}{1 + e^{2\,\left( t - {{\tau }_7} \right) \,{{\omega }_7}}}\nonumber \\
  \null & \null & \null \nonumber\\
    H_2 (t)& = & \ft 12 \,\left[ {{{\omega }_5} + 2\,{\sin^2 \left [\theta_6(t) \right]}\,
     \left( {{\omega }_6} - {\cos^2  \left[ \theta_7(t) \right ]}\,{{\omega }_7} \right)
     }\right ]\nonumber\\
     \null & \null & \null \nonumber\\
    \null & = & \frac{\left( 1 + e^{2\,\left( t - {{\tau }_6} \right) \,{{\omega }_6}} +
       e^{2\,\left( t - {{\tau }_7} \right) \,{{\omega }_7}} \right) \,
     {{\omega }_5} + 2\,\left( \left( 1 +
          e^{2\,\left( t - {{\tau }_7} \right) \,{{\omega }_7}} \right) \,
        {{\omega }_6} - e^{2\,\left( t - {{\tau }_7} \right) \,{{\omega }_7}}\,
        {{\omega }_7} \right) }{2\,
    \left( 1 + e^{2\,\left( t - {{\tau }_6} \right) \,{{\omega }_6}} +
      e^{2\,\left( t - {{\tau }_7} \right) \,{{\omega }_7}} \right) } \nonumber\\
      \null & \null & \null \nonumber\\
    H_3(t) & = & \frac{-{{\omega }_5}}{2} + {\cos^2 \left[ \theta _6(t)\right] }\,
   \left( -{{\omega }_6} + {\cos^2 \left[ \theta_7(t)\right] }\,{{\omega }_7} \right) \nonumber\\
   \null & \null & \null \nonumber\\
    \null & = & \frac{-{{\omega }_5}}{2} + \frac{e^
      {2\,\left( t - {{\tau }_6} \right) \,{{\omega }_6}}\,
     \left( \left( -e^{2\,t\,{{\omega }_7}} -
          e^{2\,{{\tau }_7}\,{{\omega }_7}} \right) \,{{\omega }_6} +
       e^{2\,t\,{{\omega }_7}}\,{{\omega }_7} \right) }{\left( 1 +
       e^{2\,\left( t - {{\tau }_6} \right) \,{{\omega }_6}} +
       e^{2\,\left( t - {{\tau }_7} \right) \,{{\omega }_7}} \right) \,
     \left( e^{2\,t\,{{\omega }_7}} + e^{2\,{{\tau }_7}\,{{\omega }_7}} \right) }\nonumber\\
    \null & \null & \null \nonumber\\
     H_4(t) & = &- \frac{1}{4}{ \left( \cos \left[2\,{\sqrt{2}}\,{{\theta }_{24}}(t)\right]\,
      {{{\omega }}_{24}} \right) }  \, =  \,\frac{1}{4} {\left( -1 + \frac{2}
       {1 + e^{\left( t - {{\tau }_{24}} \right) \,{{\omega }_{24}}}} \right) \,
    {{\omega }_{24}}} \nonumber\\
\label{Hvetti}
\end{eqnarray}
Considering now the five roots involved in this calculation:
\begin{equation}
  \begin{array}{cclcccl}
    \varpi_{24} & = & \{0,0,0,2\} & ; & \varpi_7 & = & \{1,0,-1,0\} \\
    \varpi_6 & = & \{0,1,1,0\} & ; & \varpi_5 & = & \{-1,-1,1,1\} \\
    \varpi_4 & = & \{1,-1,0,0\} & \null & \null & \null & \null \
  \end{array}
\label{5rutte}
\end{equation}
we can evaluate the five projections of the Cartan fields in the
direction of the five relevant roots and we get:
\begin{eqnarray}
h_{\varpi_{24}}(t) & \equiv \, \int \, \overrightarrow{\varpi}_{24}\cdot \overrightarrow{H}(t)
&= \,
 -\log (1 + e^{\left( t - {{\tau }_{24}} \right) \,{{\omega }_{24}}}) +
  \frac{t\,{{\omega }_{24}}}{2} \nonumber\\
h_{\varpi_{7}}(t) & \equiv \, \int \, \overrightarrow{\varpi}_{7}\cdot \overrightarrow{H}(t)
&=
 -\log (1 + e^{2\,\left( t - {{\tau }_7} \right) \,{{\omega }_7}}) +
  \frac{\log (1 + e^{2\,\left( t - {{\tau }_6} \right) \,{{\omega }_6}} +
      e^{2\,\left( t - {{\tau }_7} \right) \,{{\omega }_7}})}{2} \nonumber\\
      && -
  t\,{{\omega }_6} + t\,{{\omega }_7} \nonumber\\
h_{\varpi_{6}}(t) & \equiv \, \int \, \overrightarrow{\varpi}_{6}\cdot \overrightarrow{H}(t)
&=\frac{\log (1 + e^{2\,\left( t - {{\tau }_7} \right) \,{{\omega }_7}})}{2} -
  \log (1 + e^{2\,\left( t - {{\tau }_6} \right) \,{{\omega }_6}} +
    e^{2\,\left( t - {{\tau }_7} \right) \,{{\omega }_7}}) + t\,{{\omega }_6}
     \nonumber\\
h_{\varpi_{5}}(t) & \equiv \, \int \, \overrightarrow{\varpi}_{5}\cdot \overrightarrow{H}(t) &=
\frac{1}{4}\left[4\,\log (1 + e^{2\,\left( t - {{\tau }_7} \right) \,{{\omega }_7}}) -
    2\,\log (1 + e^{\left( t - {{\tau }_{24}} \right) \,{{\omega }_{24}}})\right. \nonumber\\
    && \left. -
    2\,t\,{{\omega }_5} - 4\,t\,{{\omega }_7} + t\,{{\omega }_{24}}\right ] \nonumber\\
h_{\varpi_{4}}(t) & \equiv \, \int \, \overrightarrow{\varpi}_{4}\cdot \overrightarrow{H}(t)
&=
\frac{1}{2}\left [-\log (1 + e^{2\,\left( t - {{\tau }_7} \right) \,{{\omega }_7}}) +
    \log (1 + e^{2\,\left( t - {{\tau }_6} \right) \,{{\omega }_6}}  +
      e^{2\,\left( t - {{\tau }_7} \right) \,{{\omega }_7}})\right. \nonumber\\
    &&\left. -
    2\,t\,{{\omega }_5} - 4\,t\,{{\omega }_6} + 2\,t\,{{\omega }_7}\right] \nonumber\\
\label{pensata}
\end{eqnarray}
In the first of eq.(\ref{pensata}) we can observe the basic building
block for the smooth realization of the cosmic billiard behaviour. It
is is given by the function:
\begin{equation}
  G(t|\omega,\tau) \equiv  -\log (1 + e^{\left( t - {{\tau }} \right) \,{{\omega }}}) +
  \frac{t\,{{\omega }}}{2}
\label{Gtfun}
\end{equation}
For $t-\tau << 0$, that is  for asymptotically early times the
behaviour of $G(t|\omega,\tau)$ is the following one:
\begin{equation}
   G(t|\omega,\tau) \simeq \frac{t\,{{\omega }}}{2}
\label{-tinfy}
\end{equation}
which corresponds to the motion of a fictitious ball with constant
velocity $v=\omega/2$. For asymptotically late times, namely for
$t-\tau >>0$, we have instead:
\begin{equation}
   G(t|\omega,\tau) \simeq -\frac{t\,{{\omega }}}{2}
\label{-tinfy2}
\end{equation}
which corresponds to the motion of a fictitious ball with inverted constant
velocity $v=-\omega/2$.
The inversion, namely the \textit{bounce} occurs in the region
$t-\tau \sim 0$. Hence it appears that the integration constants
$\tau_i$ introduced in our solution have precisely the meaning of instant of times at which bounces occur.
Furthermore each  bounce occurs precisely on the \textit{the wall} orthogonal to each  root around which we have
made compact rotations while using the compensator algorithm. On the
other hand, the components of the normal form solution
(\ref{Ynormalform}) in the CSA direction have the interpretation of components of the velocity vector of the
fictitious cosmic ball in the asymptotically early times prior to the
first cosmic bounce.  Each new rotation introduces a new bounce.
\iffigs
\begin{figure}
\begin{center}
\epsfxsize =6cm
{\epsffile{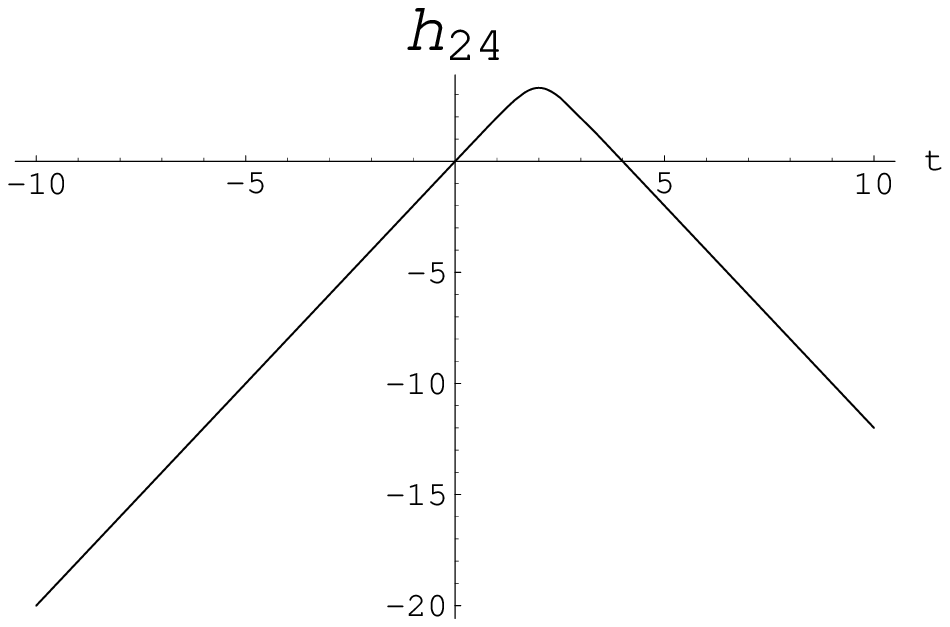}}
\epsfxsize =6cm
{\epsffile{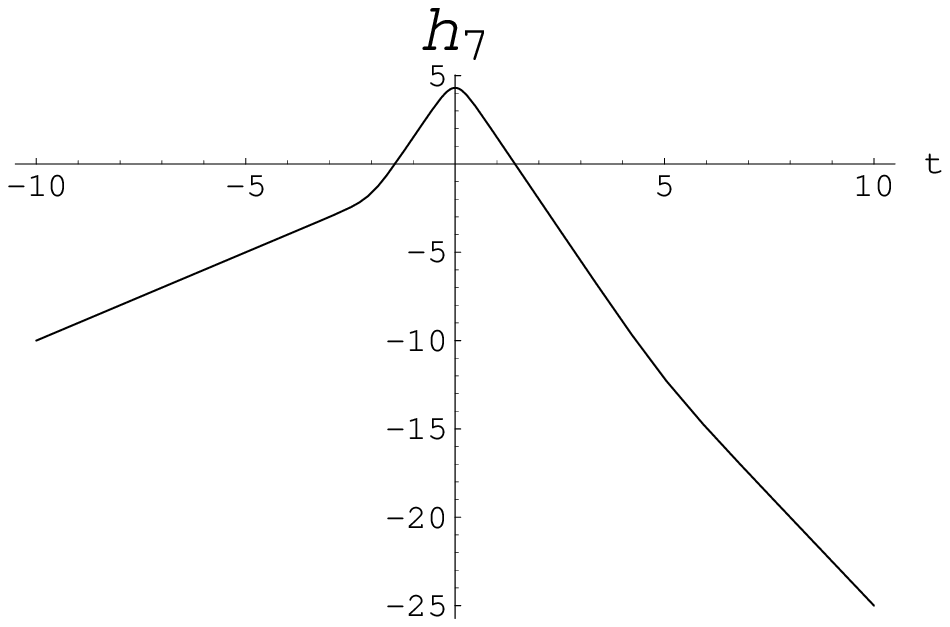}}
\vskip 0.2cm
\epsfxsize =6cm
{\epsffile{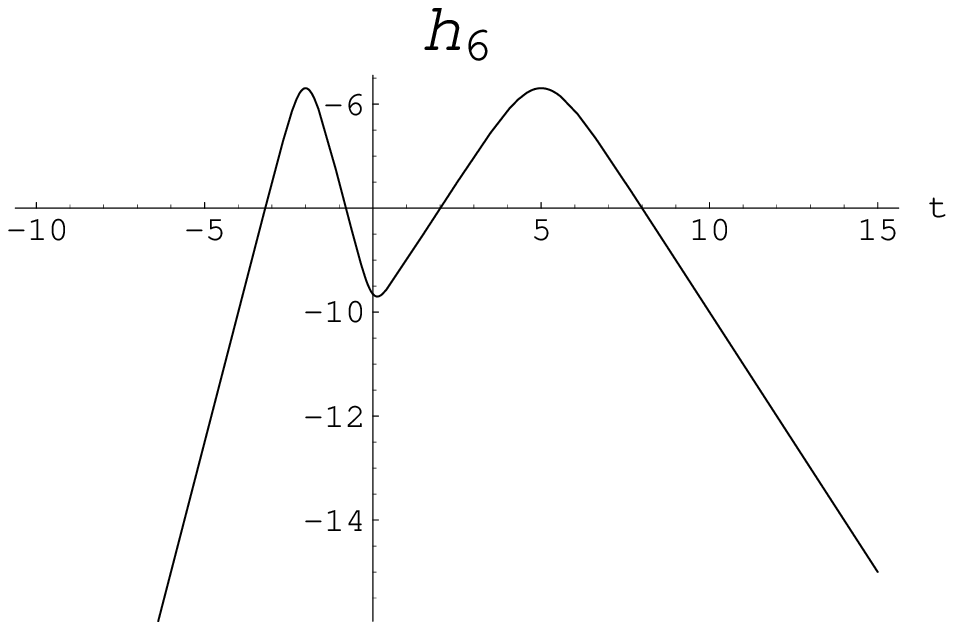}}
\epsfxsize =6cm
{\epsffile{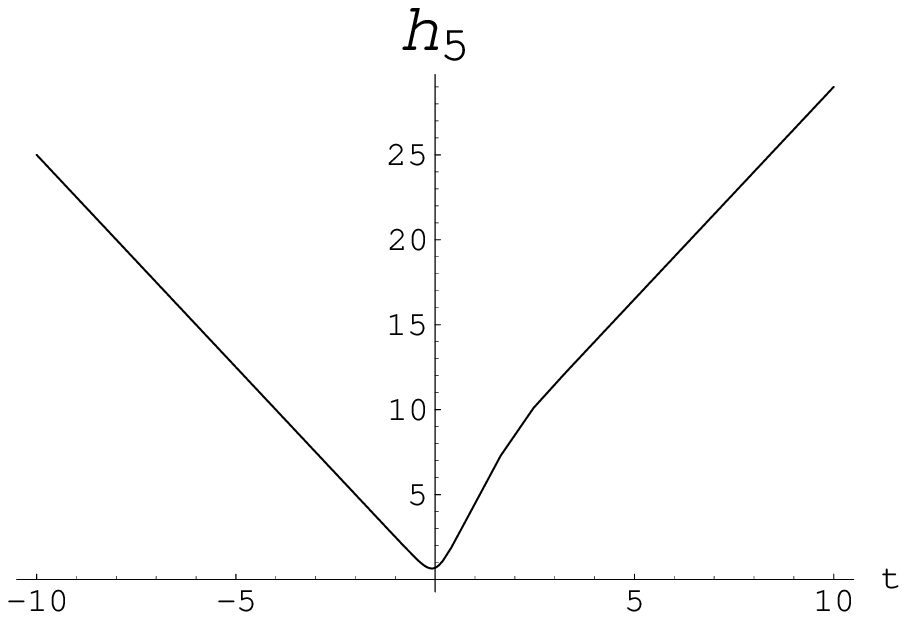}}
\vskip 0.2cm
\epsfxsize =6cm
{\epsffile{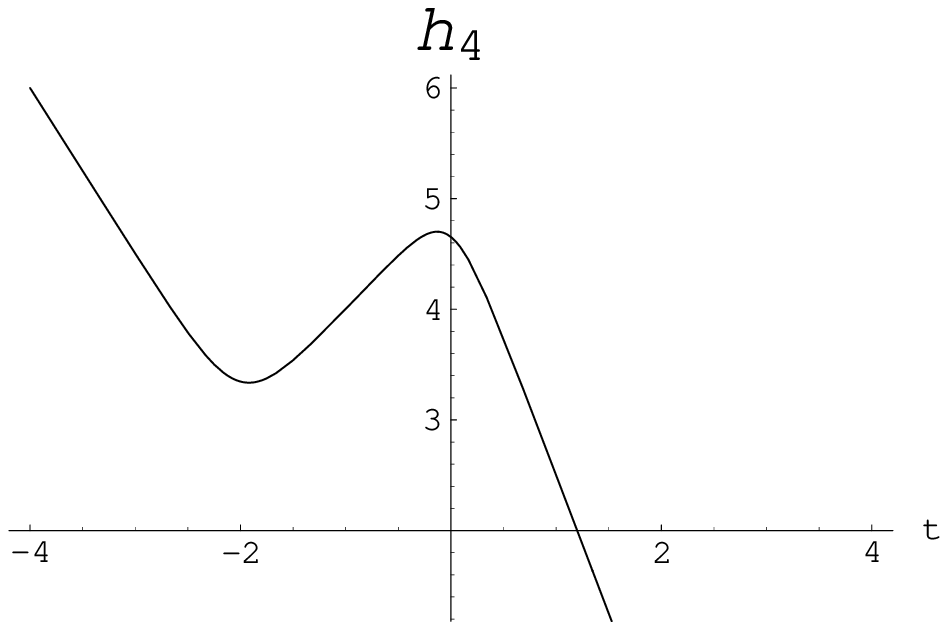}}
\caption{Plots of the Cartan fields  $h_i(t) \equiv {\vec{\varpi}}_i \cdot {\vec{h}(t)} $,
as functions of the time in the case of three rotations $\theta_{24}(t)$,
$\theta_7(t)$, $\theta_6(t)$ and with the following choice of parameters:
$\omega_{24}=4$, $\omega_7 = 3.5$, $\omega_6 = 2.5$, $\omega_5 = 0$,
$\tau_6 = -2$, $\tau_7=0$, $\tau_{24} = 2$. It is evident from
the plots that there are three bounces, exactly at $t = -2$  , $t=0$ and $t= 2 $. \label{plottucci}  }
\hskip 2cm \unitlength=1.1mm
\hskip 1.5cm \unitlength=1.1mm
\end{center}
\end{figure}
\fi
This is illustrated in fig. \ref{plottucci} where the Cartan fields
along the five relevant roots are plotted for the solution with three
rotations namely in the case of eq.(\ref{pensata}). Here we clearly
see three bounces, due to the three rotations introduced.
\section{Uplifting of $\mathrm{F}_{4(4)}$ solutions to $\mathrm{E}_{7(-5)}$ and painted walls}
\label{uplift}
Now that we have obtained explicit solutions of the first order system
(\ref{F41st}) by means of the compensator method, we can appreciate
the role of the \textit{paint group}, $\mathrm{G_{paint}}=\mathrm{SO(3)^3}$ since rotations of this latter applied
to the $\mathrm{F}_{4(4)}$ solution generate non trivial solutions
of the differential system (\ref{E7st}).
\par
To illustrate the mechanism with an explicit and manageable example  we consider the $\mathrm{F}_{4(4)}$ solution
based on the three rotations angles $\theta_{24}(t),\theta_{7}(t),\theta_6(t)$
for which we have already written the time dependence of the Cartan
fields in eq.(\ref{Hvetti}) and we complete it by writing also the
time dependence of the root components of the tangent vector.
In this case the only non vanishing root fields are $\Phi_{12}$ and
$\Omega_{3,4,8}$ respectively associated with the long root $\alpha^\ell_{12} = 2 y_4$ and with the
short roots $\alpha^s_{3,4,8} = y_2 +y_3,\, y_1 -y_3,\, y_1 +y_2 $. The time dependence of these
fields in the considered solution is given by:
\begin{eqnarray}
\Phi[\alpha_{12}^\ell](t) & = & -\left( \frac{e^
       {\frac{\left( t -
            {{\tau }_{24}} \right)
            \,{{\omega }_{24}}}{2}
         }\,{{\omega }_{24}}}
      {{\sqrt{2}}\,
      \left( 1 +
        e^
         {\left( t -
            {{\tau }_{24}} \right)
            \,{{\omega }_{24}}}
        \right) } \right) \nonumber\\
\Omega[\alpha_3^s](t) & = & \frac{-2\,e^
     {\left( t - {{\tau }_6}
         \right) \,{{\omega }_6}}\,{\sqrt{1 +
        e^
         {2\,\left( t -
            {{\tau }_7} \right) \,
           {{\omega }_7}}}}\,
    \left( \left( e^
          {2\,t\,{{\omega }_7}} +
         e^
          {2\,{{\tau }_7}\,
            {{\omega }_7}} \right)
         \,{{\omega }_6} -
      e^{2\,t\,{{\omega }_7}}\,
       {{\omega }_7} \right) }
    {\left( 1 +
      e^
       {2\,\left( t -
           {{\tau }_6} \right) \,
         {{\omega }_6}} +
      e^
       {2\,\left( t -
           {{\tau }_7} \right) \,
         {{\omega }_7}} \right) \,
    \left( e^
       {2\,t\,{{\omega }_7}} +
      e^
       {2\,{{\tau }_7}\,
         {{\omega }_7}} \right) } \nonumber\\
  \Omega[\alpha_4^s](t)       & = &\frac{-2\,e^
     {\left( t - {{\tau }_7}
         \right) \,{{\omega }_7}}\,
    {{\omega }_7}}{\left( 1 +
      e^
       {2\,\left( t -
           {{\tau }_7} \right) \,
         {{\omega }_7}} \right) \,
    {\sqrt{1 +
        \frac{1 +
           e^
            {2\,
            \left( t -
            {{\tau }_7} \right) \,
            {{\omega }_7}}}{e^
           {2\,
            \left( t -
            {{\tau }_6} \right) \,
            {{\omega }_6}}}}}}\nonumber\\
  \Omega[\alpha_8^s](t)          & = &\frac{2\,e^
     {\left( -t + {{\tau }_6}
          \right) \,{{\omega }_6} +
        \left( t - {{\tau }_7}
          \right) \,{{\omega }_7}}\
     ,{{\omega }_7}}{{\sqrt{
       \left( 1 +
         e^
          {2\,
            \left( t -
            {{\tau }_7} \right) \,
            {{\omega }_7}} \right)
        \,\left( 1 +
         e^
           {2\,
            \left( -t +
            {{\tau }_6} \right) \,
            {{\omega }_6}}\,
          \left( 1 +
            e^
            {2\,
            \left( t -
            {{\tau }_7} \right) \,
            {{\omega }_7}} \right)
         \right) }}}
\label{ordadoro}
\end{eqnarray}
and it is displayed in fig.\ref{splotti}
\iffigs
\begin{figure}
\begin{center}
\epsfxsize =6cm
{\epsffile{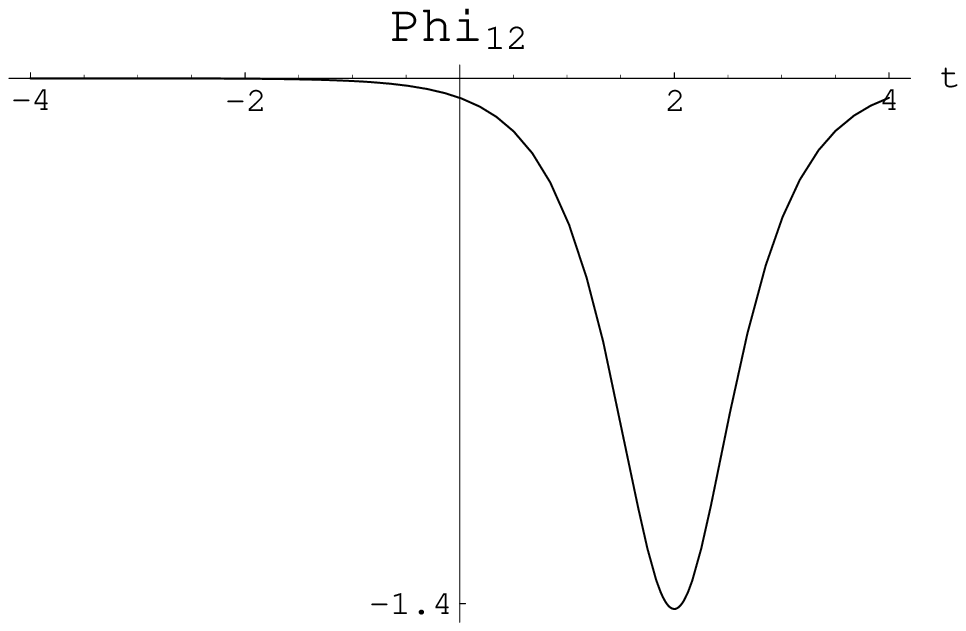}}
\epsfxsize =6cm
{\epsffile{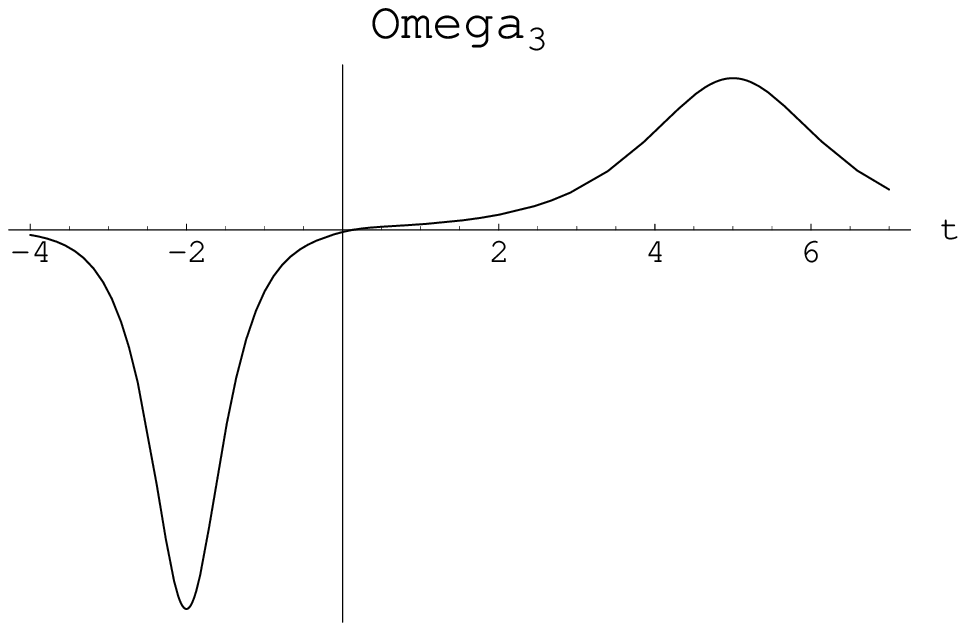}}
\vskip 0.2cm
\epsfxsize =6cm
{\epsffile{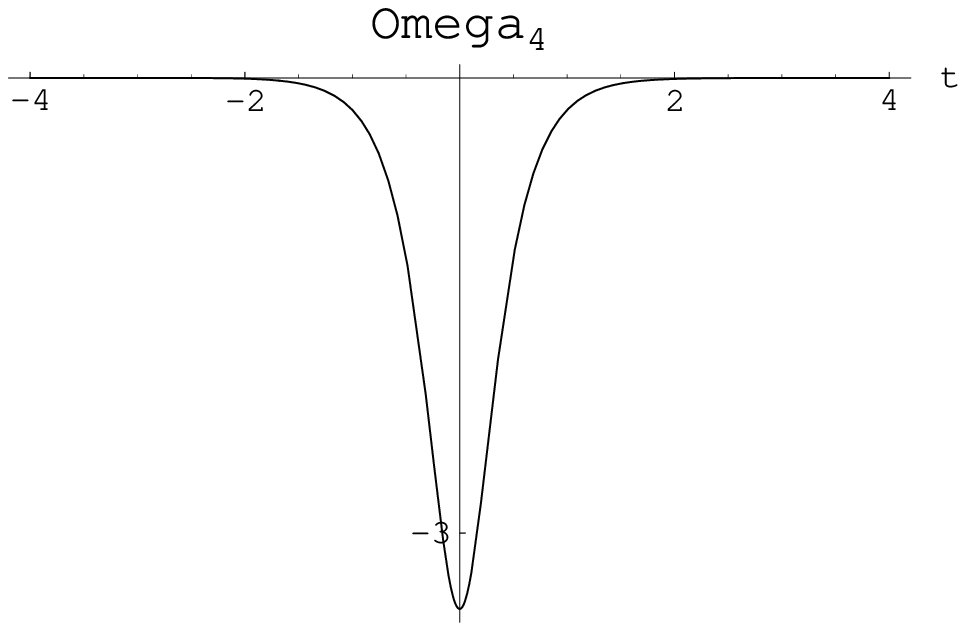}}
\epsfxsize =6cm
{\epsffile{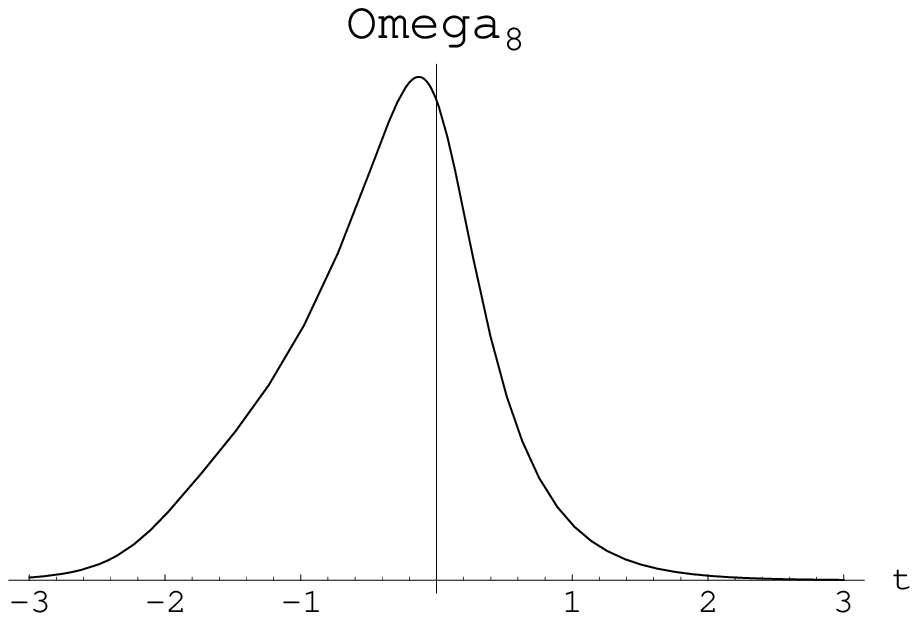}}
\caption{Plots of the root fields  $\Phi[\alpha_{12}^\ell](t)$ and $\Omega[\alpha_{3,4,8}^s](t)$
as functions of the time in the case of three rotations $\theta_{24}(t)$,
$\theta_7(t)$, $\theta_6(t)$ and with  the following choice of parameters:
$\omega_{24}=4$, $\omega_7 = 3.5$, $\omega_6 = 2.5$, $\omega_5 = 0$,
$\tau_6 = -2$, $\tau_7=0$, $\tau_{24} = 2$. It is evident from
the plots that the dynamical wall causing the bounce  at $t = 2$ is $\Phi[\alpha_{12}^\ell](t)$ , while the walls for
the $t=0$ bounce are provided by $\Omega[\alpha_4^s](t)$ and $\Omega[\alpha_8^s](t)$.
Finally the wall for the $t=-2$ bounce is provided by
$\Omega[\alpha_3^s](t)$. \label{splotti}  }
\hskip 2cm \unitlength=1.1mm
\hskip 1.5cm \unitlength=1.1mm
\end{center}
\end{figure}
\fi
\par
Uplifting this solution to an $\mathrm{E}_{7(-5)}$ solution is done by
identifying the Cartan and the long root fields of the two systems
and then by identifying:
\begin{eqnarray}
  \Omega[\alpha_3^s](t) & = & \Omega_0[\alpha_3^s](t)\nonumber\\
  \Omega[\alpha_4^s](t) & = & \Omega_0[\alpha_4^s](t)\nonumber\\
  \Omega[\alpha_8^s](t) & = & \Omega_0[\alpha_8^s](t)\nonumber\\
\label{agnizio1}
\end{eqnarray}
Next we can rotate the so obtained solution with any element of the
nine parameter paint group $\mathrm{SO(3)^3}$ whose generators are
the first nine operators $\mathbb{H}_i$ described in
eq.s(\ref{Hbasis}). For instance we can apply a rotation of a
constant angle $\psi_5$ along the $5$-th generator, namely along
$E^+_{\beta_2}$. The result putting all-together is given by the
Cartan fields in eq.(\ref{Hvetti}) and by the following root fields:
\begin{eqnarray}
\Phi[\alpha_{12}^\ell](t) & = & -\left( \frac{e^
       {\frac{\left( t -
            {{\tau }_{24}} \right)
            \,{{\omega }_{24}}}{2}
         }\,{{\omega }_{24}}}
      {{\sqrt{2}}\,
      \left( 1 +
        e^
         {\left( t -
            {{\tau }_{24}} \right)
            \,{{\omega }_{24}}}
        \right) } \right) \nonumber\\
\Omega_0[\alpha_3^s](t)  & = & \frac{-2\,e^{\left( t - {{\tau }_6} \right) \,{{\omega }_6}}\,
    {\sqrt{1 + e^{2\,\left( t - {{\tau }_7} \right) \,{{\omega }_7}}}}\,
    \cos (\frac{{{\psi }_5}}{2\,{\sqrt{2}}})\,
    \left( \left( e^{2\,t\,{{\omega }_7}} +
         e^{2\,{{\tau }_7}\,{{\omega }_7}} \right) \,
       {{{\omega }}_6} - e^{2\,t\,{{\omega }_7}}\,{{\omega }_7}
      \right) }{\left( 1 + e^
       {2\,\left( t - {{\tau }_6} \right) \,{{\omega }_6}} +
      e^{2\,\left( t - {{\tau }_7} \right) \,{{\omega }_7}} \right) \,
    \left( e^{2\,t\,{{\omega }_7}} +
      e^{2\,{{\tau }_7}\,{{\omega }_7}} \right) } \nonumber\\
\Omega_Z[\alpha_3^s](t)  & = & \frac{-2\,e^{\left( t - {{\tau }_6} \right) \,{{\omega }_6}}\,
    {\sqrt{1 + e^{2\,\left( t - {{\tau }_7} \right) \,{{\omega }_7}}}}\,
    \sin (\frac{{{\psi }_5}}{2\,{\sqrt{2}}})\,
    \left( \left( e^{2\,t\,{{\omega }_7}} +
         e^{2\,{{\tau }_7}\,{{\omega }_7}} \right) \,
       {{{\omega }}_6} - e^{2\,t\,{{\omega }_7}}\,{{\omega }_7}
      \right) }{\left( 1 + e^
       {2\,\left( t - {{\tau }_6} \right) \,{{\omega }_6}} +
      e^{2\,\left( t - {{\tau }_7} \right) \,{{\omega }_7}} \right) \,
    \left( e^{2\,t\,{{\omega }_7}} +
      e^{2\,{{\tau }_7}\,{{\omega }_7}} \right) } \nonumber\\
\Omega_0[\alpha_4^s](t)      & = &\frac{-2\,e^{\left( t - {{\tau }_7} \right) \,{{\omega }_7}}\,
    {{\omega }_7}}{\left( 1 +
      e^{2\,\left( t - {{\tau }_7} \right) \,{{\omega }_7}} \right) \,
    {\sqrt{1 + \frac{1 + e^
            {2\,\left( t - {{\tau }_7} \right) \,{{\omega }_7}}}{e^
           {2\,\left( t - {{\tau }_6} \right) \,{{\omega
           }_6}}}}}}\nonumber\\
      \Omega_0[\alpha_8^s](t)     & = & \frac{2\,e^{\left( -t + {{\tau }_6} \right) \,{{\omega }_6} +
       \left( t - {{\tau }_7} \right) \,{{\omega }_7}}\,
    \cos (\frac{{{\psi }_5}}{2\,{\sqrt{2}}})\,{{{\omega }}_7}}
    {{\sqrt{\left( 1 + e^
          {2\,\left( t - {{\tau }_7} \right) \,{{\omega }_7}} \right) \,
       \left( 1 + e^
           {2\,\left( -t + {{\tau }_6} \right) \,{{\omega }_6}}\,
          \left( 1 + e^
             {2\,\left( t - {{\tau }_7} \right) \,{{\omega }_7}} \right)
         \right) }}}\nonumber\\
     \Omega_Z[\alpha_8^s](t)    & = & \frac{2\,e^{\left( -t + {{\tau }_6} \right) \,{{\omega }_6} +
       \left( t - {{\tau }_7} \right) \,{{\omega }_7}}\,
    \sin (\frac{{{\psi }_5}}{2\,{\sqrt{2}}})\,{{{\omega }}_7}}
    {{\sqrt{\left( 1 + e^
          {2\,\left( t - {{\tau }_7} \right) \,{{\omega }_7}} \right) \,
       \left( 1 + e^
           {2\,\left( -t + {{\tau }_6} \right) \,{{\omega }_6}}\,
          \left( 1 + e^
             {2\,\left( t - {{\tau }_7} \right) \,{{\omega }_7}} \right)
         \right) }}}
\label{psi5rotsola}
\end{eqnarray}
Inserting eq.s (\ref{Hvetti}) and eq.s(\ref{psi5rotsola}) into the
differential equations (\ref{E7st}) one can patiently verify that
they are all satisfied for any value of the angle $\psi_5$.
\par
We could continue with more complicated rotations, but the lesson
taught by this example should  already be sufficiently clear. In this
solution the time dependence of $\Omega_Z[\alpha_8^s](t)$ and
$\Omega_0[\alpha_8^s](t)$ is exactly the same and the ratio of these
two fields is the constant factor $\tan \left[ \frac{{{\psi
}_5}}{2\,{\sqrt{2}}}\right] $. Similarly for the fields
$\Omega_Z[\alpha_3^s](t)$ and $\Omega_0[\alpha_3^s](t)$. Hence it
appears that the dynamical walls which raise and lower and cause the
bounces of the cosmological factors are just those displayed by the
Tits Satake projection of the supergravity scalar manifold, namely,
the quaternionic manifold $\mathrm{F_{4(4)}/Usp(6) \times SU(2)}$, rather than
$\mathrm{E_{7(-5)}/SO(12) \times SO(3)}$ in $D=3$, and after dynamical
oxidation to $D=4$, the special K\"ahler manifold  $\mathrm{Sp(6,R)/SU(3) \times
U(1)}$ rather than $\mathrm{SO^\star(12) /SU(6) \times U(1)}$.
Indeed as we have pointed out in \cite{noiKMpape} and recalled in
table \ref{tavolona} taken from \cite{noiKMpape}, the Tits Satake
projection commutes with the c-map produced by the dimensional
reduction \`a la Ehlers and we have the correspondence:
\begin{equation}
\begin{array}{rcl}
\mbox{adj}(\mathbb{U}_{D=3}) &=&
\mbox{adj}(\mathbb{U}_{D=4})\oplus\mbox{adj}(\mathrm{SL(2,\mathbb{R})_E})\oplus
W_{(2,W)} \\
\null &\Downarrow & \null\\
\mbox{adj}(\mathbb{U}^{TS}_{D=3}) &=&
\mbox{adj}(\mathbb{U}^{TS}_{D=4})\oplus\mbox{adj}(\mathrm{SL(2,\mathbb{R})_E})\oplus
W_{(2,W^{TS})} \\
\end{array}
\label{gendecompo2}
\end{equation}
where $\mathrm{SL(2,\mathbb{R})_E}$ is the Ehlers group coming from
the dimensional reduction of pure gravity and $W$ denotes the
symplectic representation to which vector fields are assigned in
$D=4$.
\begin{table}
  \centering
  {\scriptsize \begin{tabular}{|c|c|c|c|c|c|}
\hline
 \# Q.s & & D=4 & D=3 & D=2 & D=1 \\
  \hline
 $\mathcal{N}=8$ & $\mathbb{U}$ & $\mathrm{E_{7(7)}}$ & $\mathrm{E_{8(8)}}$ &
 $\mathrm{E_9}$ & $\mathrm{E_{10}}$ \\
 & $\mathbb{H}$ & $\mathrm{SU}(8)$ & $\mathrm{SO}(16)$ & $\mathrm{KE}_9$ & $\mathrm{KE}_{10}$\\ \hline
 $\mathcal{N}=6$ & $\mathbb{U}$ & $\mathrm{SO^\star(12)}$ & $\mathrm{E_{7(-5)}}$ & $\mathrm{E_{7(-5)}}^\wedge$ &
 $\mathrm{E_{7(-5)}}^{\wedge\wedge}$ \\  & $\mathbb{H}$ & $\mathrm{SU(6) \times U(1)}$ & $\mathrm{SO(12) \times SO(3)}$ &
 $\mathrm{KE}_{7(-5)}^\wedge$ &  $\mathrm{KE}_{7(-5)}^{\wedge\wedge}$ \\ \cline{2-6} & $\mathbb{U}^{TS}$ &
 $\mathrm{Sp(6,\mathbb{R})}$ & $\mathrm{F_{4(4)}}$ & $\mathrm{F_{4(4)}}^\wedge$ & $\mathrm{F_{4(4)}}^{\wedge\wedge}$ \\
& $\mathbb{H}^{TS}$ & $ \mathrm{SU(3) \times U(1)}$ & $\mathrm{Usp(6) \times SU(2)}$ &
 $\mathrm{KF}_{4(4)}^\wedge$ &$\mathrm{KF}_{4(4)}^{\wedge\wedge}$ \\ \hline
 $\mathcal{N}=5$ &  $\mathbb{U}$ & $\mathrm{SU}(5,1)$ & $\mathrm{E}_{6(-14)}$ & $\mathrm{E}_{6(-14)}^\wedge$ &
 $\mathrm{E}_{6(-14)}^{\wedge\wedge}$ \\ & $\mathbb{H}$ & $\mathrm{SU}(5)\times \mathrm{U}(1)$ &
 $\mathrm{SO}(10)\times \mathrm{SO}(2)$ & $\mathrm{KE}_{6(-14)}^\wedge$ & $\mathrm{KE}_{6(-14)}^{\wedge\wedge}$ \\
 \cline{2-6} & $\mathbb{U}^{TS}$ & $\mathrm{SU(1,1)}$ & $bc_2$ & $\mathrm{A}_{4}^{(2)}$ & $\mathrm{A}_{4}^{(2)\wedge}$ \\
 & $\mathbb{H}^{TS}$ & $\mathrm{U(1)}$ & -- & $\mathrm{KA}_{4}^{(2)}$ & $\mathrm{KA}_4^{(2)\wedge}$ \\ \hline
 $\mathcal{N}=4$ & $\mathbb{U}$ & $\mathrm{SO}(6,n)\times\mathrm{SU}(1,1)$ & $\mathrm{SO}(8,n+2)$ &
 $\mathrm{SO}(8,n+2)^\wedge$ & $\mathrm{SO}(8,n+2)^{\wedge\wedge}$ \\ & $\mathbb{H}$ &
 $\mathrm{SO}(6)\times\mathrm{SO}(n)\times\mathrm{U}(1)$ & $\mathrm{SO}(8)\times\mathrm{SO}(n+2)$ &
 $\mathrm{KSO}(8,n+2)^\wedge$ & $\mathrm{KSO}(8,n+2)^{\wedge\wedge}$ \\ \cline{2-6} $n < 6$ &
 $\mathbb{U}^{TS}$ & $\mathrm{{SO(n,n)}\times SU(1,1)}$ & $\mathrm{{SO(n+2,n+2)}}$ & $\mathrm{SO(n+2,n+2)}^\wedge$ &  $\mathrm{SO(n+2,n+2)}^{\wedge\wedge}$ \\
  & $\mathbb{H}^{TS}$ & $\mathrm{SO(n)\times SO(n) \times U(1)}$ & $\mathrm{SO(n+2)\times SO(n+2) }$ &
  $\mathrm{KSO(n+2,n+2)}^\wedge$ & $\mathrm{SO(n+2,n+2)}^{\wedge\wedge}$ \\ \hline
 $\mathcal{N}=4$ & $\mathbb{U}$ & $\mathrm{SO}(6,n)\times\mathrm{SU}(1,1)$ & $\mathrm{SO}(8,8)$ &
 $\mathrm{SO}(8,8)^\wedge$ & $\mathrm{SO}(8,8)^{\wedge\wedge}$ \\ $n=6$ & $\mathbb{H}$ &
 $\mathrm{SO}(6)\times\mathrm{SO}(6)\times\mathrm{U}(1)$ & $\mathrm{SO}(8)\times\mathrm{SO}(8)$ &
 $\mathrm{KSO}(8,8)^\wedge$ & $\mathrm{KSO}(8,8)^{\wedge\wedge}$ \\ \hline
 $\mathcal{N}=4$ & $\mathbb{U}$ & $\mathrm{SO}(6,n)\times\mathrm{SU}(1,1)$ & $\mathrm{SO}(8,n+2)$ &
 $\mathrm{SO}(8,n+2)^\wedge$ & $\mathrm{SO}(8,n+2)^{\wedge\wedge}$ \\ & $\mathbb{H}$ &
 $\mathrm{SO}(6)\times\mathrm{SO}(n)\times\mathrm{U}(1)$ & $\mathrm{SO}(8)\times\mathrm{SO}(n+2)$ &
 $\mathrm{KSO}(8,n+2)^\wedge$ & $\mathrm{KSO}(8,n+2)^{\wedge\wedge}$ \\ \cline{2-6} $n > 6$ &
 $\mathbb{U}^{TS}$ & $\mathrm{SO(6,6) \times SU(1,1)}$ & $\mathrm{SO(8,8)}$ & $\mathrm{SO(8,8)}^\wedge$ &  $\mathrm{SO(8,8)}^{\wedge\wedge}$ \\
  & $\mathbb{H}^{TS}$ & $\mathrm{SO(6) \times SO(6) \times U(1)} $ & $\mathrm{SO(8) \times SO(8)} $ & $\mathrm{KSO(8,8)}^\wedge$ & $\mathrm{KSO(8,8)}^{\wedge\wedge}$  \\ \hline
  $\mathcal{N}=3$ &  $\mathbb{U}$ & $\mathrm{SU}(3,n)$ & $\mathrm{SU}(4,n+1)$ & $\mathrm{SU}(4,n+1)^\wedge$
  & $\mathrm{SU}(4,n+1)^{\wedge\wedge}$ \\ & $\mathbb{H}$ & $\mathrm{SU(3) \times SU(n) \times U(1)}$ & $\mathrm{SU(4) \times SU(n+1) \times U(1)}$
  & $\mathrm{KSU}(4,n+1)^\wedge$ &
  $\mathrm{KSU}(4,n+1)^{\wedge\wedge}$ \\
  \hline  $\mathcal{N}=2$ & geom. & $\mathcal{SK}$ & $\mathcal{Q}$ & $\mathcal{Q}^\wedge$ &
  $\mathcal{Q}^{\wedge\wedge}$ \\ \cline{2-6} & $\mathrm{TS}[\mbox{geom.}]$ & $\mathrm{TS}[\mathcal{SK}]$&
  $\mathrm{TS}[\mathcal{Q}]$ & $\mathrm{TS}[\mathcal{Q}^\wedge]$ & $\mathrm{TS}[\mathcal{Q}^{\wedge\wedge}]$ \\ \hline
\end{tabular}
}
  \caption{In this table we present the duality algebras $\mathbb{U}_D$ in $D=4,3,2,1$, for various
  values of the number of supersymmetry charges. We also mention the corresponding Tits Satake projected algebras
   (where they are well defined) that are relevant
  for the discussion of the cosmic billiard dynamics}\label{tavolona}
\end{table}
What is actually preserved in the c-map is the paint group $\mathrm{G_{paint}}$.
\par
Hence the dynamical walls are those associated with the Tits Satake
projected model but they come, in the true supergravity theory, in
\textit{painted copies}, for instance, within the context of our
example, the copy $\Omega_0$ and the copy $\Omega_Z$. The paint group
rotates these copies into each other. The explicit form taken by the
diagram (\ref{gendecompo2}) in the worked out example studied by the
present paper is:
\begin{equation}
\begin{array}{rcl}
\mbox{adj}(\mathrm{E_{7(-5)}}) &=&
\mbox{adj}(\mathrm{SO^\star(12)})\oplus\mbox{adj}(\mathrm{SL(2,\mathbb{R})_E})\oplus
(\mathbf{2},\mathbf{32}_s)  \\
\null &\Downarrow & \null\\
\mbox{adj}(\mathrm{F_{4(4)}}) &=&
\mbox{adj}(\mathrm{Sp(6,\mathbb{R})}\oplus\mbox{adj}(\mathrm{SL(2,\mathbb{R})_E})\oplus
(\mathbf{2},\mathbf{14})  \
\end{array}
\label{Tsdecompo6}
\end{equation}
The representation $\mathbf{14}$ of $\mathrm{Sp(6,\mathbb{R})}$ is that of an antisymmetric symplectic
traceless tensor:
\begin{equation}
\begin{array}{ccc}
 \mbox{dim}_{\mathrm{Sp(6,\mathbb{R})}} & \begin{array}{c}
  \vspace{-0.6cm} \\ \widetilde{\yng(1,1)}\
 \end{array} & \, = \, \mathbf{14}
\end{array}
\label{14repre}
\end{equation}
On the other hand the invariance of the paint group through
dimensional reduction and oxidation can be easily checked as
follows. First of all we note that $\mathrm{G_{paint}}=\mathrm{SO(3)^3}$ is
both a subgroup of $\mathrm{SO(12)}$ and of $\mathrm{SU(6)}$ as it is
easily verified through the subgroup chain:
\begin{equation}
\begin{array}{ccccccc}
  \mathrm{SO(12)} & \supset & \mathrm{SU(6)} & \supset & \mathrm{SU(4)} & \times & \mathrm{SU(2)}  \\
    \downarrow & \null & \downarrow & \null & \simeq & \null & \simeq \\
  \mathrm{SO(12)} & \supset & \mathrm{SU(6)} & \supset & \mathrm{SO(6)} & \times &
  \mathrm{SO(3)}\\
  \downarrow & \null & \downarrow & \null & \bigcup & \null & \downarrow \\
   \mathrm{SO(12)} & \supset & \mathrm{SU(6)} & \supset & \mathrm{SO(4)} \times \mathrm{SO(2)} & \times &
  \mathrm{SO(3)} \\
   \downarrow & \null & \downarrow & \null & \simeq & \null & \downarrow \\
 \mathrm{SO(12)} & \supset & \mathrm{SU(6)} & \supset & \mathrm{SO(3)\times SO(3)} \times \mathrm{SO(2)} & \times &
  \mathrm{SO(3)} \\
\end{array}
\end{equation}
Secondly we note that the non maximally split coset manifold
(\ref{so12starmanif}) appearing in $D=4$ has dimension $30$ and rank
$3$. This means that out of the $30$ positive roots there are three,
$\beta_1$, $\beta_2$ and $\beta_3$ that are orthogonal to the $3$ non--compact
Cartan generators. Together with the three compact Cartan generators
they make up the same $\mathrm{SO(3)^3}$ \textit{paint} Lie algebra as in the
$D=3$ case. Furthermore the remaining $27$ non compact roots which
together with the $3$ non compact Cartans span the solvable Lie
algebra of $Solv \left(\mathrm{ SO^\star(12) /SU(6) \times U(1)} \right)
$ are accounted for in the following way. The Tits Satake projection
of $\mathrm{SO^\star(12)}$ is the maximally split Lie algebra
$\mathrm{Sp(6,\mathbb{R}})$.
This latter is non simply laced and has $9$ positive roots which
distribute into $3$ long ones ($\alpha^\ell = 2\epsilon_i$ ($i=1,\dots,3$) and $6$
short ones ($\alpha^s = \epsilon_i \pm \epsilon_j $, $i <j, \,  i,j=1,2,3$).
Just as in $D=3$ the long roots of $\mathrm{Sp(6,\mathbb{R})}$ correspond to roots of
$\mathrm{SO^\star(12)}$ that are singlets under the paint group,
while the short ones correspond to roots of $\mathrm{SO^\star(12)}$
which arrange into the following $12$ dimensional representation
\begin{equation}
\mathbf{12}_{paint} \, = \,   \left( 2,2,0\right)  \oplus \left( 2,0,2\right) \oplus \left( 0,2,2\right)
\label{12painto}
\end{equation}
In $D=3$ we have $4$-copies of the representation
$\mathbf{12}_{paint}$ while in $D=4$ we just have $2$--copies of the same. It is instructive
to compare how the total number of roots is retrieved in the two
cases:
\begin{equation}
\begin{array}{lcccccl}
\mbox{ $\#$ of $\mathrm{E}_7$ roots = }  63 & = & \underbrace{3}_{compact} & +
&  \underbrace{12}_{long} & + & 4 \times \mathbf{12}_{paint} \nonumber\\
\null & = & \underbrace{3}_{compact} & +
&  \underbrace{12}_{long} & + & \underbrace{12}_{short} \times
\left( \mathbf{1+3}\right) _{\mathrm{SO(3)_{diag}}} \\
\mbox{ $\#$ of $\mathrm{SO}^\star(12)$ roots = }  30 & = & \underbrace{3}_{compact} & +
&  \underbrace{3}_{long} & + & 2 \times \mathbf{12}_{paint} \nonumber\\
\null & = & \underbrace{3}_{compact} & +
&  \underbrace{3}_{long} & + & \underbrace{6}_{short} \times
\left( \mathbf{1+3}\right) _{\mathrm{SO(3)_{diag}}} \\
\end{array}
\label{comparus}
\end{equation}
In eq.(\ref{comparus}) the second and fourth lines recall that each
of the short roots of either $\mathrm{F_{4(4)}}$ or
$\mathrm{Sp(6,\mathbb{R})}$ has $4$ preimages in the $D=4$
algebra which arrange into a triplet plus a singlet with respect to
the diagonal subgroup $\mathrm{SO(3)}_{diag}=\mathbb{G}^0_{paint}$.
This shows how the structure of the paint group filters through the
dimensional reduction. We can analyze this phenomenon also at the
level of the symplectic representation $\mathbf{W}$ to which the
vector fields are assigned. For the full $\mathcal{N}=6$ supergravity model,
this representation is the spinorial $\mathbf{32}_s$ of
$\mathrm{SO^\star(12)}$. Following the general discussion given in our recent
paper \cite{noiKMpape}, the $32$ weights of this representation
are in one to one correspondence with the roots of $\mathrm{E}_7$ which have
non vanishing grading with respect to the highest root $\psi
=\alpha[63]$ in the numbering of appendix \ref{benissimo1}. This
root set subdivides into $32=8+24$ where $8$ roots are Tits-Satake
projected into $8$ long roots of $\mathrm{F_{4(4)}}$, while $24$ are
Tits-Satake projected into $6$ short roots of the same. The
$14$-dimensional representation of $\mathrm{Sp(6,\mathbb{R})}$ is
just made by these $8+6$ roots of $\mathrm{F}_{4(4)}$ which have non vanishing
grading with respect to its own highest root $\psi_{TS}$. Indeed, as
we have noted in \cite{noiKMpape} the Tits Satake projection of the highest root
is the highest root of the target algebra.
\par
The above discussion provides the essential tools to perform the
dimensional oxidation of the solutions we have found to full fledged
solutions of supergravity models in $D=4$ or even in higher
dimensions. We do not address this issue in the present paper leaving
it for further publications where we also plan to provide a systematic
analysis of the Tits Satake projection for all supergravity theories
linking it to the properties of the compactification manifolds.
\par
We deem that the present detailed case-study has illustrated the role
of the \textbf{dimensional reduction invariant paint group} in
reducing the study of billiard dynamics to simpler maximally split
cosets.
\section{Conclusions}
\label{concludo} In this paper we have considered one of the two
necessary extensions of the analysis of \textit{smooth cosmic
billiards} initiated by us in \cite{noiconsasha}: that to
supergravity theories with lesser supersymmetry than the maximal
one. The other necessary extension is the further reduction to
$D<3$ dimensions, which we have recently addressed in
\cite{noiKMpape} by studying the universal field--theoretical
mechanism of the affine extension. As displayed in the systematic
analysis presented by us in \cite{noiKMpape}, lesser supersymmetry
involves a general new feature: cosets that are not maximally
split and correspond to non maximally non compact real sections of
their isometry algebra. For these cosets the \textit{compensator
method} devised by us in \cite{noiconsasha} cannot be directly
applied. Yet we have shown in this paper that the dynamical
problem can be reduced, also in these cases, to a problem which
can be solved with the compensator method. In fact the original
system can be reduced to a maximally split one, performing the
Tits--Satake projection of the original Lie algebra. The solutions
of the projected system (that can be easily found with the
compensator method) are also solutions of the complete one.
Moreover, we also showed that many other solutions can be obtained
from these by global rotations of a suitable compact group that we
named \emph{paint group}. Although we do not have the general
integral for these cases, we showed
how to obtain a large class of solutions that, probably, are the most relevant from the physical point of view.\\
Tits Satake projection of the original Lie algebra has emerged as
a central token in discussing cosmic billiards for lesser
supersymmetry. We have illustrated its role by an in depth
analysis of a specific example that of $\mathcal{N}=6$
supergravity. Through this case-study we were able to extrapolate
the main general features that apply to all supergravity models
and which we plan to study systematically in a future publication.
In particular we have elucidated the key role of the
$\mathbb{G}_{\mathrm{paint}}$ group, a notion not yet introduced
in the literature and leading to the idea of \textit{painted
billiard walls}. The main property of
$\mathbb{G}_{\mathrm{paint}}$ is that it commutes with the c-map,
namely with dimensional reduction. Hence it filters through
dimensional oxidation and can be retrieved in higher dimensional
supergravity.
\par
The main research line that streams from our results is the analysis
of the Tits Satake projection and of its kernel (the paint group) in more general contexts,
in particular in the context of generic special K\"ahler geometry, of
which our case study is also an example (see for instance \cite{myparis} for a
review). Furthermore keeping in mind the generic interpretation of
the scalar manifold $\mathcal{M}_{scalar}$ as \textit{moduli space} for the geometry of the \textit{compactification
manifold}, a Calabi Yau $\mathcal{M}_{CY}$ in the case where $\mathcal{M}_{scalar}$ is special
K\"ahlerian, it is challenging to obtain the interpretation of the
Tits Satake projection at the level of the compact manifold geometry.
This, as already stressed, we plan to do in the immediate future.
\par
It is at the same time quite interesting to consider the interplay
between the Tits Satake projection and the gauging of supergravity
models which is also on agenda.
\par
As we have illustrated in this paper we can easily obtain smooth
realizations  of the cosmic billiard with several bounces. The number
of these bounces, however,  is finite, as long as we deal with finite
algebras, namely as long as we discuss higher dimensional configurations
from a $D=3$ perspective. This is so because bounces are created, as we have shown,  by
compact group rotations along different generators and there is a
finite number of them if the number of roots is finite. In order to
see infinite bounces and may be chaos we have to have infinitely many
roots, namely we have to look at higher dimensional supergravity from
a $D=2$ or $D=1$ perspective. This requires to consider the affine or
hyperbolic Ka\v c--Moody extensions which we have addressed in the
recent paper \cite{noiKMpape}. Yet, as touched upon there and
readdressed in the present example here, the Tits Satake projection
commutes with the affine extension and in general with dimensional
reduction, which preserves the structure of the paint group
$\mathbb{G}_{\mathrm{paint}}$. Hence a door has been open how to
paint walls and roots also in Ka\v c-Moody algebras.
\newpage
\appendix
\section{Listing the positive roots of $\mathrm{E}_{7}$}
\label{benissimo1}
Listing of all positive roots of $\mathrm{E}_7$. The first column gives the Dynkin label,
the last gives the euclidean components of the root vectors
\begin{eqnarray*}
\begin{array}{ccccl}
 \alpha[1] & =  & \{ 1,
      0, 0, 0, 0, 0, 0\} & = &\{ 1, -1, 0,
      0, 0, 0, 0\}   \\
\alpha[2] & =  & \{ 0,
      1, 0, 0, 0, 0, 0\} & = &\{ 0, 1, -1,
      0, 0, 0, 0\}   \\
\alpha[3] & =  & \{ 0,
      0, 1, 0, 0, 0, 0\} & = &\{ 0, 0,
      1, -1, 0, 0, 0\}   \\
\alpha[4] & =  & \{ 0,
      0, 0, 1, 0, 0, 0\} & = &\{ 0, 0, 0,
      1, -1, 0, 0\}   \\
\alpha[5] & =  & \{ 0,
      0, 0, 0, 1, 0, 0\} & = &\{ 0, 0, 0, 0,
       1, -1, 0\}   \\
\alpha[6] & =  & \{ 0,
      0, 0, 0, 0, 1, 0\} & = &\{ 0, 0, 0, 0,
       1, 1, 0\}   \\
\alpha[7] & =  & \{ 0,
      0, 0, 0, 0, 0,
      1\} & = &\{ - \frac{1}{2} , - \frac{1}{2} , - \frac{1}{2} , -
\frac{1}{2} , - \frac{1}{2} , - \frac{1}{2} , \frac{1}{{\sqrt{2}}}\}   \\
\alpha[8] & =  & \{ 1,
      1, 0, 0, 0, 0, 0\} & = &\{ 1, 0, -1,
      0, 0, 0, 0\}   \\
\alpha[9] & =  & \{ 0,
      1, 1, 0, 0, 0, 0\} & = &\{ 0, 1,
      0, -1, 0, 0, 0\}   \\
\alpha[10] & =  & \{ 0,
      0, 1, 1, 0, 0, 0\} & = &\{ 0, 0, 1,
      0, -1, 0, 0\}   \\
\alpha[11] & =  & \{ 0,
      0, 0, 1, 1, 0, 0\} & = &\{ 0, 0, 0, 1,
       0, -1, 0\}   \\
\alpha[12] & =  & \{ 0,
      0, 0, 1, 0, 1, 0\} & = &\{ 0, 0, 0, 1,
       0, 1, 0\}   \\
\alpha[13] & =  & \{ 0,
      0, 0, 0, 0, 1,
      1\} & = &\{ - \frac{1}{2} , - \frac{1}{2} , - \frac{1}{2} , -
\frac{1}{2} , \frac{1}{2}, \frac{1}{2}, \frac{1}{{\sqrt{2}}}\} \
  \\
\alpha[14] & =  & \{ 1,
      1, 1, 0, 0, 0, 0\} & = &\{ 1, 0,
      0, -1, 0, 0, 0\}   \\
\alpha[15] & =  & \{ 0,
      1, 1, 1, 0, 0, 0\} & = &\{ 0, 1, 0,
      0, -1, 0, 0\}   \\
\alpha[16] & =  & \{ 0,
      0, 1, 1, 1, 0, 0\} & = &\{ 0, 0, 1, 0,
       0, -1, 0\}   \\
\alpha[17] & =  & \{ 0,
      0, 1, 1, 0, 1, 0\} & = &\{ 0, 0, 1, 0,
       0, 1, 0\}   \\
\alpha[18] & =  & \{ 0,
      0, 0, 1, 0, 1,
      1\} & = &\{ - \frac{1}{2} , - \frac{1}{2} , - \frac{1}{2} , \
\frac{1}{2}, - \frac{1}{2} , \frac{1}{2}, \
\frac{1}{{\sqrt{2}}}\}   \\
\alpha[19] & =  & \{ 0,
      0, 0, 1, 1, 1, 0\} & = &\{ 0, 0, 0, 1,
       1, 0, 0\}   \\
\alpha[20] & =  & \{ 1,
      1, 1, 1, 0, 0, 0\} & = &\{ 1, 0, 0,
      0, -1, 0, 0\}   \\
\alpha[21] & =  & \{ 0,
      1, 1, 1, 1, 0, 0\} & = &\{ 0, 1, 0, 0,
       0, -1, 0\}   \\
\alpha[22] & =  & \{ 0,
      1, 1, 1, 0, 1, 0\} & = &\{ 0, 1, 0, 0,
       0, 1, 0\}   \\
\alpha[23] & =  & \{ 0,
      0, 1, 1, 0, 1,
      1\} & = &\{ - \frac{1}{2} , - \frac{1}{2} , \frac{1}{2}, - \frac{1}{2} , - \frac{1}{2} , \frac{1}{2}, \frac{1}{{\sqrt{2}}}\} \
  \\
\alpha[24] & =  & \{ 0,
      0, 1, 1, 1, 1, 0\} & = &\{ 0, 0, 1, 0,
       1, 0, 0\}   \\
\alpha[25] & =  & \{ 0,
      0, 0, 1, 1, 1,
      1\} & = &\{ - \frac{1}{2} , - \frac{1}{2} , - \frac{1}{2} , \
\frac{1}{2}, \frac{1}{2}, - \frac{1}{2} , \
\frac{1}{{\sqrt{2}}}\}   \\
\alpha[26] & =  & \{ 1,
      1, 1, 1, 1, 0, 0\} & = &\{ 1, 0, 0, 0,
       0, -1, 0\}   \\
\alpha[27] & =  & \{ 1,
      1, 1, 1, 0, 1, 0\} & = &\{ 1, 0, 0, 0,
       0, 1, 0\}   \\
\alpha[28] & =  & \{ 0,
      1, 1, 1, 0, 1,
      1\} & = &\{ - \frac{1}{2} , \frac{1}{2}, - \frac{1}{2} , - \frac{1}{2} , - \frac{1}{2} , \frac{1}{2}, \frac{1}{{\sqrt{2}}}\} \
  \\
\alpha[29] & =  & \{ 0,
      1, 1, 1, 1, 1, 0\} & = &\{ 0, 1, 0, 0,
       1, 0, 0\}   \\
\alpha[30] & =  & \{ 0,
      0, 1, 1, 1, 1,
      1\} & = &\{ - \frac{1}{2} , - \frac{1}{2} , \frac{1}{2}, - \frac{1}{2} , \frac{1}{2}, - \frac{1}{2} , \frac{1}{{\sqrt{2}}}\} \
  \\
\alpha[31] & =  & \{ 0,
      0, 1, 2, 1, 1, 0\} & = &\{ 0, 0, 1, 1,
       0, 0, 0\}   \\
       \end{array}
       \end{eqnarray*}
\begin{eqnarray*}
\begin{array}{ccccl}
\alpha[32] & =  & \{ 1,
      1, 1, 1, 0, 1,
      1\} & = &\{ \frac{1}{2}, -
\frac{1}{2} , - \frac{1}{2} , - \frac{1}{2} , - \frac{1}{2} , \frac{1}{2}, \frac{1}{{\sqrt{2}}}\} \
  \\
\alpha[33] & =  & \{ 1,
      1, 1, 1, 1, 1, 0\} & = &\{ 1, 0, 0, 0,
       1, 0, 0\}   \\
\alpha[34] & =  & \{ 0,
      1, 1, 1, 1, 1,
      1\} & = &\{ - \frac{1}{2} , \frac{1}{2}, - \frac{1}{2} , - \frac{1}{2} , \frac{1}{2}, - \frac{1}{2} , \frac{1}{{\sqrt{2}}}\} \
  \\
\alpha[35] & =  & \{ 0,
      1, 1, 2, 1, 1, 0\} & = &\{ 0, 1, 0, 1,
       0, 0, 0\}   \\
\alpha[36] & =  & \{ 0,
      0, 1, 2, 1, 1,
      1\} & = &\{ - \frac{1}{2} , - \frac{1}{2} , \frac{1}{2}, \frac{1}{2}, -
\frac{1}{2} , - \frac{1}{2} , \frac{1}{{\sqrt{2}}}\} \
  \\
\alpha[37] & =  & \{ 1,
      1, 1, 1, 1, 1,
      1\} & = &\{ \frac{1}{2}, -
\frac{1}{2} , - \frac{1}{2} , - \frac{1}{2} , \frac{1}{2}, - \frac{1}{2} , \frac{1}{{\sqrt{2}}}\} \
  \\
\alpha[38] & =  & \{ 1,
      1, 1, 2, 1, 1, 0\} & = &\{ 1, 0, 0, 1,
       0, 0, 0\}   \\
\alpha[39] & =  & \{ 0,
      1, 1, 2, 1, 1,
      1\} & = &\{ - \frac{1}{2} , \frac{1}{2}, - \frac{1}{2} , \frac{1}{2}, -
\frac{1}{2} , - \frac{1}{2} , \frac{1}{{\sqrt{2}}}\} \
  \\
\alpha[40] & =  & \{ 0,
      1, 2, 2, 1, 1, 0\} & = &\{ 0, 1, 1, 0,
       0, 0, 0\}   \\
\alpha[41] & =  & \{ 0,
      0, 1, 2, 1, 2,
      1\} & = &\{ - \frac{1}{2} , - \frac{1}{2} , \frac{1}{2}, \frac{1}{2}, \frac{1}{2}, \
\frac{1}{2}, \frac{1}{{\sqrt{2}}}\}   \\
\alpha[42] & =  & \{ 1,
      1, 1, 2, 1, 1,
      1\} & = &\{ \frac{1}{2}, -
\frac{1}{2} , - \frac{1}{2} , \frac{1}{2}, -
\frac{1}{2} , - \frac{1}{2} , \frac{1}{{\sqrt{2}}}\} \
  \\
\alpha[43] & =  & \{ 1,
      1, 2, 2, 1, 1, 0\} & = &\{ 1, 0, 1, 0,
       0, 0, 0\}   \\
\alpha[44] & =  & \{ 0,
      1, 1, 2, 1, 2,
      1\} & = &\{ - \frac{1}{2} , \frac{1}{2}, - \frac{1}{2} , \frac{1}{2}, \frac{1}{2}, \
\frac{1}{2}, \frac{1}{{\sqrt{2}}}\}   \\
\alpha[45] & =  & \{ 0,
      1, 2, 2, 1, 1,
      1\} & = &\{ - \frac{1}{2} , \frac{1}{2}, \frac{1}{2}, - \frac{1}{2} , -
\frac{1}{2} , - \frac{1}{2} , \frac{1}{{\sqrt{2}}}\} \
  \\
\alpha[46] & =  & \{ 1,
      1, 1, 2, 1, 2,
      1\} & = &\{ \frac{1}{2}, -
\frac{1}{2} , - \frac{1}{2} , \frac{1}{2}, \frac{1}{2}, \
\frac{1}{2}, \frac{1}{{\sqrt{2}}}\}   \\
\alpha[47] & =  & \{ 1,
      1, 2, 2, 1, 1,
      1\} & = &\{ \frac{1}{2}, -
\frac{1}{2} , \frac{1}{2}, - \frac{1}{2} , -
\frac{1}{2} , - \frac{1}{2} , \frac{1}{{\sqrt{2}}}\} \
  \\
\alpha[48] & =  & \{ 1,
      2, 2, 2, 1, 1, 0\} & = &\{ 1, 1, 0, 0,
       0, 0, 0\}   \\
\alpha[49] & =  & \{ 0,
      1, 2, 2, 1, 2,
      1\} & = &\{ - \frac{1}{2} , \frac{1}{2}, \frac{1}{2}, - \frac{1}{2} , \frac{1}{2}, \
\frac{1}{2}, \frac{1}{{\sqrt{2}}}\}   \\
\alpha[50] & =  & \{ 1,
      1, 2, 2, 1, 2,
      1\} & = &\{ \frac{1}{2}, -
\frac{1}{2} , \frac{1}{2}, - \frac{1}{2} , \frac{1}{2}, \
\frac{1}{2}, \frac{1}{{\sqrt{2}}}\}   \\
\alpha[51] & =  & \{ 1,
      2, 2, 2, 1, 1,
      1\} & = &\{ \frac{1}{2}, \frac{1}{2}, \
- \frac{1}{2} , - \frac{1}{2} , -
\frac{1}{2} , - \frac{1}{2} , \frac{1}{{\sqrt{2}}}\} \
  \\
\alpha[52] & =  & \{ 0,
      1, 2, 3, 1, 2,
      1\} & = &\{ - \frac{1}{2} , \frac{1}{2}, \frac{1}{2}, \frac{1}{2}, - \frac{1}{2} , \
\frac{1}{2}, \frac{1}{{\sqrt{2}}}\}   \\
\alpha[53] & =  & \{ 1,
      1, 2, 3, 1, 2,
      1\} & = &\{ \frac{1}{2}, -
\frac{1}{2} , \frac{1}{2}, \frac{1}{2}, - \frac{1}{2} , \
\frac{1}{2}, \frac{1}{{\sqrt{2}}}\}   \\
\alpha[54] & =  & \{ 1,
      2, 2, 2, 1, 2,
      1\} & = &\{ \frac{1}{2}, \frac{1}{2}, \
- \frac{1}{2} , - \frac{1}{2} , \frac{1}{2}, \
\frac{1}{2}, \frac{1}{{\sqrt{2}}}\}   \\
\alpha[55] & =  & \{ 0,
      1, 2, 3, 2, 2,
      1\} & = &\{ - \frac{1}{2} , \frac{1}{2}, \frac{1}{2}, \frac{1}{2}, \frac{1}{2}, -
\frac{1}{2} , \frac{1}{{\sqrt{2}}}\}   \\
\alpha[56] & =  & \{ 1,
      1, 2, 3, 2, 2,
      1\} & = &\{ \frac{1}{2}, -
\frac{1}{2} , \frac{1}{2}, \frac{1}{2}, \frac{1}{2}, -
\frac{1}{2} , \frac{1}{{\sqrt{2}}}\}   \\
\alpha[57] & =  & \{ 1,
      2, 2, 3, 1, 2,
      1\} & = &\{ \frac{1}{2}, \frac{1}{2}, \
- \frac{1}{2} , \frac{1}{2}, - \frac{1}{2} , \
\frac{1}{2}, \frac{1}{{\sqrt{2}}}\}   \\
\alpha[58] & =  & \{ 1,
      2, 2, 3, 2, 2,
      1\} & = &\{ \frac{1}{2}, \frac{1}{2}, \
- \frac{1}{2} , \frac{1}{2}, \frac{1}{2}, - \frac{1}{2} , \frac{1}{{\sqrt{2}}}\}   \\
\alpha[59] & =  & \{ 1,
      2, 3, 3, 1, 2,
      1\} & = &\{ \frac{1}{2}, \frac{1}{2}, \
\frac{1}{2}, - \frac{1}{2} , - \frac{1}{2} , \
\frac{1}{2}, \frac{1}{{\sqrt{2}}}\}   \\
\alpha[60] & =  & \{ 1,
      2, 3, 3, 2, 2,
      1\} & = &\{ \frac{1}{2}, \frac{1}{2}, \
\frac{1}{2}, - \frac{1}{2} , \frac{1}{2}, - \frac{1}{2} , \frac{1}{{\sqrt{2}}}\}   \\
\alpha[61] & =  & \{ 1,
      2, 3, 4, 2, 2,
      1\} & = &\{ \frac{1}{2}, \frac{1}{2}, \
\frac{1}{2}, \frac{1}{2}, - \frac{1}{2} , - \frac{1}{2} , \frac{1}{{\sqrt{2}}}\}   \\
\alpha[62] & =  & \{ 1,
      2, 3, 4, 2, 3,
      1\} & = &\{ \frac{1}{2}, \frac{1}{2}, \
\frac{1}{2}, \frac{1}{2}, \frac{1}{2}, \frac{1}{2}, \frac{1}{{\sqrt{2}}}\} \
  \\
\alpha[63] & =  & \{ 1,
      2, 3, 4, 2, 3, 2\} & = &\{ 0, 0, 0, 0,
       0, 0, {\sqrt{2}}\}   \\
\end{array}
\end{eqnarray*}
\section{Explicit construction of the fundamental and adjoint
representation of $\mathrm{F}_{4(4)}$}
\label{appendoF4}
The semisimple complex Lie algebra $\mathrm{F}_{4}$ is defined by the Dynkin
diagram in figure \ref{F4dynk} and a set of simple roots corresponding
to such diagram was provided in eq.(\ref{simplef4}). A complete list
of the 24 positive roots was  given in table \ref{tablettas}. The
roots were further subdivided into the set of 12 long roots and 12
short roots respectively listed in table \ref{etaset} and
\ref{deltaset} where their correspondence with $\mathrm{E}_{7}$ roots was
spelled out. The adjoint representation of $\mathrm{F}_{4}$ is
$52$--dimensional, while its fundamental representation is
$26$--dimensional. This dimensionality is true for all real sections
of the Lie algebra but the explicit structure of the representation
is quite different in each real section. Here we are interested in
the maximally split real section $\mathrm{F}_{4(4)}$. For such a section we have
a maximal, regularly embedded, subgroup $\mathrm{SO(5,4)} \subset
\mathrm{F_{4(4)}}$. The decomposition of the representations with
respect to this particular subgroup is the essential instrument for their actual
construction. For the adjoint representation we have the
decomposition:
\begin{equation}
  \underbrace{{\bf 52} }_{\mbox{adj}\,F_{4(4)}}\, \stackrel{\mathrm{SO(5,4)}}{\Longrightarrow} \,
  \underbrace{{\bf 36}}_{\mbox{adj}\, \mathrm{SO(5,4)}} \oplus
 \underbrace{ {\bf 16}}_{\mbox{spinor of }\,\mathrm{SO(5,4)} }
\label{adjdecompo}
\end{equation}
while for the fundamental one we have:
\begin{equation}
  \underbrace{{\bf 26} }_{\mbox{fundamental }\,F_{4(4)}}\, \stackrel{\mathrm{SO(5,4)}}{\Longrightarrow} \,
  \underbrace{{\bf 9}}_{\mbox{vector of } \mathrm{SO(5,4)}} \oplus
 \underbrace{ {\bf 16}}_{\mbox{spinor of }\mathrm{SO(5,4)} }  \oplus
 \underbrace{ {\bf 1}}_{\mbox{singlet of }\mathrm{SO(5,4)} }
\label{fundecompo}
\end{equation}
In view of this, we fix our conventions for the  $\mathrm{SO(5,4)}$ invariant
metric as it follows
\begin{equation}
  \eta_{AB} = \mbox{diag}\left \{ +,+,+,+,+,-,-,-,-\right\}
\label{eta54}
\end{equation}
and we perform an explicit construction of the $16 \times 16$ dimensional gamma
matrices which satisfy the Clifford algebra
\begin{equation}
  \left\{ \Gamma_A \, , \, \Gamma_B\right\} = \eta_{AB} \, {\bf 1}
\label{clif45}
\end{equation}
and are \textit{all completely real}. This construction is provided by the following tensor products:
\begin{eqnarray}
\Gamma_1 & = & \sigma_1 \otimes \sigma_3  \otimes {\bf 1} \otimes {\bf 1} \nonumber\\
\Gamma_2 & = & \sigma_3 \otimes \sigma_3 \otimes {\bf 1} \otimes {\bf 1} \nonumber\\
\Gamma_3 & = & {\bf 1} \otimes \sigma_1 \otimes {\bf 1} \otimes \sigma_1 \nonumber\\
\Gamma_4 & = & {\bf 1} \otimes \sigma_1 \otimes \sigma_1 \otimes \sigma_3 \nonumber\\
\Gamma_5 & = & {\bf 1} \otimes \sigma_1 \otimes \sigma_3 \otimes \sigma_3 \nonumber\\
\Gamma_6 & = & {\bf 1} \otimes {\rm i} \sigma_2 \otimes {\bf 1} \otimes {\bf 1} \nonumber\\
\Gamma_7 & = & {\bf 1} \otimes \sigma_1 \otimes {\rm i}\, \sigma_2  \otimes \sigma_3 \nonumber\\
\Gamma_8 & = & {\bf 1} \otimes \sigma_1 \otimes {\bf 1} \otimes {\rm i} \sigma_2 \nonumber\\
\Gamma_9 & = & {\rm i} \, \sigma_2 \otimes \sigma_3 \otimes {\bf 1} \otimes {\bf 1}
\label{gamme45}
\end{eqnarray}
where by $\sigma_i$ we have denoted the standard Pauli matrices:
\begin{equation}
  \sigma_1 = \left(\begin{array}{cc}
    0 & 1 \\
    1 & 0 \
  \end{array} \right) \quad ; \quad \sigma_2 = \left(\begin{array}{cc}
    0 & -{\rm i} \\
    {\rm i} & 0 \
  \end{array}  \right) \quad ; \quad \sigma_3 = \left(\begin{array}{cc}
    1 & 0 \\
    0 & -1 \
  \end{array} \right)
\label{paulisigma}
\end{equation}
Moreover we introduce the  $C_+$ charge conjugation matrix, such that:
\begin{eqnarray}
C_+ & = &  \left(C_+\right) ^T \quad ; \quad C_+^2 ={\bf 1}
\nonumber\\
  C_+ \, \Gamma_A \, C_+ &=& \left( \Gamma_A \right )^T
\label{cpdefi}
\end{eqnarray}
In the basis of eq. (\ref{gamme45}) the explicit form of $C_+$ is
given by:
\begin{equation}
  C_+ = {\rm i} \, \sigma_2 \otimes \sigma_1 \otimes {\rm i} \,
  \sigma_2 \otimes \sigma_1
\label{Cpdefi}
\end{equation}
Then we define the usual generators $J_{AB} = - J_{BA}$ of the pseudorthogonal algebra
$\mathrm{SO(5,4)}$ satisfying the commutation relations:
\begin{equation}
 \left[  J_{AB} \, , \, J_{CD} \right]  = \eta_{BC} \, J_{AD} -
 \eta_{AC} \, J_{BD} -\eta_{BD} J_{AC} + \eta_{AD} J_{BC}
\label{stanJABcom}
\end{equation}
and we construct the spinor and the vector representations by respectively
setting:
\begin{equation}
  J^s_{CD} = \frac 14 \left[ \Gamma_{C} \, , \,  \Gamma_D \right] \quad
  ; \quad \left(  J^v_{CD} \right) _{A}^{\phantom{A}B} \, = \,
  \eta_{CA} \, \delta_D^B \, - \, \eta_{DA} \, \delta_C^B
\label{spinvecrep}
\end{equation}
In this way if $v_A$ denote the components of a vector, $\xi$
those of a real spinor and $\epsilon^{AB} = - \epsilon^{BA}$ are the
parameters of an infinitesimal $\mathrm{SO(5,4)}$ rotation we can
write the $\mathrm{SO(5,4)}$ transformation as follows:
\begin{equation}
  \delta_{\mathrm{SO(5,4)}} \, v_A = 2 \, \epsilon_{AB} \, v^B \quad ; \quad \delta_{\mathrm{SO(5,4)}} \,
  \xi = \ft 12 \, \epsilon^{AB} \, \Gamma_{AB} \xi
\label{delta45}
\end{equation}
where indices are raised and lowered with the metric (\ref{eta54}).
Furthermore we introduce the conjugate spinors via the position:
\begin{equation}
  \overline{\xi} \equiv \xi^T \, C_+
\label{conjspin}
\end{equation}
With these preliminaries, we are now a position to write the explicit
form of the $26$-dimensional fundamental representation of $\mathrm{F}_{4(4)}$ and in this way to construct
also its structure constants and hence its adjoint representation,
which is our main goal.
\par
According to eq.(\ref{adjdecompo}) the parameters of an $\mathrm{F}_{4(4)}$
representation are given by an anti-symmetric tensor $\epsilon_{AB}$
and a spinor $q$. On the other hand a \textit{vector} in the
$26$--dimensional representation is specified by a collection of
three objects, namely a scalar $\phi$, a vector $v_A$ and a spinor
$\xi$. The representation is constructed if we specify the $\mathrm{F_{4(4)}}$
transformation of these objects. This is done by writing:
\begin{equation}
  \delta_{F_{4(4)}} \, \left(\begin{array}{c}
    \phi \\
    v_A \\
    \xi \
  \end{array} \right) \, \equiv \, \left[ \epsilon^{AB} T_{AB} \, + \, \overline{q} \,Q\right] \,
  \left(\begin{array}{c}
     \phi \\
     v_A \\
     \xi \
  \end{array} \right) \, = \, \left(\begin{array}{l}
    \overline{q} \, \xi \\
    2 \, \epsilon_{AB} \, v^B \, + \, a \, \overline{q}  \, \Gamma_A \, \xi \\
    \ft 12 \, \epsilon^{AB} \, \Gamma_{AB} \, - \,3 \,  \phi \, q \, - \, \frac{1}{a} \, v^A \, \Gamma_A \xi \
  \end{array} \right)
\label{deltaf44}
\end{equation}
where $a$ is a numerical real arbitrary but non-null parameter.
Eq.(\ref{deltaf44}) defines the generators $T_{AB}$ and $Q$ as $26
\times 26$ matrices and therefore completely specifies the
fundamental representation of the Lie algebra $\mathrm{F_{4(4)}}$.
Explicitly we have:
\begin{equation}
  T_{AB} \, = \, \left(\begin{array}{c|c|c}
    0 & 0 & 0 \\
    \hline
    0 & J^v_{AB} & 0 \\
    \hline
    0 & 0 & J^s_{AB} \
  \end{array} \right)
\label{TAB}
\end{equation}
and
\begin{equation}
  Q_\alpha \, = \, \left(\begin{array}{c|c|c}
    0 & 0 & \delta_\alpha^\beta \\
    \hline
    0 & 0 & a \,  \left( \Gamma_A \right )_\alpha^\beta \\
    \hline
    -3\, \delta_\alpha^\beta & -\frac{1}{a} \, \left( \Gamma_B \right)_\alpha^\beta & 0 \
  \end{array} \right)
\label{Qgen}
\end{equation}
and the Lie algebra commutation relations are evaluated to be the
following ones:
\begin{eqnarray}
 \left[  T_{AB} \, , \, T_{CD} \right] & = &\eta_{BC} \, T_{AD} -
 \eta_{AC} \, T_{BD} -\eta_{BD} T_{AC} + \eta_{AD} T_{BC} \nonumber\\
\left[  T_{AB} \, , \, Q \right] & = & \ft 12 \, \Gamma_{AB} \, Q
\nonumber\\
\left[  Q_\alpha \, , \, Q_\beta \right] & = & - \frac{1}{12} \,  \left( C_+ \Gamma^{AB}
\right) _{\alpha\beta} \, T_{AB}
\label{F4comme}
\end{eqnarray}
Eq.(\ref{F4comme}), together with eq.s(\ref{gamme45}) and eq.(\ref{cpdefi}) provides an explicit numerical
construction of
the structure constants of the maximally split $\mathrm{F_{4(4)}}$ Lie
algebra. What we still have to do is to identify the relation between
the tensorial basis of generators  in eq. (\ref{F4comme}) and the
Cartan-Weyl basis in terms of Cartan generators and step operators.
To this effect let us enumerate the $52$ generators of $\mathrm{F}_{4(4)}$
in the tensorial representation according to the following table:
\begin{equation}
  \begin{array}{|rl|rl|rl|rl|}
  \hline
 \Omega_{1} = & T_{12}  & \Omega_{2}= & T_{13} & \Omega_{3}= & T_{14} & \Omega_{4}= & T_{15}  \\
 \Omega_{5} = & T_{16} & \Omega_{6}= & T_{17} & \Omega_{7}= & T_{18} & \Omega_{8}= & T_{19}  \\
 \Omega_{9} = & T_{23} & \Omega_{10} =& T_{24} & \Omega_{11}= & T_{25} & \Omega_{12}= & T_{26}  \\
 \Omega_{13} = & T_{27} & \Omega_{14}= & T_{28} & \Omega_{15}= & T_{29} & \Omega_{16}= & T_{34}  \\
 \Omega_{17} = & T_{35} & \Omega_{18}= & T_{36} & \Omega_{19}= & T_{37} & \Omega_{20}= & T_{38}  \\
 \Omega_{21} = & T_{39} & \Omega_{22}= & T_{45} & \Omega_{23}= & T_{46} & \Omega_{24}= & T_{47}  \\
 \Omega_{25} = & T_{48} & \Omega_{26}= & T_{49} & \Omega_{27}= & T_{56} & \Omega_{28}= & T_{57}  \\
 \Omega_{29} = & T_{58} & \Omega_{30}= & T_{59} & \Omega_{31}=& T_{67} & \Omega_{32}= & T_{68}  \\
 \Omega_{33} = & T_{69} & \Omega_{34}= & T_{78} & \Omega_{35}= & T_{79} & \Omega_{36}= & T_{89}  \\
 \Omega_{37} = & Q_{1} & \Omega_{38}= & Q_{2} & \Omega_{39}= & Q_{3} & \Omega_{40}= & Q_{4}  \\
 \Omega_{41} = & Q_{5} & \Omega_{42}= & Q_{6} & \Omega_{43}= & Q_{7} & \Omega_{44}= & Q_{8 }  \\
 \Omega_{45} = & Q_{9} & \Omega_{46}= & Q_{10} & \Omega_{47}= & Q_{11} & \Omega_{48}= & Q_{12}  \\
 \Omega_{49} = & Q_{13} & \Omega_{50}= & Q_{14} & \Omega_{51}= & Q_{15} & \Omega_{52}= & Q_{16}  \\
\hline
  \end{array}
\label{conversione}
\end{equation}
Then, as Cartan subalgebra we take the linear span of the following
generators:
\begin{equation}
  CSA \, \equiv \, \mbox{span}\left( \Omega_5 \, , \, \Omega_{13} \, ,
  \,\Omega_{20} \, , \, \Omega_{26} \right)
\label{CSAF44}
\end{equation}
and furthermore we specify the following basis:
\begin{equation}
  \begin{array}{rclcrcl}
    \mathcal{H}_1 & = & \Omega_5 + \Omega_{13} & ; & \mathcal{H}_2 & = & \Omega_5 - \Omega_{13} \\
    \mathcal{H}_3 & = & \Omega_{20} + \Omega_{26} & ; & \mathcal{H}_4 & = & \Omega_{20} - \Omega_{26} \
  \end{array}
\label{CSAbasa}
\end{equation}
With respect to this basis the step operators corresponding to the
positive roots of $\mathrm{F_{4(4)}}$ as ordered and displayed in table
\ref{tablettas} are those enumerated in
table \ref{tablettona}.
\begin{table}
  \centering
{\scriptsize  $$
\begin{array}{|ll|l|c|}
\hline
\mbox{name} & \mbox{Dynkin lab}&\mbox{comp. root} & \mbox{step operator
$E^{\varpi} \, = \, $} \nonumber\\
\hline
\varpi[ 1 ] &
 \{ 1, 0, 0,
      0\} & -{y_1} - {y_2} - {y_3}  +
{y_4}& \left( -{{\Omega }_3} - {{\Omega }_8} + {{\Omega }_{23}} - {{\Omega }_{33}}\right)  \nonumber\\
\varpi[ 2 ] &
 \{ 0, 1, 0,
      0\} & 2\, {y_3}& \left( {{\Omega }_{16}} - {{\Omega }_{21}} + {{\Omega }_{25}} + {{\Omega }_{36}}\right)
\nonumber\\
\varpi[ 3 ] &
 \{ 0, 0, 1,
      0\} & {y_2} - {y_3}\ & \left( {{\Omega }_{37}} + {{\Omega }_{39}} + {{\Omega }_{41}} - {{\Omega }_{43}} +
  {{\Omega }_{45}} - {{\Omega }_{47}} + {{\Omega }_{49}} + {{\Omega }_{51}}\right)
\nonumber\\
\varpi[ 4 ] &
 \{ 0, 0, 0,
      1\} & {y_1} - {y_2}\ & \left( {{\Omega }_{11}} + {{\Omega }_{28}}\right)
\nonumber\\
\varpi[ 5 ] &
 \{ 1, 1, 0,
      0\} & -{y_1} - {y_2} + {y_3}  +
{y_4}& -\frac{1}{\sqrt{2}}\left( -{{\Omega }_2} + {{\Omega }_7} + {{\Omega }_{18}} + {{\Omega }_{32}}\right)  \nonumber\\
\varpi[ 6 ] &
 \{ 0, 1, 1,
      0\} & {y_2} + {y_3}\ & -\frac{1}{\sqrt{2}}\left( -{{\Omega }_{38}} - {{\Omega }_{40}} + {{\Omega }_{42}} - {{\Omega }_{44}} +
  {{\Omega }_{46}} - {{\Omega }_{48}} - {{\Omega }_{50}} - {{\Omega }_{52}}\right)
\nonumber\\
\varpi[ 7 ] &
 \{ 0, 0, 1,
      1\} & {y_1} - {y_3}\ & -\left( -{{\Omega }_{37}} - {{\Omega }_{39}} + {{\Omega }_{41}} - {{\Omega }_{43}} +
  {{\Omega }_{45}} - {{\Omega }_{47}} - {{\Omega }_{49}} - {{\Omega }_{51}}\right)
\nonumber\\
\varpi[ 8 ] &
 \{ 1, 1, 1,
      0\} & -{y_1} + {y_4}\ & -\frac{1}{2}\left( {{\Omega }_{38}} + {{\Omega }_{40}} - {{\Omega }_{42}} + {{\Omega }_{44}} +
  {{\Omega }_{46}} - {{\Omega }_{48}} - {{\Omega }_{50}} - {{\Omega }_{52}}\right)
\nonumber\\
\varpi[ 9 ] &
 \{ 0, 1, 2,
      0\} & 2\, {y_2} & -\frac{1}{2}\left( {{\Omega }_1} + {{\Omega }_6} + {{\Omega }_{12}} - {{\Omega }_{31}}\right)
\nonumber\\
\varpi[ 10 ] &
 \{ 0, 1, 1,
      1\} & {y_1} + {y_3}\ & -\frac{1}{\sqrt{2}}\left( {{\Omega }_{38}} + {{\Omega }_{40}} + {{\Omega }_{42}} - {{\Omega }_{44}} +
  {{\Omega }_{46}} - {{\Omega }_{48}} + {{\Omega }_{50}} + {{\Omega }_{52}}\right)
\nonumber\\
\varpi[ 11 ] &
 \{ 1, 1, 2,
      0\} & -{y_1} + {y_2} - {y_3}  +
{y_4} & -\frac{1}{2\sqrt{2}}\left( {{\Omega }_{10}} + {{\Omega }_{15}} - {{\Omega }_{24}} + {{\Omega }_{35}}\right)  \nonumber\\
\varpi[ 12 ] &
 \{ 1, 1, 1,
      1\} & -{y_2} + {y_4}\ & -\frac{1}{2}\left( -{{\Omega }_{38}} - {{\Omega }_{40}} - {{\Omega }_{42}} + {{\Omega }_{44}} +
  {{\Omega }_{46}} - {{\Omega }_{48}} + {{\Omega }_{50}} + {{\Omega }_{52}}\right)
\nonumber\\
\varpi[ 13 ] &
 \{ 0, 1, 2,
      1\} & {y_1} + {y_2}\ & -\frac{1}{\sqrt{2}}\left(  {{\Omega }_4} + {{\Omega }_{27}}\right)
\nonumber\\
\varpi[ 14 ] &
 \{ 1, 2, 2,
      0\} & -{y_1} + {y_2} + {y_3}  +
{y_4} & -\frac{1}{4}\left( -{{\Omega }_9} + {{\Omega }_{14}} + {{\Omega }_{19}} + {{\Omega }_{34}} \right) \nonumber\\
\varpi[ 15 ] &
 \{ 1, 1, 2,
      1\} & -{y_3} + {y_4}\ & -\frac{1}{2}\left( {{\Omega }_{22}} - {{\Omega }_{30}}\right)
\nonumber\\
\varpi[ 16 ] &
 \{ 0, 1, 2,
      2\} & 2\, {y_1}
& -\frac{1}{2}\left( {{\Omega }_1} - {{\Omega }_6} + {{\Omega }_{12}} + {{\Omega }_{31}} \right) \nonumber\\
\varpi[ 17 ] &
 \{ 1, 2, 2,
      1\} & {y_3} + {y_4}\ & -\frac{1}{2\sqrt{2}}\left( {{\Omega }_{17}} + {{\Omega }_{29}}\right)
\nonumber\\
\varpi[ 18 ] &
 \{ 1, 1, 2,
      2\} & {y_1} - {y_2} - {y_3}  +
{y_4} & -\frac{1}{2\sqrt{2}}\left( {{\Omega }_{10}} + {{\Omega }_{15}} + {{\Omega }_{24}} - {{\Omega }_{35}}\right)  \nonumber\\
\varpi[ 19 ] &
 \{ 1, 2, 3,
      1\} & {y_2} + {y_4}\ & -\frac{1}{2\sqrt{2}}\left(  {{\Omega }_{38}} - {{\Omega }_{40}} + {{\Omega }_{42}} + {{\Omega }_{44}} +
  {{\Omega }_{46}} + {{\Omega }_{48}} + {{\Omega }_{50}} - {{\Omega }_{52}}\right)
\nonumber\\
\varpi[ 20 ] &
 \{ 1, 2, 2,
      2\} & {y_1} - {y_2} + {y_3}  +
{y_4} & -\frac{1}{4}\left( -{{\Omega }_9} + {{\Omega }_{14}} - {{\Omega }_{19}} - {{\Omega }_{34}}\right)  \nonumber\\
\varpi[ 21 ] &
 \{ 1, 2, 3,
      2\} & {y_1} + {y_4}\ & -\frac{1}{2\sqrt{2}}\left( -{{\Omega }_{38}} + {{\Omega }_{40}} + {{\Omega }_{42}} + {{\Omega }_{44}} +
  {{\Omega }_{46}} + {{\Omega }_{48}} - {{\Omega }_{50}} + {{\Omega }_{52}}\right)
\nonumber\\
\varpi[ 22 ] &
 \{ 1, 2, 4,
      2\} & {y_1} + {y_2} - {y_3}  +
{y_4} & -\frac{1}{4}\left( {{\Omega }_3} + {{\Omega }_8} + {{\Omega }_{23}} - {{\Omega }_{33}}\right)  \nonumber\\
\varpi[ 23 ] &
 \{ 1, 3, 4,
      2\} & {y_1} + {y_2} + {y_3}  +
{y_4} & -\frac{1}{4\sqrt{2}}\left( {{\Omega }_2} - {{\Omega }_7} + {{\Omega }_{18}} + {{\Omega }_{32}}\right)  \nonumber\\
\varpi[ 24 ] &
 \{ 2, 3, 4,
      2\} & 2\, {y_4} & -\frac{1}{8}\left( {{\Omega }_{16}} + {{\Omega }_{21}} + {{\Omega }_{25}} - {{\Omega }_{36}}\right)
\nonumber\\
\hline
\end{array}
$$}
\caption{Listing of the step operators corresponding to the positive roots
of $\mathrm{F}_{4(4)}$.}
  \label{tablettona}
\end{table}
The steps operators corresponding to negative roots are obtained from
those associate with positive ones  via the following relation:
\begin{equation}
  E^{-\varpi} = - \mathcal{C} \, E^{\varpi} \, \mathcal{C}
\label{CEC}
\end{equation}
where the $26 \times 26$ symmetric matrix $\mathcal{C}$ is defined in
the following way:
\begin{equation}
  \mathcal{C} \, = \, \left(\begin{array}{c|c|c}
    \mathbf{1 }& 0 & 0 \\
    \hline
    0 & \eta & 0 \\
    \hline
    0 & 0 & C_+ \
  \end{array} \right)
\label{Ccalmat}
\end{equation}
A further comment is necessary about the normalizations of the step
operators $E^{\varpi}$ which are displayed in table \ref{tablettona}.
They have been fixed with the following criterion. Once we have
constructed the algebra, via the generators (\ref{TAB}),(\ref{Qgen}),
we have the Lie structure constants encoded in eq.(\ref{F4comme}) and
hence we can diagonalize the adjoint action of the Cartan generators
(\ref{CSAbasa}) finding which linear combinations of the remaining
generators correspond to which root. Each root space is
one-dimensional and therefore we are left with the task of choosing an
absolute normalization for what we want to call the step operators:
\begin{equation}
  E^{\varpi} = \lambda_\varpi \, \left( \mbox{linear combination of $\Omega$.s} \right)
\label{pacco}
\end{equation}
The values of $\lambda_\varpi$ are now determined by the following
non trivial conditions:
\begin{enumerate}
  \item The differences $\mathbb{H}^i = \left( E^{\varpi_i} -
  E^{-\varpi_i}\right)
  $ should close a subalgebra $\mathbb{H} \subset F_{4(4)}$, the
  maximal compact subalgebra $\mathrm{SU(2)_R \times Usp(6)}$
  \item The sums $\mathbb{K}^i = \frac{1}{\sqrt{2}}\left( E^{\varpi_i}+
  E^{-\varpi_i}\right)$ should span a $28$-dimensional
  representation of $\mathbb{H}$, namely the aforementioned
  $\mathbf{(2,14)}$ of $\mathrm{SU(2)_R \times Usp(6)}$
\end{enumerate}
We arbitrarily choose the first four $\lambda_\varpi$ associated with
simple roots and then all the others are determined. The result is
that displayed in table \ref{tablettona}.
Using the Cartan generators defined by eq.s (\ref{CSAbasa}) and the
step operators enumerated in table \ref{tablettona} one can calculate
the structure constants of $\mathrm{F}_{4(4)}$ in the Cartan-Weyl basis, namely:
\begin{eqnarray}
\left[ \mathcal{H}_i \, , \, \mathcal{H}_j\right ]  & =  & 0\nonumber\\
\left[ \mathcal{H}_i \, , \, E^\varpi \right]  & =  & \varpi^i \, E^\varpi \nonumber\\
\left[ E^\varpi \, , \, E^{-\varpi} \right]  & =  & \varpi \, \cdot \,  \mathcal{H}
\nonumber\\
\left[ E^{\varpi_i} \, , \, E^{\varpi_j} \right]  & =  &
\mathcal{N}_{\varpi_i , \varpi_j} \, E^{\varpi_i + \varpi_j}
\label{cartaweila}
\end{eqnarray}
in particular one obtains the explicit numerical value of the
coefficients $\mathcal{N}_{\varpi_i , \varpi_j}$, which, as it is well
known, are the only ones not completely specified by the components of
the root vectors in the root system. The result of this computation,
following from eq.(\ref{F4comme}) is that encoded in eq.s
(\ref{nalbeLL}, \ref{nalbeLS}, \ref{nalbeSS}) of the main text.
\par
As a last point we can investigate the properties of the maximal
compact subalgebra $\mathrm{SU(2)} \times \mathrm{Usp(6)} \subset
\mathrm{F_{4(4)}}$. As we know a basis of generators for this
subalgebras is provided by:
\begin{equation}
  H_i = \left( E^{\varpi_i} - E^{-\varpi_i} \right)
  \quad ; \quad (i=1 , \dots , 24)
\label{Hiusp6su2}
\end{equation}
but it is not a priori clear which are the generators of $\mathrm{SU(2)_R}$ and
which of $\mathrm{Usp(6)}$. By choosing a basis of Cartan generators
of the compact algebra and diagonalizing their adjoint action this
distinction can be established.
The generators of $\mathrm{SU(2)_R}$ are the following linear combinations:
\begin{eqnarray}
J_X & = & \frac {1}{4\sqrt{2}} \left(H_1 -H_{14} + H_{20} -H_{22} \right)  \nonumber\\
J_Y & = & \frac {1}{4\sqrt{2}}\left(H_5 + H_{11} - H_{18} + H_{23} \right) \nonumber\\
J_Z & = & \frac {1}{4\sqrt{2}} \left(-H_2 + H_{9} - H_{16} -H_{24} \right)
\label{su2generiF4}
\end{eqnarray}
close the standard commutation relations:
\begin{equation}
  \left[ J_i \, , \, J_j \right] = \epsilon_{ijk} \, J_k
\label{standonecom}
\end{equation}
and commute with all the generators of $\mathrm{Usp(6)}$. These
latter are displayed as follows.
\begin{equation}
  \begin{array}{ccc}
    \mathcal{H}_1^{(Usp6)} & = & - \frac{{H_2}}{2} - \frac{{H_9}}{2} + \frac{{H_{16}}}{2} - \frac{{H_{24}}}{2} \\
    \mathcal{H}_2^{(Usp6)} & = & - \frac{{H_2}}{2} + \frac{{H_9}}{2} + \frac{{H_{16}}}{2} + \frac{{H_{24}}}{2} \\
    \mathcal{H}_3^{(Usp6)} & = & \frac{{H_2}}{2} + \frac{{H_9}}{2} + \frac{{H_{16}}}{2} - \frac{{H_{24}}}{2} \
  \end{array}
\label{usp6carti}
\end{equation}
are the Cartan generators. On the other hand the nine pairs of generators
which are rotated one into the other by the Cartans with eigenvalues
equal to the roots of the compact algebra are the following ones
\begin{equation}
  \begin{array}{|rcl|rcl|}
  \hline
    W_1 & = & {H_{10}} & Z_1 & = & {H_7} \\
    W_2 & = & {H_4} & Z_2 & = & -{H_{13}} \\
    W_3 & = & {H_6} & Z_3 & = & -{H_3} \\
    W_4 & = & -{H_1} + {H_{14}} + {H_{20}} - {H_{22}} & Z_4 & = & - {H_5} - {H_{11}} - {H_{18}} + {H_{23}} \\
    W_5 & = & {H_{21}} & Z_5 & = & - {H_8} \\
    W_6 & = & {H_1} + {H_{14}} + {H_{20}} + {H_{22}} & Z_6 & = & {H_5} - {H_{11}} - {H_{18}} - {H_{23}} \\
    W_7 & = & - {H_1} - {H_{14}} + {H_{20}} + {H_{22}} & Z_7 & = & {H_5} - {H_{11}} + {H_{18}} + {H_{23}} \\
    W_8 & = & {H_{17}} & Z_8 & = & {H_{15}} \\
    W_9 & = & {H_{12}} & Z_9 & = & {H_{19}} \\
    \hline
  \end{array}
\label{wzgeneri}
\end{equation}

\end{document}